\documentclass[]{aa}
\usepackage[utf8]{inputenc}
\usepackage[english]{babel}
\usepackage{graphicx}
\usepackage{graphbox}
\usepackage{xcolor}
\usepackage[export]{adjustbox}
\usepackage{soul}
\usepackage{booktabs} 
\usepackage{amssymb,amsmath}
\usepackage{array}
\usepackage{siunitx}
\usepackage{physics}
\usepackage{natbib}
\usepackage{textgreek}
\usepackage{upgreek}
\usepackage{placeins}
\usepackage[autostyle]{csquotes}
\usepackage{paralist}
\usepackage[varg]{txfonts}
\usepackage[normalem]{ulem}
\usepackage{url}
\usepackage[unicode=true]{hyperref}
\bibpunct{(}{)}{;}{a}{}{,} 
\hypersetup{
    colorlinks=true,
    linkcolor={blue!50!black},
    citecolor={blue!50!black}, 
    urlcolor={blue!80!black}   
}
\usepackage{etoolbox}
\makeatletter
\renewcommand*\aa@pageof{, page \thepage{} of \pageref*{LastPage}}
\makeatother

\begin{document}

\title{A solar rotation signature in cosmic dust: frequency analysis of dust particle impacts on the Wind spacecraft}
\titlerunning{A solar rotation signature in cosmic dust}

\author{L.~R.~Baalmann\inst{\ref{eth}}, S.~Hunziker\inst{\ref{eth}}, A.~Péronne\inst{\ref{eth}}, J.~W.~Kirchner\inst{\ref{eth2},\ref{wsl}}, K.-H.~Glassmeier\inst{\ref{tub},\ref{mpi}}, D.~M.~Malaspina\inst{\ref{ucb1},\ref{ucb2}}, L.~B.~Wilson~III\inst{\ref{goddard}}, C.~Strähl\inst{\ref{eth}}, S.~Chadda\inst{\ref{ucb2},\ref{ucb3}}, and V.~J.~Sterken\inst{\ref{eth}}}

\institute{ETH Zürich, Institute for Particle and Astroparticle Physics, 8093 Zürich, Switzerland\label{eth}\\
e-mail: \href{mailto:lbaalmann@phys.ethz.ch}{\texttt{lbaalmann@phys.ethz.ch}} \and
ETH Zürich, Department of Environmental Systems Science, 8092 Zürich, Switzerland\label{eth2} \and
Swiss Federal Research Institute WSL, 8903 Birmensdorf, Switzerland\label{wsl} \and
Technische Universität Braunschweig, Institute of Geophysics and Extraterrestrial Physics,  38106 Braunschweig, Germany\label{tub} \and
Max Planck Institute for Solar System Research, 37077 Göttingen, Germany\label{mpi} \and
University of Colorado, Boulder, Astrophysical and Planetary Sciences Department, Boulder, CO 80309, USA\label{ucb1} \and
University of Colorado, Boulder, Laboratory for Atmospheric and Space Physics, Boulder, CO 80309, USA\label{ucb2} \and
NASA Goddard Space Flight Center, Greenbelt, MD 20771, USA\label{goddard} \and
University of Colorado, Boulder, Department of Physics, Boulder, CO 80309, USA\label{ucb3}}

\date{Received 22 March 2024 / Accepted 6 July 2024}

\abstract{}{Dust particle impacts on the Wind spacecraft were detected with its plasma wave instrument Wind/WAVES. Frequency analysis on the resulting dust impact time series has revealed spectral peaks indicative of a solar rotation signature. We investigated whether this solar rotation signature is embedded in the interplanetary or in the interstellar dust (ISD) and whether it is caused by co-rotating interaction regions (CIRs), by the sector structure of the interplanetary magnetic field (IMF), or by external effects.}{We performed frequency analysis on different subsets of the data to investigate the origin of these spectral peaks, comparing segments of Wind's orbit when the spacecraft moved against or with the ISD inflow direction and comparing the time periods of the ISD focusing phase and the ISD defocusing phase of the solar magnetic cycle. A superposed epoch analysis of the number of dust impacts during CIRs was used to investigate the systematic effect of CIRs. Case studies of time periods with frequent or infrequent occurrences of CIRs were performed and compared to synthetic data of cosmic dust impacts affected by CIRs. We performed similar case studies for time periods with a stable or chaotic IMF sector structure. The superposed epoch analysis was repeated for a time series of the spacecraft floating potential.}{Spectral peaks were found at the solar rotation period of ${\sim}27\,\si{d}$ and its harmonics at $13.5\,\si{d}$ and $9\,\si{d}$. This solar rotation signature may affect both interplanetary and interstellar dust. The appearance of this signature correlates with the occurrence of CIRs but not with the stability of the IMF sector structure. The CIRs cause, on average, a reduction in the number of dust impact detections. Periodic changes of the spacecraft's floating potential were found to partially counteract this reduction by enhancing the instrument's sensitivity to dust impacts; these changes of the floating potential are thus unlikely to be the cause of the solar rotation signature.}{} 
\maketitle

\section{Introduction}\label{sec:intro}

Cosmic dust is highly prevalent in the Solar System \citep[e.g.][]{brownlee1985}. Depending on its origin, one distinguishes between interplanetary dust particles \citep[IDPs; e.g.][]{gruen+2001} and interstellar dust \citep[ISD; e.g.][]{gruen+1993,sterken+2019}. 
As the name suggests, IDPs originate in the Solar System. They are, for example, evaporated from comets, created through collisions of asteroids, or ejected from active moons. In contrast, ISD enters the heliosphere from the local interstellar medium. As such, ISD grains have hyperbolic orbits and are generally faster than the local escape speed \citep{gruen+1994}, whereas IDPs are typically bound to elliptic orbits around the Sun (although IDPs can also reach sufficient speeds to escape the Solar System as \textbeta-meteoroids; \citealp[e.g.][]{zook+1975,wehry+1999,czechowski+2010}). Because the polarity of the interplanetary magnetic field (IMF) changes periodically with the $22\,\si{yr}$ solar magnetic cycle, ISD is alternately focused towards or defocused away from the ecliptic plane with a $22\,\si{yr}$-periodicity \citep{gustafson+1979,sterken+2012}.

In situ, both IDPs and ISD are often jointly measured with dedicated dust detectors. The discovery of ISD in the Solar System, for example, was made with the dedicated dust detector onboard \textit{Ulysses} \citep{gruen+1993}. However, dust impacts on spacecraft can also be registered with instruments not intended for this purpose, such as plasma wave instruments. 

Measurements of dust impacts with plasma wave antennas have been made, for example but not limited to, at Saturn on board Voyager 2 \citep{aubier+1983, gurnett+1983} and \textit{Cassini} \citep[e.g.][]{kurth+2006}, during cometary approaches by the International Cometary Explorer \citep{gurnett+1986} and by Deep Space 1 \citep{tsurutani+2004}, as well as at Mars on board Mars Atmosphere and Volatile Evolution \citep{andersson+2015}, at Earth by Cluster II \citep{vaverka+2017} and the Magnetospheric Multiscale missions \citep{vaverka+2018}, by both spacecraft of the Solar Terrestrial Relations Observatory \citep[STEREO; e.g.][and references thereof]{meyervernet+2009}, and in close vicinity to the Sun by Parker Solar Probe \citep{page+2020} and Solar Orbiter \citep{zaslavsky+2021}.

Of particular interest for dust detections with plasma wave instruments is the Wind spacecraft (see \citealp{wilson+2021} for a comprehensive review). The Wind mission was launched in 1994 to investigate the solar wind and its plasma processes in near-Earth space. In its almost 30 years of service, Wind's electric field instrument, Wind/WAVES \citep{bougeret+1995}, has indirectly measured impacts of cosmic dust on the spacecraft. 
This was first reported by \citet{malaspina+2014}, who found that the daily number of dust impacts is correlated with Wind's orbital direction of motion around the Sun: More dust impacts were measured when the spacecraft moved against the preferential inflow direction of ISD than when it moved with the ISD inflow direction.

Some signals recorded by the two STEREO spacecraft were identified as impacts of nanometre-sized dust particles \citep[\enquote{nanodust};][]{zaslavsky+2012}; however, this was later contested \citep[e.g.][]{kellogg+2018}. \citet{kellogg+2016} analysed the waveforms of the dust impacts on Wind in detail and concluded that the instrument is not sensitive to impacts of nanodust. They also found that the ISD impacts measured by Wind and STEREO were consistent when Wind and STEREO A/B were close to each other. 
A database of dust impacts on Wind was published by \citet{malaspina+wilson2016} and is updated every few years. 

This publication reports on the discovery of solar rotation signatures in dust impact data measured by plasma wave antennas (Sect.~\ref{sec:spectrum}). The discovery of these signatures leads to the following investigations:
\begin{itemize}
    \item In Sect.~\ref{sec:ana_isd-idp} we discuss whether the solar rotation signatures stem from the interstellar or the interplanetary dust population, or from both.
    \item In Sect.~\ref{sec:ana_cirs} we investigate whether the solar rotation signatures are imprinted on the dust detections by co-rotating interaction regions (CIRs).
    \item In Sect.~\ref{sec:ana_hcs} we discuss whether the solar rotation signatures are imprinted on the dust detections by the IMF sector structure or crossings of the heliospheric current sheet.
    \item In Sect.~\ref{sec:ana_instrumental} we consider whether the solar rotation signatures are not dust signatures at all but caused by external effects.
\end{itemize}

One physical mechanism that may cause the solar rotation signatures is a local reduction or enhancement of dust particles whenever a CIR passes by the spacecraft. A similar dust depletion mechanism has been proposed in the past to occur close to the Sun during coronal mass ejections \citep[CMEs;][]{ragot+2003}; this has been indirectly observed by \citet{stenborg+2023} using Parker Solar Probe. Numerical simulations have found that this effect can cause either a reduction or an enhancement of the local dust density at $1\,\si{AU}$ \citep{wagner+2009,obrien+2018}; a depletion of observed dust impacts on Wind coinciding with CMEs was discovered by \citet{stcyr+2017}. A superposed epoch analysis of dust impacts measured by Wind during CIRs is performed in Sect.~\ref{sec:depletion} to investigate the effect of CIRs on the local dust environment measured by Wind.  

Another possible mechanism that may cause the solar rotation signatures could be a periodic deflection of dust particles by the alternating IMF sector structure, which has been proposed as the origin of Jovian dust streams by \citet{hamilton+1993}. \citet{hsu+2010} and \citet{flandes+2011} report that both the IMF sector structure and CIRs act on Saturnian and Jovian dust streams, causing strong enhancements of nanodust particle measurements with the dust detectors on board \textit{Cassini} and \textit{Ulysses}, respectively. However, the particles of Jovian and Saturnian dust streams are assumed to have radii of roughly ${\sim}10\,\si{nm}$ \citep{zook+1996,hsu+2011}, which is smaller than the approximately submicron-sized range to which Wind is assumed to be sensitive \citep{malaspina+2014}.
Spectral signatures of the solar rotation have also been found in the dust impact data measured by the STEREO spacecraft (Chadda et al., in prep.).

\section{Dynamics of cosmic dust in the Solar System}\label{sec:background}

Cosmic dust in the Solar System is primarily affected by three forces: solar gravity, solar radiation pressure (SRP), and the Lorentz force (Sect.~\ref{sec:forces}). Of these forces, only the Lorentz force, through variations in the IMF, changes on timescales that can cause the solar rotation signatures. Therefore, the most salient properties of the IMF are presented in Sect.~\ref{sec:imf}, including periodic changes induced by the IMF sector structure (Sect.~\ref{sec:hcs}) and by CIRs (Sect.~\ref{sec:cir}).

\subsection{Primary forces acting on cosmic dust in the Solar System}\label{sec:forces}

In heliocentric coordinates, both solar gravity and SRP are radially directed and thus are often combined into one term: 
\begin{equation}
    \va*{F}_{\mathrm{G \& SRP}} = - \frac{(1-\beta)\, G M_{\odot}\, m_{\mathrm{d}}}{r^2} \vu*{e}_r \ ,
\end{equation}
where $G$ is the gravitational constant, $M_{\odot}$ the solar mass, $m_{\mathrm{d}}$ the mass of the dust particle, $r$ the heliocentric distance, $\vu*{e}_r$ the unit vector radially pointing away from the Sun, and $\beta\equiv\abs{\va*{F}_{\mathrm{SRP}}}/\abs{\va*{F}_{\mathrm{G}}}$ is the ratio of SRP to solar gravity. The \textbeta-ratio is, for a given dust particle, a constant that depends on properties of the particle such as its composition, mass, and morphology \citep{sterken+2019}. For a given species of dust particles, the \textbeta-ratio is often given as a function of the particle mass or size (\citealp[e.g.][Fig.~3]{gustafson1994}; see also Fig.~\ref{fig:beta}). The solar irradiance, and through it the SRP, only changes by about $0.1\%$ over the solar $11\,\si{yr}$-cycle \citep{froehlich+1998}; this variance is disregarded in this study. Similarly, Poynting-Robertson drag is only relevant for long-term dynamics of IDPs and is, thus, not considered \citep{robertson1937,altobelli2004}.

The third relevant force acting on dust particles is the Lorentz force, which in terms of its acceleration is given by
\begin{equation}\label{eq:lorentz}
    \va*{a}_{\mathrm{L}} = \frac{q_{\mathrm{d}}}{m_{\mathrm{d}}} \qty(\va*{v}_{\mathrm{d}}-\va*{v}_{\mathrm{sw}}) \times \va*{B}_{\mathrm{sw}} \ ,
\end{equation}
where $q_{\mathrm{d}}/m_{\mathrm{d}}$ is the dust particle's charge-to-mass ratio, $\va*{v}_{\mathrm{d}}$ is the velocity of the dust particle, $\va*{v}_{\mathrm{sw}}$ is the velocity of the expanding solar wind, and $\va*{B}_{\mathrm{sw}}$ is the magnetic field vector of the IMF. 

Of these three forces, only the Lorentz force can cause short-term modulations of the dust environment that can lead to the observed solar rotation signatures (Sect.~\ref{sec:spectrum}). Assuming that the mass of a dust particle is constant, these variations in the Lorentz force must be caused by a change of the dust particle's surface charge, the solar wind speed, or the IMF.

The surface charge, $q_{\mathrm{d}}$, of a submicrometer dust particle in the IMF corresponds to a constant surface potential of roughly $+5\,\si{V}$ \citep{mukai1981}, and is higher for nanodust particles due to the \enquote*{small particle effect} \citep{watson1973}. Although nanodust has charging timescales of multiple days \citep{ma+2013}, for submicrometer dust particles the charge can vary at $1\,\si{AU}$ on a timescale of minutes \citep{sterken+2022}. The major contributions to the charging environment in the inner Solar System are solar UV radiation, which is assumed to be reasonably constant, and the solar wind. 
Therefore, knowing the spatial and time evolution of the solar wind and, especially, the IMF is essential for understanding the dynamics of cosmic dust in the heliosphere.

\subsection{The large-scale interplanetary magnetic field}\label{sec:imf}

The large-scale IMF is typically described by an archimedean spiral; the magnetic field lines are \enquote*{frozen} into the radially expanding solar wind and are wound up through the Sun's rotation \citep{parker1958}. This \enquote*{Parker spiral} has no polar component; its radial component is dominant close to the Sun, and its azimuthal component is dominant at larger heliocentric distances: at $1\,\si{AU}$ the angle between the IMF and the radial direction is about $\ang{45}$, whereas close to Jupiter this angle has increased to roughly $\ang{80}$ \citep{owens+2013}. Close to the solar equatorial plane, the Parker spiral lies almost parallel to the equatorial plane. At higher latitudes, however, its radial component is significantly tilted with respect to the equatorial plane; the IMF gains a component along the $z$-axis of the solar ecliptic coordinate system.\footnote{The ecliptic plane, in which Earth and the Wind spacecraft revolve around the Sun, is itself inclined by $\ang{7}$ with respect to the solar equatorial plane; therefore, the Parker spiral can have a minor $z$-component even within the ecliptic plane. This is ignored within this study.} At large heliocentric distances, where the azimuthal component is dominant over the radial, this $z$-component is negligible.

In simplified terms, the solar magnetic field can be described by a magnetic dipole that is at solar minimum roughly aligned with the solar rotation axis; thus, the field lines of the IMF point towards the Sun in one solar hemisphere and away from it in the other. Accordingly, a large component of the IMF points into the azimuthal direction in one hemisphere and into the anti-azimuthal direction in the other. Charged particles moving parallel to the ecliptic plane, such as ISD, thus experience a Lorentz force that in both hemispheres points either towards the ecliptic plane or away from it, depending on which polarity lies in which hemisphere. Thus, ISD particles are either focused towards or defocused away from the ecliptic plane \citep{morfill+1979,landgraf2000,sterken+2012}. 

The polarity of the IMF flips with the solar $11\,\si{yr}$ cycle, causing alternating focusing phases and defocusing phases of ISD within the solar magnetic $22\,\si{yr}$-cycle \citep{sterken+2012}. Other periodic changes of the IMF correspond to its sector structure (Sect.~\ref{sec:hcs}) and to CIRs (Sect.~\ref{sec:cir}), both of which are associated with the solar rotation period, which is ${\sim}27\,\si{d}$ for an observer on or at Earth.

\subsection{The heliospheric current sheet and interplanetary magnetic field sector structure}\label{sec:hcs}

The interface between the regions of opposite IMF polarities is referred to as the \enquote*{heliospheric current sheet} (HCS). If the solar magnetic field were perfectly described by a dipole that is aligned with the solar rotation axis, the HCS would be a flat plane identical to the solar equatorial plane. However, even at solar minimum when the solar magnetic field is well-approximated by a dipole, the dipole axis and the rotation axis are slightly tilted with respect to each other. This warps the HCS into what is commonly called the \enquote*{ballerina skirt} \citep[e.g.][]{jokipii+1981}. Within the ecliptic plane, this ballerina skirt is apparent as sectors of opposing magnetic polarity; the HCS constitutes the sector boundaries. This \enquote*{sector structure} co-rotates with the Sun \citep[e.g.][]{owens+2013}.

When the solar magnetic field is approximated reasonably well by a magnetic dipole close to solar minimum, the IMF features a two-sector structure. As the solar cycle progresses and higher multipole moments of the magnetic field increase in strength, the IMF often changes from a two-sector structure to a four-sector structure. These four sectors are generally not identically sized \citep[e.g.][]{richardson2018}. Due to further warping of the HCS, an orbiting object such as Earth or Wind can experience more than four sector boundary crossings or even skim the HCS for an extended period of time \citep{owens+2013}.

The IMF sector structure is known to imprint the solar rotation period on some cosmic dust particles like Jovian and Saturnian dust streams: because the IMF features opposite polarities on each side of the HCS, the Lorentz force points towards the ecliptic plane on one side of the sector boundary and away from it on the other, affecting charged particles such as ISD and IDPs. This was observed by \textit{Ulysses} for Jovian dust streams: in one sector of the IMF a collimated stream of dust particles would move northward through the ecliptic plane, and in the other sector it would move southward through the ecliptic plane; thus, a spacecraft residing in the ecliptic plane would encounter streams of particles twice per solar rotation period. At higher latitudes, the spacecraft would only encounter the northernmost or southernmost point of inflection of the particle stream, which would occur only once per solar rotation \citep{hamilton+1993}.

The timing of the Jovian dust streams observed by \textit{Ulysses} was furthermore associated with CIRs (see Sect.~\ref{sec:cir}) by \citet{flandes+2011}; Table~1 of that reference indicates that successive Jovian dust streams may be separated by roughly half a solar rotation period at low jovigraphic latitudes and by a full solar rotation period at higher jovigraphic latitudes, which agrees with the scenario proposed by \citet{hamilton+1993}. 

Continuously updated lists of sector boundary crossings of Earth and of the IMF sector structure are provided by \citet{svalgaard2023,svalgaard2023_struc}. These lists are used to investigate whether the solar rotation signatures are caused by the IMF sector structure in Sect.~\ref{sec:ana_hcs}.

\subsection{Co-rotating interaction regions}\label{sec:cir}

How tightly the IMF is coiled around the Sun depends on the speed of the solar wind, $v_{\mathrm{sw}}$. Close to the ecliptic plane, the slow solar wind is emitted from the Sun's streamer belt with $v_{\mathrm{sw}}\approx 400\,\si{km/s}$ at $1\,\si{AU}$, resulting in an angle between the IMF and the radial direction of about $\ang{45}$ at $1\,\si{AU}$. A fast solar wind is emitted from coronal holes, reaching speeds of $v_{\mathrm{sw}}\approx 750\,\si{km/s}$ and angles of about $\ang{30}$ at $1\,\si{AU}$ between the IMF and the radial direction; the Parker spiral is wound less tightly for the fast solar wind compared to the slow solar wind. Thus, the slow and the fast solar wind must interface, generating \enquote*{stream interaction regions} (SIRs). If SIRs persists for multiple solar rotations, they are referred to as \enquote*{co-rotating interaction regions} \citep{richardson2018}.

The SIRs consist of compression regions, where the magnetic field strength and plasma density are increased, followed by rarefaction regions, where the magnetic field strength and plasma density relax to their unperturbed values \citep{richardson2018}. Therefore, inside a SIR the Lorentz force is enhanced, accelerating dust particles far more than the unperturbed slow solar wind does. This has been reported by \citet{hsu+2010} for Saturnian dust stream particles. As mentioned before, the Jovian dust streams have also been associated with CIRs by \citet{flandes+2011}.

\citet{jian+2006} measured 365 encounters of Wind with SIRs from 1995 until 2004, about half of which persisted for multiple months (CIRs). This list was updated until 2009 by \citet{jian2009}. \citet{hajra+2022} measured 290 CIRs encountered by Earth from 2008 until 2019 and confirm that CIRs are most common during the declining phase of the solar cycle and rarest close to solar maximum. These lists are used to investigate whether the solar rotation signatures are caused by CIRs in Sect.~\ref{sec:ana_cirs}.

\section{Dust impact measurements on Wind}\label{sec:theory}

A brief overview over the Wind spacecraft and the Wind/WAVES instrument with which dust impacts are measured is given in Sect.~\ref{sec:windwaves}. Section~\ref{sec:amplitude:main} describes how the measured signal amplitude of a dust impact depends on the dust particle's mass and speed, which is relevant for the physical interpretation of the data. The dust impact database and corrections for some instrumental effects are introduced in Sect.~\ref{sec:database}. A brief overview of long-term patterns of the daily number of dust impacts is presented in Sect.~\ref{sec:overview}. 

\subsection{The Wind spacecraft and the WAVES instrument}\label{sec:windwaves}

\begin{figure}
    \centering
    \includegraphics[width=\linewidth]{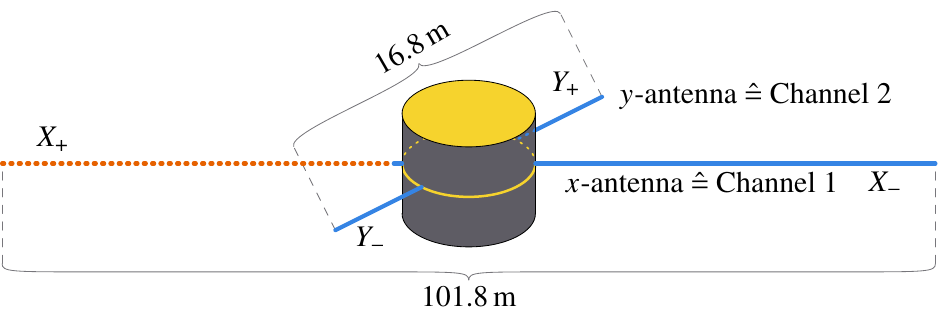}
    \caption{Schematic drawing of the actively used Wind/WAVES antennas (thick blue lines) on the Wind spacecraft (grey and gold cylinder), not drawn to scale. The majority of the $X_+$-arm was cut in 2000 and 2002 (dotted red line).}
    \label{fig:wind}
\end{figure}

Since its launch on 1 November 1994, Wind has performed numerous maneuvers in various regions of near-Earth space up to geocentric distances of $0.01\,\si{AU}$ \citep{wilson+2021}. Since June 2004 it has been orbiting $\mathrm{L}_{1}$. 

The Wind spacecraft is spin-stabilised with a period of approximately $3\,\si{s}$; the spin axis points towards ecliptic south. Within the spin plane lie two dipole antennas, referred to as the $x$- and the $y$-antenna, or as Channel~1 and Channel~2, respectively. The dipoles' arms are referred to as $X_{\pm}$ and $Y_{\pm}$ (see Fig.~\ref{fig:wind}). These dipole antennas are oriented perpendicular to each other and to the spacecraft surface. Both dipoles consist of $0.3\,\si{mm}$-thin wire; the $x$-dipole had, at launch, a length of $101.8\,\si{m}$ tip-to-tip, whereas the $y$-dipole was much shorter with a tip-to-tip length of $16.8\,\si{m}$ \citep{malaspina+wilson2016}. 

The $X_{+}$-arm was shortened by two dust impacts, first on 3 August 2000 and again on 25 September 2002 \citep{malaspina+2014}. The $X_{-}$-arm and the $Y_{\pm}$-arms remain undamaged to this day. The antenna cuts introduced significant asymmetries in the $x$-dipole's differential voltage measurements, bringing it functionally closer to a monopole than a dipole antenna, though the measurements are still differential between the two antenna arms \citep{kellogg+2016}. For this reason, it is advisable to use only the dust impacts observed by the $y$-antenna when the signal amplitude is of interest (see Sect.~\ref{sec:amplitude}). If only the number of impacts is of relevance, the impacts recorded by either antenna can be used. However, the response by the $x$-antenna to dust impacts differs before and after the antenna breaks; the two antennas do not have the same sensitivity thresholds and do not register the same amplitude distributions (see Fig.~\ref{fig:distr-breaks}). This must be accounted for.

Dust impacts on the spacecraft were measured with the fast time domain sampler (TDSF) of the Wind/WAVES instrument \citet{bougeret+1995}. The TDSF sampling rate is not held constant; since April 2011 it was periodically changed every six days to a coarser sampling rate for a duration of roughly two days. When the TDSF is set to this coarse sampling rate, no dust impacts can be measured. These periodically occurring gaps must be accounted for, as is described in Appendix~\ref{app:data_ratechange} and evaluated in Appendix~\ref{app:eval_6dgap}.

Several other instrumental effects and external influences that affected the measurements of dust impacts occurred before 2005 (see Appendix~\ref{app:lims_summary}). Furthermore, the many different orbits of the Wind spacecraft, especially when changing the distance to Earth, have a noticeable effect on the observations of dust impacts (see Appendix~\ref{app:data_dist_dep}). For data analyses on time scales longer than a few weeks (e.g.\ the frequency analyses performed in Sect.~\ref{sec:fourier}) it is, therefore, advisable to use only the dust impacts observed since 2005, when Wind continuously orbited $\mathrm{L}_1$. 

The physical processes that allow dust impact measurements via plasma wave antennas were investigated by \citet{shen+2021,shen+2023}. These processes and the resulting signals differ for dipole and monopole antennas. Most notably, dipole antennas measure a considerably stronger amplitude when a dust particle impacts closer to the antenna's base, unlike monopole antennas, which are not as sensitive to the distance between the impact site and the antenna's base \citep{shen+2023}. Furthermore, the measured amplitude depends on the floating potential of the spacecraft: for a monopole antenna, a higher floating potential would not strongly affect the amplitude of the impact signal's main peak, whereas a higher floating potential may severely reduce the amplitude of the measured main peak for a dipole antenna \citep{shen2023}.

\subsection{Mass and speed dependence of the impact signals}\label{sec:amplitude:main}

The amplitude of the signal that is generated by a dust impact depends on the charge, $Q$, released by the impact. This charge depends on the mass, $m$, and the impact speed, $v$, of the impacting particle with respect to the spacecraft \citep{dietzel+1973}:
\begin{equation}\label{eq:Qmv}
    Q = \gamma\, m\, v^{\alpha} \ .
\end{equation}
A typical value for the power of the impact speed dependence is $\alpha=3.5$ \citep[][Ch.~9.2]{balogh+2001}, depending on the target material impacted by the dust particle \citep{auer2001,collette+2014}. For dipole antennas, the measured signal amplitude additionally depends strongly on the distance between the impact site and the antenna \citep{shen+2023}. The constant of proportionality, $\gamma$, is generally not known. 
Because only signal amplitudes above a threshold of $4\,\si{mV}$ were considered when compiling the dataset \citep{malaspina+2014}, a faster relative speed between the spacecraft and the dust particles can lead to more detections of dust impacts: more charge is released for faster impacts, shifting the measurement threshold to lower masses.

\subsubsection{Impact speed of interstellar dust particles}\label{sec:amplitude:speed}

For ISD the impact speed depends on the orbital position of the spacecraft with respect to the ISD inflow direction and on the heliocentric speed of the ISD particle. Because ISD particles of different masses experience SRP at different strengths, i.e.\ because the \textbeta-ratio depends on the particle mass, the heliocentric speed of an ISD particle also depends on its mass. 

\begin{figure}
    \centering
    \includegraphics[width=\linewidth]{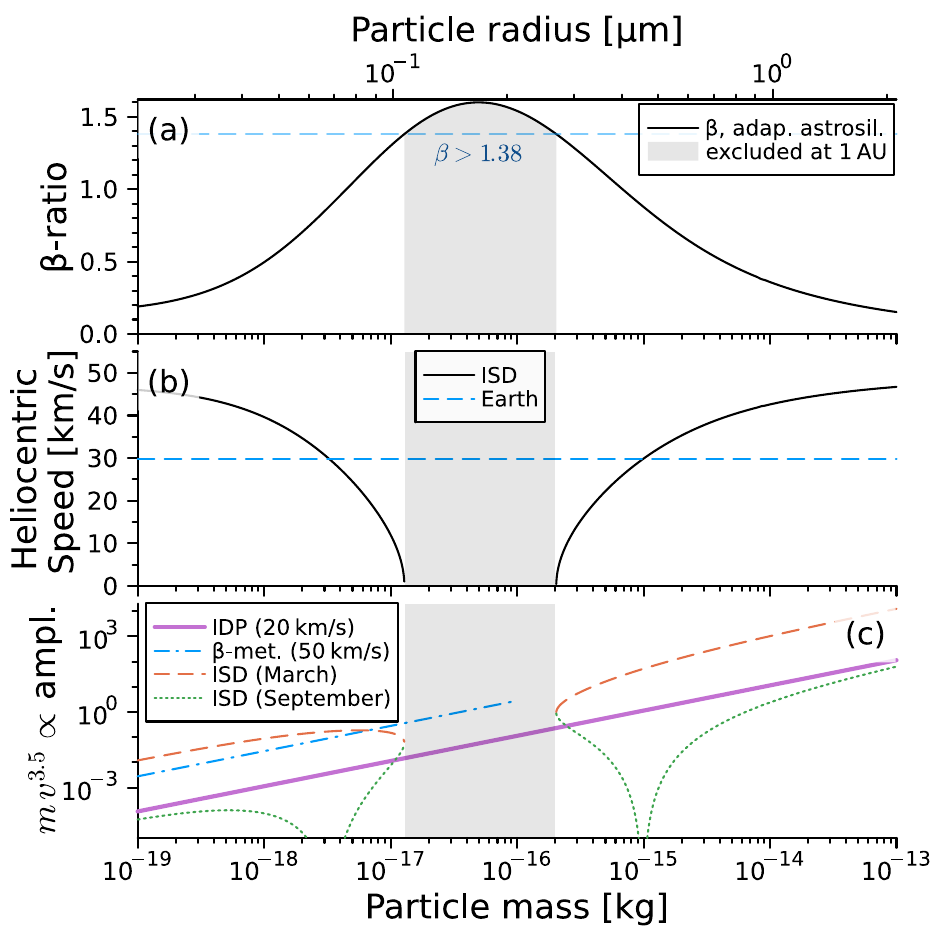}
    \caption{Relevant dust properties versus particle mass. 
    \textit{Top panel:} \textbeta-ratio versus particle mass following the astronomical silicates of \citet{gustafson1994} adapted to have a maximum of $\beta=1.6$ \citep{sterken+2013}.\\ 
    \textit{Middle panel:} Subsequent heliocentric speed of ISD particles (solid black curve) dependent on the particle mass using Eq.~(\ref{eq:visd}); Earth's orbital speed (dashed blue line) for comparison.\\ 
    \textit{Bottom panel:} Signal amplitude of a single dust particle, taken as $m\,v^{3.5}$ as per Eq.~(\ref{eq:Qmv}), for an IDP with an impact speed of $20\,\si{km/s}$ (solid magenta line), a \textbeta-meteoroid of $50\,\si{km/s}$ (dash-dotted blue line), and an ISD particle in March (dashed red curve) and in September (dotted green curve). No line was plotted for \textbeta-meteoroids above $m>10^{-16}\,\si{kg}$ because \textbeta-meteoroids are constrained to lower masses \citep{wehry+1999,moorhead2021}.\\
    The secondary horizontal axis of the top panel gives the particle radius, assuming spherical and compact particles with a density of $2500\,\si{kg/m^3}$ \citep{sterken+2013}. ISD with $\beta>1.38$ (dashed blue horizontal in the top panel) cannot reach Earth's orbit; it is excluded (grey-shaded area).}
    \label{fig:beta}
\end{figure}

This is depicted in Fig.~\ref{fig:beta}a and Fig.~\ref{fig:beta}b, showing the \textbeta-ratio and the heliocentric speed of an ISD particle in dependence of the particle's mass, respectively. The \textbeta-curve has been calculated using the astronomical silicates of \citet{gustafson1994} rescaled to have a maximum of $\beta=1.6$ \citep{sterken+2013}. For other materials, such as pure silicates, the \textbeta-curve can feature much lower maxima \citep[e.g.][]{kimura+1999}. The heliocentric speed of the dust particle has been calculated with
\begin{equation}\label{eq:visd}
    v_{\mathrm{ISD}} = \sqrt{\frac{2GM_{\odot}(1-\beta)}{a_{\otimes}}+v_{\infty}^2} \ ,
\end{equation}
where $G$ is the gravitational constant, $M_{\odot}$ is the solar mass, $\beta$ is the mass-dependent \textbeta-ratio, $a_{\otimes}=1\,\si{AU}$ is the orbital radius of Earth, and $v_{\infty}=26\,\si{km/s}$ is the ISD speed at infinity. The impact speed of the particle, under the simplifying assumption that the incoming ISD velocity vector is parallel to the ecliptic plane, is
\begin{equation}\label{eq:vrel}
    v_{\mathrm{rel}} = \sqrt{\qty(v_{\mathrm{ISD}}+v_{\otimes}\sin\phi)^2 + \qty(v_{\otimes}\cos\phi)^2} \ ,
\end{equation}
where $v_{\otimes}\approx 29.78\,\si{km/s}$ is the orbital speed of Earth and $\phi$ is the phase of the orbit ($\sin\phi=\pm1$ in March and September, respectively). The ISD inflow vector is not parallel to the ecliptic plane but instead inclined by about $\ang{5}$, coinciding with the interstellar Helium inflow direction \citep[][]{landgraf1998,strub+2015,swaczyna+2018}; this is neglected here.

Depending on the particle's \textbeta-ratio, ISD can reach a heliocentric speed of up to $49.5\,\si{km/s}$ ($\beta=0$), which results in an impact speed of almost $80\,\si{km/s}$ in March and only about $20\,\si{km/s}$ in September. This variation of the impact speed by a factor of four becomes a factor of 128 in amplitude as per Eq.~(\ref{eq:Qmv}). Generally, the ISD particles in the relevant mass regime, $m\ll 10^{-9}\,\si{kg}$, have higher \textbeta-ratios, $\beta>0$, and, thus, lower heliocentric speeds (Fig.~\ref{fig:beta}). Particles with $\beta\approx0.9$ have a heliocentric speed that is comparable to Earth's orbital speed. These particles would have an impact speed of about $60\,\si{km/s}$ in March and about $0\,\si{km/s}$ in September, i.e.\ they are not measurable in September. For $\beta>1.38$, ISD particles can no longer reach $1\,\si{AU}$, which corresponds to ISD particles within the mass range of $m_{\mathrm{ISD}}\in[1.3\times 10^{-17}, 2.0\times 10^{-16}]\,\si{kg}$ that cannot be observed with the Wind spacecraft throughout the year. Particles with $0.9<\beta<1.38$ have a slower heliocentric speed than Earth; in September, their relative velocity vectors point in the opposite direction compared to ISD with $\beta <0.9$. 

In terms of the particle radius, assuming spherical and compact particles with a density of $2500\,\si{kg}$ \citep{sterken+2015}, the \textbeta-gap for $\beta>1.38$ corresponds to the radius range of $a\in[0.11,0.27]\,\si{\micro m}$.

\subsubsection{Signal amplitude of individual particle impacts}\label{sec:amplitude:single}

Using the mass and the mass-dependent speed of an ISD particle, the amplitude that this individual particle would generate upon impacting the Wind spacecraft can be calculated, not accounting for the unknown proportionality constant, $\gamma$, of Eq.~(\ref{eq:Qmv}). Figure~\ref{fig:beta}c shows how the signal amplitude of an impacting particle changes with its mass. For IDPs, a relative speed of $20\,\si{km/s}$ is assumed \citep{gruen+1985}. The heliocentric speed of \textbeta-meteoroids can be significantly faster, commonly reaching $40\,\si{km/s}$ (\citealp[][Fig.~3.15]{wehry2002}; see also \citealp{wehry+2004,zaslavsky+2021}); this results in impact speeds of about $\sqrt{\qty(40\,\si{km/s})^2+\qty(30\,\si{km/s})^2}=50\,\si{km/s}$ because the approximately anti-sunward motion of \textbeta-meteoroids is perpendicular to Wind's orbit (however, see \citealp{wehry+1999} for \textbeta-meteoroids that are strongly deflected from anti-sunward trajectories). The signal amplitude for impacts of ISD was calculated with impact speeds as per Eqs.~(\ref{eq:visd}, \ref{eq:vrel}) for an impact in March and September, assuming the \textbeta-curve that is displayed in Fig.~\ref{fig:beta}a. 

As Fig.~\ref{fig:beta}c indicates, at a given mass ISD generates a much higher signal amplitude in March than an IDP. In September, the signal amplitude of ISD is slightly lower than for an IDP at very low or very high masses ($\beta\approx 0$), and considerably lower at intermediate masses ($\beta> 0$). Because the impact signals must exceed a certain amplitude threshold to be measured by the instrument, this should result in more detections of ISD impacts in March than in September: for a given mass, the amplitude of an ISD particle is much higher in March than it is in September; thus, in March less massive ISD particles can exceed the amplitude threshold. Furthermore, the particle flux increases with the relative speed; therefore, not only is the amplitude threshold exceeded by particles of lower mass in March compared to September, but more particles impact the spacecraft in March than in September. 

There is an exception to this trend: the relative speed of ISD in September exceeds the relative speed of IDPs for $1.32<\beta<1.38$ (see Fig.~\ref{fig:beta}c). For $\beta\approx 0.9$ ISD in September has almost the same heliocentric velocity as Earth; the relative speed is close to zero. The heliocentric speed of ISD in September decreases further the closer the particle is to the \textbeta-gap, $\beta\approx 1.38$, increasing the relative speed. For $1.32<\beta<1.38$ the relative speed exceeds $20\,\si{km/s}$; for $\beta\approx 1.38$ the heliocentric speed of ISD is close to zero and the relative speed stems mostly from Earth's orbital speed. The mass intervals corresponding to $1.32<\beta<1.38$ are small, $m_{\mathrm{ISD}}\in[1.06, 1.28]\times 10^{-17}\,\si{kg}\cup [2.05, 2.52]\times 10^{-16}\,\si{kg}$, compared to the vast mass intervals with $\beta<1.32$, and, thus, contain comparably few particles.

\begin{figure}
    \centering
    \includegraphics[width=\linewidth]{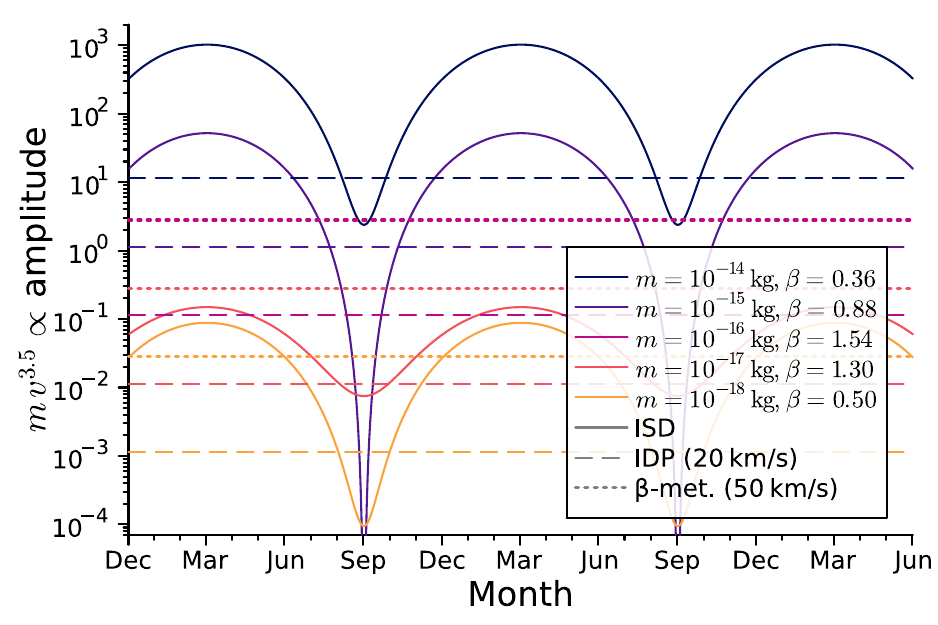}
    \caption{Signal amplitude of a single dust particle, taken as $m\,v^{3.5}$ as per Eq.~(\ref{eq:Qmv}), for an interstellar (solid curves) and interplanetary (dashed horizontal lines) dust particle and a \textbeta-meteoroid (dotted horizontal lines) of mass $m=10^{-18}\,\si{kg}$ (yellow lines), $m=10^{-17}\,\si{kg}$ (red lines), $m=10^{-16}\,\si{kg}$ (purple lines), $m=10^{-15}\,\si{kg}$ (violet lines), and $m=10^{-14}\,\si{kg}$ (indigo lines). The ISD with $m=10^{-16}\,\si{kg}$ (solid purple curve) cannot reach $1\,\si{AU}$ due to its high \textbeta-ratio; thus, this curve is not visible. No lines were plotted for \textbeta-meteoroids above $m>10^{-16}\,\si{kg}$. For more details, see Sect.~\ref{sec:amplitude:single}.}
    \label{fig:sim-amp-vs-time}
\end{figure}

The time dependence of the signal amplitudes is investigated further with the aid of Fig.~\ref{fig:sim-amp-vs-time}, which indicates how the signal amplitude of an impacting particle of mass $m$ changes with the orbital position of the spacecraft, given by the month of the calendar year. As before, a relative speed of $20\,\si{km/s}$ and $50\,\si{km/s}$ is assumed for IDPs and \textbeta-meteoroids, respectively. No curve is visible for ISD with $m_{\mathrm{ISD}}=10^{-16}\,\si{kg}$, which cannot reach $1\,\si{AU}$ due to its high \textbeta-ratio, and no curves were drawn for \textbeta-meteoroids above $10^{-16}\,\si{kg}$ due to the cutoff of their mass distribution \citep{wehry+1999,moorhead2021}. 

In March the ISD impact speed of up to $v_{\mathrm{rel}}\lesssim 80\,\si{km/s}$ is considerably higher than the assumed IDP impact speed of ca.\ $20\,\si{km/s}$. Therefore, IDPs that generate signals of comparable amplitude must be much more massive. For example, to generate the same signal amplitude as an ISD particle of $m_{\mathrm{ISD}}=10^{-15}\,\si{kg}$ in March, an IDP would have to be more than thirty times as massive, $m_{\mathrm{IDP}}\approx 4.6\times 10^{-14}\,\si{kg}$. In September, the same ISD particle would generate a signal amplitude more than ten orders of magnitude lower than in March; it would be undetectable.

The assumed impact speed of \textbeta-meteoroids ($50\,\si{km/s}$) is much higher than for IDPs ($20\,\si{km/s}$), resulting in signal amplitudes that are higher by a factor of ${\sim}25$. At a mass of $m=10^{-17}\,\si{kg}$ the signal amplitude of a \textbeta-meteoroid can be even higher than for an ISD particle impact in March. 

To summarise, due to the impact-speed-dependent amplitude threshold of the instrument, more ISD impacts are expected to be measurable when the spacecraft moves against the ISD inflow direction in March compared to when it moves with the ISD inflow in September. Section~\ref{sec:amplitude} will show that this seasonal variation of ISD impact detections is not apparent at all signal amplitudes, most likely due to the dissimilar mass distributions of ISD, IDPs, and \textbeta-meteoroids in the inner Solar System.

\subsubsection{Mass distribution of interstellar dust and interplanetary dust particles}\label{sec:amplitude:distribution}

\begin{figure}
    \centering
    \includegraphics[width=\linewidth]{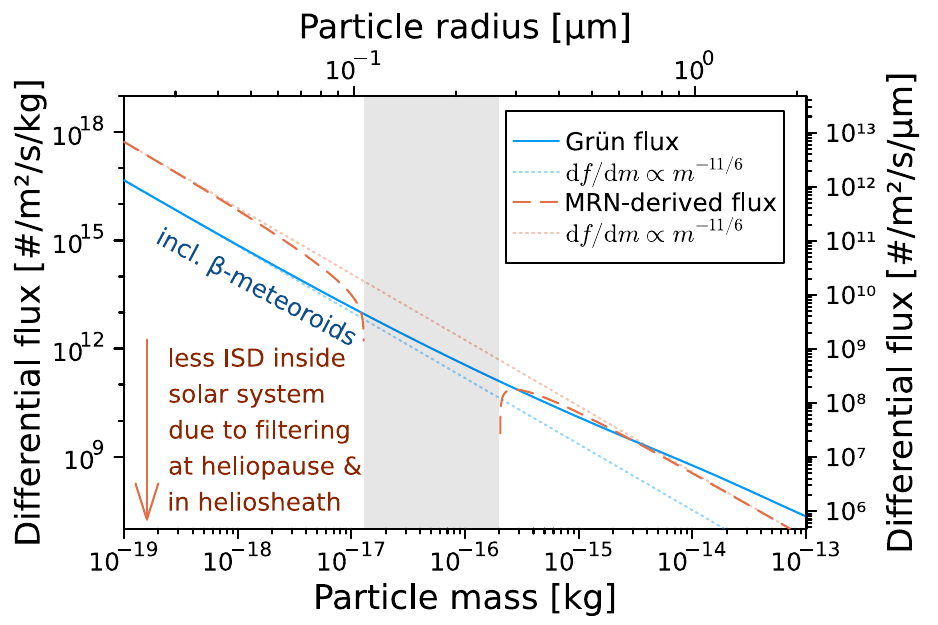}
    \caption{Differential flux of IDPs following the Grün flux (solid blue curve) and ISDs derived from the MRN flux (dashed red curve). The faint dotted lines indicate the deviation from the power law $\mathrm{d}f/\mathrm{d}m\propto m^{11/6}$.}
    \label{fig:massdist}
\end{figure}

The mass distribution of IDPs and \textbeta-meteoroids can be described by the \enquote*{Grün flux} \citep{gruen+1985}: in terms of a differential mass distribution, it follows a power law, $\mathrm{d}f/\mathrm{d}m\propto m^{-11/6}$, at low masses but features an excess with respect to that power law at masses above $m_{\mathrm{IDP}}>10^{-17}\,\si{kg}$. This has been graphed in Fig.~\ref{fig:massdist}. The contribution of the \textbeta-meteoroids to the Grün flux is mostly constrained to masses below $m_{\beta}<10^{-16}\,\si{kg}$; \citet{wehry+1999} find that the mass distribution of \textbeta-meteoroids is shifted to lower masses compared to IDPs, whereas ISD was rare at these low masses at the time of this study, during the defocusing phase of the solar magnetic cycle.

The differential mass distribution of ISD outside the heliosphere, the \enquote*{MRN distribution} \citep{mathis+1977}, follows the same power law, $\mathrm{d}f/\mathrm{d}m\propto m^{-11/6}$. In terms of the particle radius, this power law corresponds to $\mathrm{d}f/\mathrm{d}a\propto a^{-3.5}$. This has been graphed in Fig.~\ref{fig:massdist}, following the approach of \citet{draine+1984}, assuming a hydrogen number density in the ISM of $n_{\mathrm{H}}=0.1\,\si{cm^{-3}}$, a dust particle density of $2500\,\si{kg/m^3}$, and the heliocentric speed calculated via Eq.~(\ref{eq:visd}).

However, the ISD distribution is further modulated as ISD enters the heliosphere \citep{sterken+2013}, causing a notable deprivation of ISD particles below $m_{\mathrm{ISD}}\lesssim 10^{-15}\,\si{kg}$ compared to the power law and an outright exclusion at masses below $m_{\mathrm{ISD}}\lesssim 10^{-19}\,\si{kg}$ \citep{krueger+2015}, corresponding to particle radii of $a_{\mathrm{ISD}}\lesssim 0.45\,\si{\micro m}$ and $a_{\mathrm{ISD}}\lesssim 20\,\si{nm}$, respectively. This modulation is an active research topic (e.g.\ Hunziker et al., in prep.; Baalmann et al., in prep.), and, thus, the mass distribution of ISD inside the heliosphere is not known with high accuracy. 

To summarise, ISD is practically nonexistent in the inner Solar System at masses below $m_{\mathrm{ISD}}<10^{-19}\,\si{kg}$ due to filtering at the heliopause \citep{slavin+2012}, is excluded by SRP in the \textbeta-gap of $m_{\mathrm{ISD}}\in[1.3\times 10^{-17}, 2.0\times 10^{-16}]\,\si{kg}$, and is vanishingly rare at masses above $m_{\mathrm{ISD}}>10^{-13}\,\si{kg}$ \citep[cf.][]{krueger+2019} due to the MRN power law distribution and the limits on cosmic dust abundances from remote observations \citep[e.g.][]{frisch+1999}. During the defocusing phase, small ISD is furthermore defocused away from the ecliptic plane; in the focusing phase, it is focused towards it.

In contrast, IDPs follow the same power law at low masses but are not modulated by the heliopause, do not feature a \textbeta-gap, and show an excess compared to the power law at masses above $m_{\mathrm{IDP}}>10^{-17}\,\si{kg}$; \textbeta-meteoroids are mostly constrained to $m_{\beta}<10^{-16}\,\si{kg}$. While very small IDPs should also be affected by the solar magnetic cycle, these particles most likely have insufficient mass to be measurable as they impact Wind.

\subsection{Dust impact database}\label{sec:database}

\begin{figure}
    \centering
    \includegraphics[width=\linewidth]{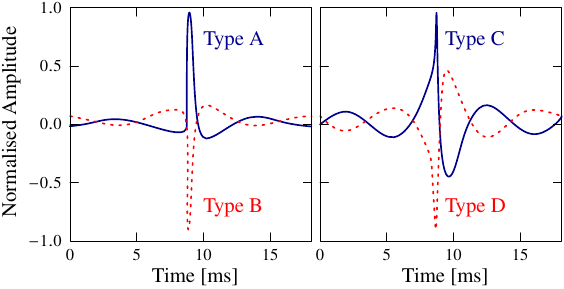}
    \caption{Stylised waveforms for the four morphological phenotypes of dust impact signals measured by the TDSF. Only impacts that generated single-spike signals (left panel, types A and B) are taken into account for the analyses. Reproduced with permission after \citet[][Fig.~2]{malaspina+wilson2016}; copyright of the original figure by John Wiley and Sons.}
    \label{fig:waveforms}
\end{figure}

The methodology used to compile the dataset of dust impacts on Wind is described by \citet{malaspina+wilson2016}. Dust impacts are identified by cross-correlating the time-resolved amplitude signal with four predetermined typical phenotypes of dust impact waveforms (types A to D; see Fig.~\ref{fig:waveforms}): if the cross-correlation between the signal and a given morphological phenotype exceeds a given threshold, the signal is identified as a dust impact. 

Morphological types A and B feature a single main peak in their waveform, whereas morphological types C and D follow the primary peak with an overshoot. The physical mechanism that causes these overshoots is not fully understood. Waveforms of type~B (D) are nearly identical to type~A (C) with a flipped sign, indicating that the corresponding dust impacts occurred closer to the opposite arm of the respective dipole antenna. In its current version, spanning the time period from 1995 until the end of August 2023, the dataset is publicly available on CDAWeb.\footnote{\texttt{WI\_L3-DUSTIMPACT\_WAVES} on \url{https://cdaweb.gsfc.nasa.gov}}

The dataset contains the millisecond-precise timestamp of each dust impact; the peak signal amplitudes of the $x$-antenna and of the $y$-antenna, which are also referred to as Channel~1 and Channel~2, respectively; the morphological type for each channel; and a location flag that denotes whether Wind was positioned within Earth's magnetosphere, the lunar wake, or neither.

A number of corrections were made to account for instrumental effects and similar limitations (see Appendix~\ref{app:lims_summary}). These are explained in detail in Appendix~\ref{app:reduction} and briefly summarised here:
\begin{itemize}
    \item Only the data since 1 January 2005 were used, when Wind continuously orbited $\mathrm{L}_1$. Earlier data show a geocentric distance dependence, and were affected by different data transfer rates (Appendix~\ref{app:data_dist_dep}). Both breaks of the $x$-dipole occurred before 2005. 
    \item The Wind/WAVES instrument was insensitive to dust impacts for about five months in 2013 and one month in 2014 (Appendix~\ref{app:data_removal}). Spurious events measured during these time intervals were discarded within the scope of this investigation.
    \item A periodic change of the TDSF sampling rate was introduced in April 2011, making the Wind/WAVES instrument insensitive to dust impacts for $45\,\si{h}\,36\,\si{min}$ every six days (Appendix~\ref{app:data_ratechange}). The full two days of data during which these measurement gaps occurred were removed. 
    \item The time series of dust impact signals of morphological types C and D have unexplained features, predominantly before 2005 \citep{malaspina+wilson2016}; furthermore, the amplitude distribution of dust impacts of types C and D differs from those of types A and B. Therefore, only signals of morphological types A and B were taken into account. Dust impacts of types C and D are briefly investigated in Appendix~\ref{app:type_cd}.
    \item When investigating the amplitude distribution of the dust impact signals (Sect.~\ref{sec:amplitude}), only dust impacts measured by the $y$-antenna were taken into account (see Fig.~\ref{fig:distr-breaks}).
\end{itemize}

\subsection{Overview of dust impacts at $\mathrm{L}_1$}\label{sec:overview}

\begin{figure*}
    \centering
    \includegraphics[width=\linewidth]{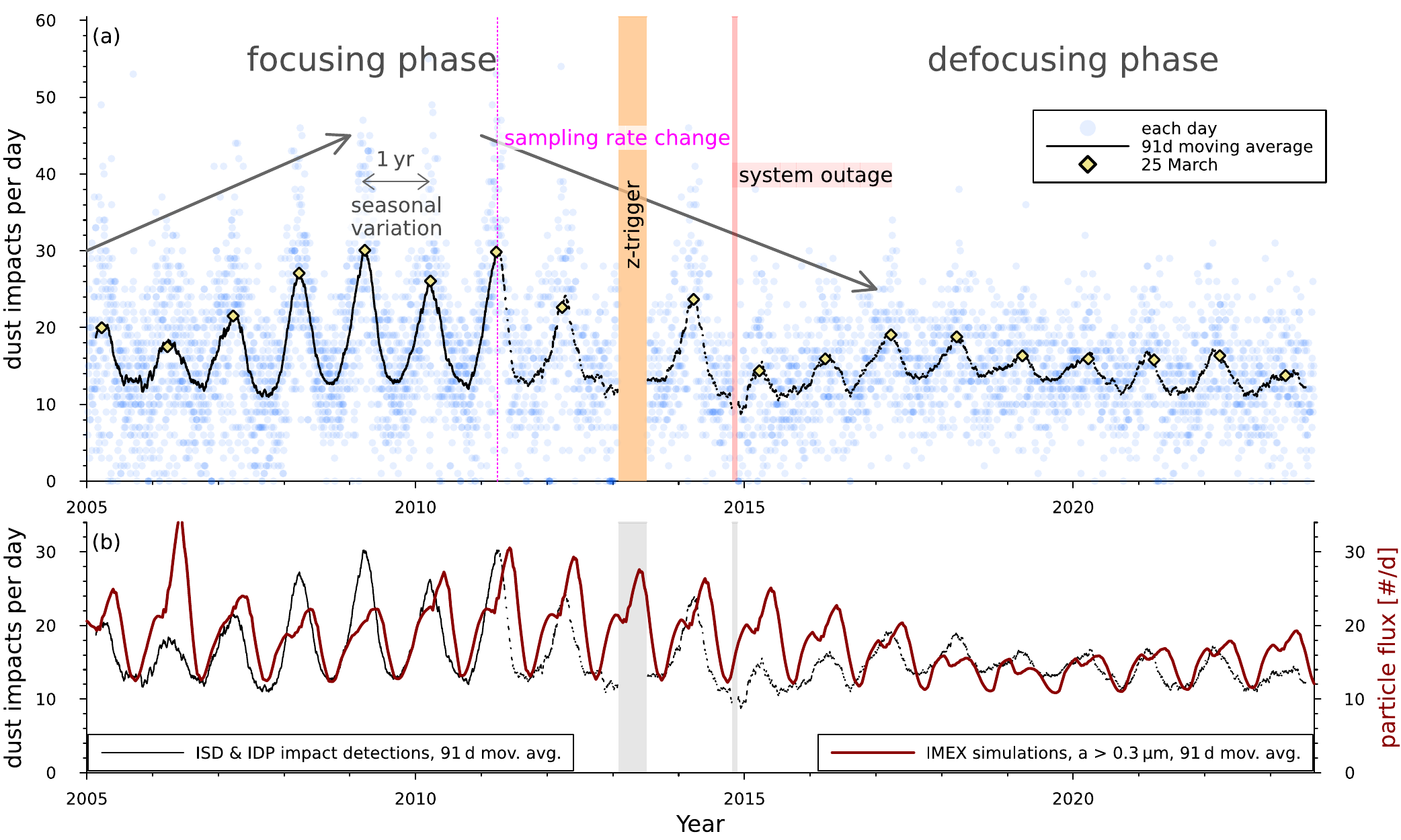}
    \caption{Dust impacts observed by Wind versus time. 
    \textit{Top panel:} Daily number of dust impacts on Wind at $\mathrm{L}_1$ that generated signals of morphological types A and B, given for each day (blue discs) and as a centred moving average with a width of $91\,\si{d}$ (black line). For comparison, yellow diamonds mark the moving average flux on 25 March of each year, which is close to each annual maximum. The sampling rate was periodically changed beginning in April 2011, marked by a vertical dotted pink line, making the instrument insensitive to dust impacts for roughly two days every six days. Larger time periods where the instrument was inoperable in regard to measuring dust impacts occurred in 2013 (shaded orange, \enquote*{$z$-trigger}) and in late 2014 (shaded red, \enquote*{system outage}). The increase and decrease of dust impacts due to the solar magnetic cycle is indicated by diagonal grey arrows. The dataset has been corrected for these instrumental effects as per Sect.~\ref{sec:database} (see Appendix~\ref{app:reduction}).\\
    \textit{Bottom panel:} Daily number of dust impacts observed by Wind as a $91\,\si{d}$ centred moving average (black curve, identical to the top panel) compared to the numerically simulated particle flux of ISD at Wind's orbital position, vertically offset by an assumed constant IDP flux of $10\,\si{\#/d}$. (For more information on the simulations, see Sect.~\ref{sec:simulations}.)}
    \label{fig:overview}
\end{figure*}

An overview of the observed dust impacts on Wind at $\mathrm{L}_1$ that generated signals of morphological types A and B, spanning the time interval from 1 January 2005 to 31 August 2023, is given in Fig.~\ref{fig:overview}a. Because the number of dust impacts shows a strong stochastic day-to-day variation, it is expedient to view a moving average of the data. A window width of a quarter year has proven to be a reasonable compromise between smoothing out the random variations without inadvertently suppressing long-term patterns. 

On top of a roughly constant level of about 12 impacts per day, these long-term patterns are dominated by a seasonal variation: in March of each year, dust impact detections occur at a higher rate compared to September of the same year. This corresponds to the orbital position of Wind with respect to the inflow direction of ISD; the spacecraft moves against the ISD inflow in March and with it in September of each year \citep[cf.][Fig.~4]{malaspina+2014}. A higher relative speed between the spacecraft and the ISD in March of every year directly results in a higher observed particle flux. Furthermore, as per Eq.~(\ref{eq:Qmv}), a faster relative speed can increase the signal amplitude that is measured by the instrument by multiple orders of magnitude, allowing for the detection of lower-mass particles that would otherwise not produce signal amplitudes above the instrument's detection threshold (Fig.~\ref{fig:sim-amp-vs-time}). 

IDPs revolve around the Sun on mostly prograde elliptical orbits and are not expected to show this seasonal variation. However, the plane of symmetry of IDPs is slightly inclined by $\ang{3.7}\pm\ang{0.6}$ with respect to the ecliptic plane \citep{leinert+1976}. Furthermore, Earth and Wind encounter meteoroid streams typically once or twice per orbit per stream. These annual effects are assumed to be negligible compared to the seasonal variation stemming from Wind's relative velocity with respect to the ISD inflow direction. Therefore, the seasonal variation is attributed to ISD.

The seasonal variation itself varies on a decadal time scale: it is strongest around the solar minimum of December 2008 and weakest around the solar minimum of December 2019, corresponding to the focusing and defocusing phases of the solar magnetic cycle. The lower envelope of the seasonal minima roughly forms the previously noted horizontal level of about 12 impacts per day, which was identified as the interplanetary component of the dust population, whereas the seasonal and decadal variations are mainly associated with the interstellar component (see Sect.~\ref{sec:simulations}). It is, unfortunately, not yet possible to identify an individual impacting particle as interplanetary or interstellar.

\subsubsection{Comparison with numerical simulations}\label{sec:simulations}

Fig.~\ref{fig:overview}b compares the observed daily number of dust impacts with a numerically simulated cosmic dust particle flux at Wind's orbital position. The numerical simulations were performed with the IMEX code \citep{sterken+2012,strub+2019}, which models ISD under the influence of solar gravity, SRP, and the Lorentz force, assuming a Parker spiral IMF that is modulated with the solar magnetic $22\,\si{yr}$-cycle. Simulated ISD particles were launched at a distance of $50\,\si{AU}$ upstream of the Sun; the model does not yet include the outer heliosphere and, thus, does not yet include the filtering effects by the heliosheath \citep[e.g.][]{slavin+2012,sterken+2013}.

ISD trajectories were computed for spherical compact particles with a density of $2500\,\si{kg/m^3}$ and radii of $a\in\{0.30, 0.41, 0.54, 0.73\}\,\si{\micro m}$, assuming adapted astronomical silicates with a maximum \textbeta-ratio of $\beta_{\max}=1.6$ \citep{gustafson1994,sterken+2013}; see also Fig.~\ref{fig:beta}. The resulting flux densities at the modelled particle sizes were integrated over a MRN-like power law size distribution with an ISM hydrogen number density of $n_{\mathrm{H}}=0.1\,\si{cm^{-3}}$ \citep{mathis+1977}; confer Sect.~\ref{sec:amplitude:distribution}. 

The assumption of a MRN-like power law size distribution motivated the cutoff of the modelled size distribution below $a_{\min}=0.3\,\si{\micro m}$: slightly smaller particles, $a\in[0.1, 0.3]\,\si{\micro m}$ cannot reach the Wind spacecraft because they are excluded by SRP (see Sect.~\ref{sec:amplitude:distribution}), assuming the same \textbeta-curve as in Sect.~\ref{sec:amplitude:speed}, and even smaller particles, $a<0.1\,\si{\micro m}$, were excluded as a simplification of the heliosheath filtering. Although \textit{Ulysses} observed ISD particles down to masses of ${\sim}10^{-18}\,\si{kg}$, corresponding to particle radii of a few tens of nanometres, the observed ISD size distribution was strongly depleted at these particle radii compared to a MRN distribution \citep{krueger+2015}.

The particle flux was calculated by multiplying the integrated flux density with the surface area of the detector, which was assumed to be the apparent cross-section of Wind's cylindrical body, $1.8\,\si{m}\times 2.4\,\si{m}=4.32\,\si{m^2}$ \citep[see][]{hervig+2022}. A constant IDP flux of $10\,\si{\#/d}$ was assumed, resulting in the simulated particle flux that is shown in Fig.~\ref{fig:overview}b.

These numerical simulations are not intended for a detailed comparison with the observed dust impacts: for example, the modelled particle flux does not take into account the signal amplitude threshold of the Wind/WAVES instrument and, thus, is not identical to the expected number of dust impact detections. Nevertheless, although these numerical simulations are simplified, they successfully reproduce the decadal variation due to the solar magnetic $22\,\si{yr}$-cycle and the seasonal variation due to the spacecraft's orbit.

\subsubsection{Dust impact observations at different signal amplitudes}\label{sec:amplitude}

The long-term patterns in the daily number of dust impacts that are presented in Fig.~\ref{fig:overview} are not equally present at all signal amplitudes. As per Eq.~(\ref{eq:Qmv}), the signal amplitude that is generated by an impacting dust particle depends on the particle's mass and impact speed. However, because the mass and the impact speed cannot be disentangled from the signal amplitude, it is not known to which mass range the instrument is sensitive.

\begin{figure}
    \centering
    \includegraphics[width=\linewidth]{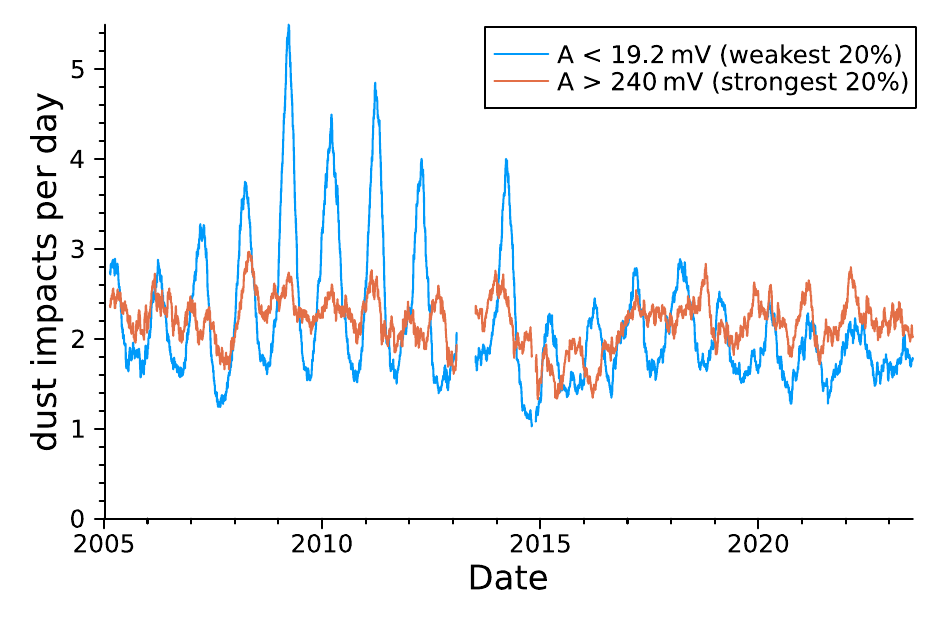}
    \caption{Time series for all dust impacts of morphological types A and B with a $y$-antenna signal amplitude below $19.2\,\si{mV}$ (blue) and above $240\,\si{mV}$ (red), graphed as a centred moving average with a width of $91\,\si{d}$. The two amplitude selections each correspond to $20\%$ of all impacts.}
    \label{fig:10pc_timeseries}
\end{figure}

Empirically, the long-term patterns in the daily number of dust impacts look starkly different for the weakest and the strongest signal amplitudes. This is shown in Fig.~\ref{fig:10pc_timeseries}, which presents the time series for all dust impacts with signal amplitudes below $19.2\,\si{mV}$ and above $240\,\si{mV}$, corresponding to the weakest and strongest $20\%$ of all signals measured by the $y$-antenna (Channel 2) with morphological types A and B. 

As Fig.~\ref{fig:10pc_timeseries} shows, the seasonal variation that was noted for the full dataset (see Fig.~\ref{fig:overview}a) is readily apparent for the weakest signal amplitudes, showing an annual maximum of daily dust impacts in March and an annual minimum in September. This seasonal variation is stronger before 2015, during the focusing phase of the solar magnetic cycle, than since 2015, in the defocusing phase. In contrast, for the dust impacts that generated the strongest signal amplitudes, Fig.~\ref{fig:10pc_timeseries} shows a near-constant if noisy level of about two daily impacts with no clear evidence of the seasonal variation. 
Because the seasonal variation is attributed to ISD, this implies that a significant fraction of the measured impacts with the weakest amplitudes are caused by ISD, whereas the strongest signal amplitudes are presumably mainly caused by impacts of IDPs. 

\begin{figure}
    \centering
    \includegraphics[width=\linewidth]{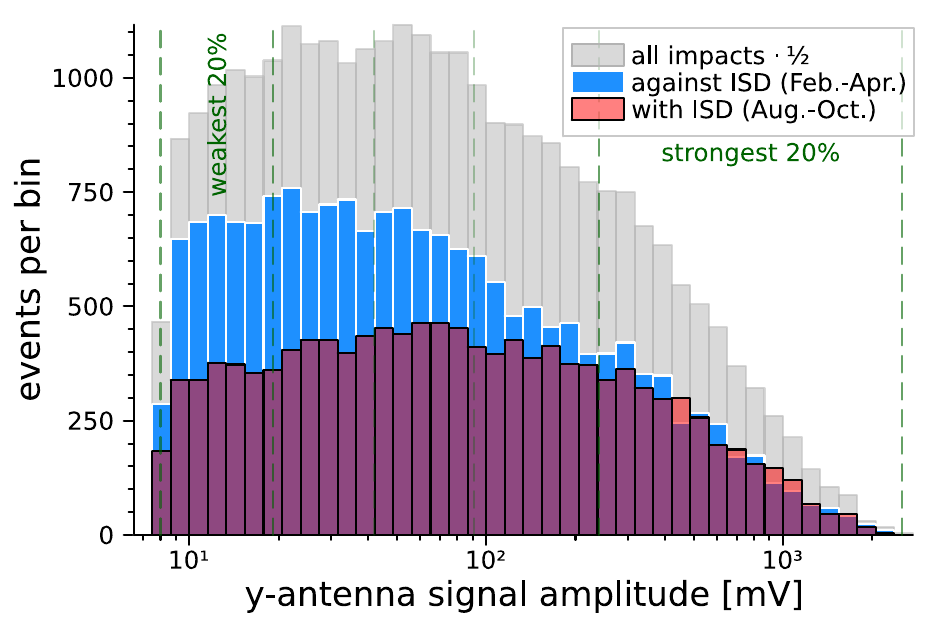}
    \caption{Amplitude distributions of all impacts (grey bars) with morphological types A and B measured by the $y$-antenna, impacts measured only when the spacecraft moved against the ISD inflow direction during February, March, and April (shown with blue filled bars outlined in white), and impacts measured only when the spacecraft moved with the ISD inflow direction during August, September, and October (shown with red filled bars outlined in black). The distribution of all impacts has been scaled by a factor of $0.5$ to tighten the histogram's vertical axis. The vertical dashed green lines indicate every $20^{\mathrm{th}}$ percentile of the signal amplitude; the weakest and the strongest $20\%$ of all impact signals were selected when generating Fig.~\ref{fig:10pc_timeseries}.}
    \label{fig:distr-ch2-againstwith}
\end{figure}

This is supported by the distribution of the signal amplitudes of the impacts measured by the $y$-antenna (Channel 2), which is displayed in Fig.~\ref{fig:distr-ch2-againstwith}. In the orbital segment where the spacecraft moves against the ISD inflow direction (February to April), an excess of weak-amplitude signals is observed compared to the orbital segment where the spacecraft moves with the ISD inflow direction (August to October). This excess is identified as ISD. Above an amplitude of $A\gtrsim 200\,\si{mV}$ the distributions for both orbital segments are similar; Fig.~\ref{fig:10pc_timeseries} shows almost no seasonal variation for signal amplitudes $A>240\,\si{mV}$.

This gives rise to the following hypothesis: the measured ISD-to-IDP ratio decreases with increasing signal amplitude. In the amplitude range above $A\gtrsim 200\,\si{mV}$ virtually no impacts from ISD were registered, whereas in the amplitude range below $A\lesssim 20\,\si{mV}$ the number of impacts during February to April is roughly twice as high as during August to October. 

Because ISD is much faster than IDPs in March and generally slower in September, this implies that the excess at weak amplitudes in March is caused by low-mass ISD that does not generate sufficiently strong signal amplitudes to be measurable in September; furthermore, a lower particle flux due to a lower relative speed also reduces the number of impacts per time. That the signal amplitude distribution looks similar at strong amplitudes for both investigated orbital segments of Fig.~\ref{fig:distr-ch2-againstwith} implies that very little ISD of sufficiently high mass and velocity to generate these signal amplitudes exists. However, because IDPs are generally slower in March than ISD, this implies that the strong-amplitude impacts are generated by even higher-mass IDPs, i.e.\ there must be a considerable excess of the IDP mass distribution at high masses compared to ISD. An alternative explanation could be a second population of IDPs that is considerably faster but not considerably less massive than the first IDP population. A candidate can be \textbeta-meteoroids, which often feature slightly lower masses but much faster speeds than other IDPs (Sects.~\ref{sec:amplitude:single} \&~\ref{sec:amplitude:distribution}). 

In particular, for the mass range around $m\approx10^{-17}\,\si{kg}$ the signal amplitudes of \textbeta-meteoroids are similar to ISD in March (Sect.~\ref{sec:amplitude:single}). At slightly higher masses, ISD is excluded by the \textbeta-gap but IDPs and \textbeta-meteoroids are not (Sect.~\ref{sec:amplitude:distribution}). This may indicate that the dust impact observations are predominantly caused by ISD and \textbeta-meteoroids of this mass regime.

In order to investigate this further, it would be essential to know the impact speed and mass of a measured particle impact, necessitating a mission at $1\,\si{AU}$ with a more elaborate dust detector. Knowledge of the direction of origin of an impacting dust particle would allow for better differentiation between ISD, \textbeta-meteoroids, and other IDPs. A dedicated dust detector on the Lunar Gateway (\citealp{wozniakiewicz+2021,arnet2023}; Sterken et al., in prep.)\ or the proposed SunCHASER mission \citep{posner+2021,cho+2023} would therefore be of great benefit.
Nevertheless, this analysis indicates that the ISD impacts observed by Wind are primarily associated with weak-amplitude signals, and that strong-amplitude impacts are predominantly caused by IDPs.

\section{Frequency analysis}\label{sec:fourier}

Frequency analysis has been performed on the dataset of dust impacts at $\mathrm{L}_1$, which covers the time range from 2005 to the end of August 2023. This dataset contains two long-term gaps in the data and many periodically occurring $2\,\si{d}$-long gaps since April 2011 (see Sect.~\ref{sec:database}).

Therefore, the resulting time series is no longer evenly sampled, and the necessary assumptions for the discrete Fourier transform (DFT) are not met. The most common alternative to DFT for unevenly spaced data, the Lomb-Scargle transform \citep{lomb1976,scargle1982}, can suffer from phase and amplitude artefacts \citep[e.g.][]{foster1995,cumming+1999}. Instead, the method proposed by \citet[][Suppl.~Mat.]{kirchner+2013}, which is based on the date-compensated discrete Fourier transform \citep[DCDFT;][]{ferrazmello1981}, is used. Consistent results are also obtained from Kirchner and Neal's implementation of the Weighted Wavelet Z transform \citep{foster1996}, which additionally suppresses spectral aliasing that can arise from uneven sampling. These methods are presented and evaluated in Appendix~\ref{app:methodeval}.

Within this investigation the periodogram amplitude was calculated by Kirchner \& Neal's DCDFT-based method at each frequency on an equidistant frequency grid. This grid corresponds to the natural frequencies of a DFT oversampled by a factor of 15, increasing the number of frequencies at which the spectrum is evaluated by that factor of 15 compared to a DFT. The data were linearly detrended before calculating the spectrum. The $95\%$ significance thresholds of each respective spectrum were estimated by the $95^{\textrm{th}}$ percentile of periodogram amplitudes for randomly resampled time series (see Appendix~\ref{app:noiselevel}): the probability of a random noise peak at the respective frequency reaching this threshold is $5\%$; thus, the statistical significance of a peak that reaches the threshold is $95\%$, and higher peaks are more significant.

\subsection{Discovery of solar rotation signatures}\label{sec:spectrum}

\begin{figure}
    \centering
    \includegraphics[width=\linewidth]{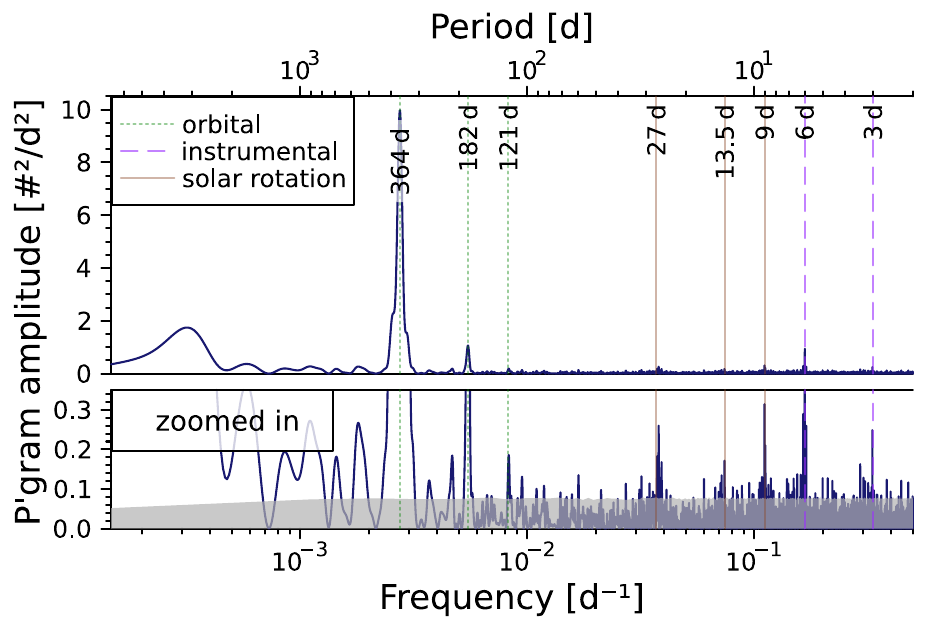}
    \caption{Periodogram of the daily number of dust impacts with morphological types A and B recorded at $\mathrm{L}_1$. The periodogram was evaluated at the DFT's natural frequencies and oversampled by a factor of 15. The three most relevant frequencies and their harmonics are marked by vertical lines: the $365\,\si{d}$-period corresponding to the orbital period of the spacecraft (dotted green lines), the $6\,\si{d}$-period introduced by instrumental effects (dashed purple lines), and the solar rotation period at $27\,\si{d}$ (solid brown lines). The bottom panel shows a zoomed-in view of the full spectrum with the estimated 95\% significance thresholds indicated by a grey shaded area.}
    \label{fig:spectrum}
\end{figure}

The periodogram of the daily dust impacts with morphological types A and B recorded at $\mathrm{L}_1$ is presented in Fig.~\ref{fig:spectrum}. For easier comparison, the most relevant periods and their harmonics have been marked. 

The spectrum's most powerful peak lies at $364\,\si{d}$; its first harmonic at $182\,\si{d}$ also has a considerable power. These features correspond to the revolution of the Wind spacecraft around the Sun and the seasonal variability reported in Sect.~\ref{sec:overview}: once per orbit, the spacecraft moves against the inflow direction of ISD, and once per orbit it moves with this direction. Because no dependence on the direction of motion of the spacecraft is known for IDPs, these spectral features are caused almost entirely by ISD.

Another powerful spectral feature comes from the periodic change of the sampling rate, which leads to the removal of two days of dust data every six days since April 2011 (see Appendix~\ref{app:data_ratechange}). This spectral peak is observed at $5.99\,\si{d}$, and its first harmonic lies at $3.02\,\si{d}$. These two periods have been marked in the figure by dashed lines.

Three more spectral peaks are evident in the data and are interpreted as signatures of the solar rotation. This marks the first discovery of solar rotation signatures in cosmic dust data. The most powerful and most clearly defined peak lies at $9.01\,\si{d}$ and is identified as the second harmonic of a $27\,\si{d}$ period. Another clearly defined peak lies at $13.5\,\si{d}$ and is identified as the first harmonic of the same. There is no notable spectral peak at precisely $27\,\si{d}$; however, a broader peak at $26.4\,\si{d}$ is present. The frequency resolution of the periodogram of Fig.~\ref{fig:spectrum} is $1.5\times 10^{-4}\,\si{d}^{-1}$, which corresponds to roughly $0.11\,\si{d}$ in the period range at $27\,\si{d}$; the displacement of the peak from $27\,\si{d}$ to $26.4\,\si{d}$ is thus not an effect of the frequency resolution. 

The origin of the solar rotation signatures is investigated further by determining:
\begin{itemize}
    \item whether the solar rotation signatures are caused by the interstellar or by the interplanetary dust population (Sect.~\ref{sec:ana_isd-idp});
    \item whether the solar rotation signatures are created through CIRs (Sect.~\ref{sec:ana_cirs});
    \item whether the solar rotation signatures are created through the IMF sector structure or crossings of the heliospheric current sheet (Sect.~\ref{sec:ana_hcs}); and
    \item whether the solar rotation signatures are not genuine signatures of the dust environment but created through periodic external effects (Sect.~\ref{sec:ana_instrumental}).
\end{itemize}

\subsection{Interstellar versus interplanetary dust}\label{sec:ana_isd-idp}

There are two primary selection methods that can help distinguish an ISD population from an IDP population. First, much more ISD is detected when the spacecraft moves against the ISD inflow direction in March than when it moves with the ISD inflow direction in September. Thus, if the solar rotation signatures are more powerful in the early calendar year than in the late calendar year, summed over all years, they should be caused, at least partially, by ISD. If they have the same power throughout the year, they should be caused predominantly by IDPs (see Sect.~\ref{sec:withagainst}).

Second, the ecliptic plane contains much more ISD during the focusing phase of the solar magnetic cycle than during the defocusing phase, whereas the population of measurable IDPs appears to be not as strongly affected by the solar magnetic cycle (see Fig.~\ref{fig:10pc_timeseries} and Sect.~\ref{sec:amplitude}). Thus, if the solar rotation signatures are much stronger during the focusing phase than during the defocusing phase, they should be caused, at least partially, by ISD. If they are of comparable power in both phases, they should be caused predominantly by IDPs (see Sect.~\ref{sec:focusdefocus}).

\subsubsection{Against versus with the interstellar dust inflow direction}\label{sec:withagainst}

\begin{figure}
    \centering
    \includegraphics[width=\linewidth]{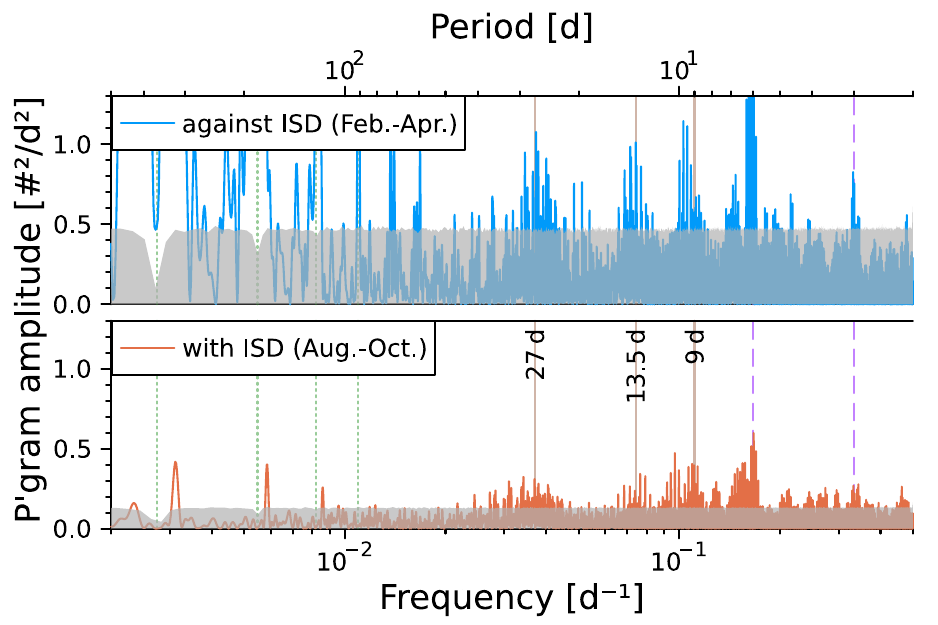}
    \caption{Periodograms of the daily number of dust impacts with morphological types A and B for the three months of each year that correspond either to the orbital segment against the ISD inflow (February to April, blue curve, top panel) or with the ISD inflow direction (August to October, red curve, bottom panel). The two panels are scaled identically; the estimated $95\%$ significance thresholds are indicated by grey shaded areas. Frequencies of interest are marked by vertical lines as in Fig.~\ref{fig:spectrum}.}
    \label{fig:againstwith}
\end{figure}

Figure~\ref{fig:againstwith} displays a comparison of the power spectra of the daily number of dust impacts with morphological types A and B during the orbital segments against or with the ISD inflow direction, respectively. For the respective dataset, all data outside a quarter orbit, corresponding to the months of February to April when going against the ISD inflow and August to October when going with the ISD inflow, were removed before calculating the spectra. This receiver function, which is one during the selected quarter orbit and zero during the remaining three-quarters of the orbit, is visible in the harmonics of the $365\,\si{d}$-period that decline in power with decreasing period. The convolution of the dust impact time series with this receiver function significantly diffuses the spectral peaks. It also causes the estimated $95\%$ significance threshold to drop to a lower value at a period of $365\,\si{d}$ and its harmonics.

Both spectra show evidence of the solar rotation signatures. All three solar rotation peaks are more powerful by about a factor of three when going against the ISD inflow than when going with the ISD inflow, which indicates that the solar rotation signatures are not exclusively present in IDPs but are, at least in large part, caused by ISD.

\subsubsection{Focusing versus defocusing phase}\label{sec:focusdefocus}

\begin{figure}
    \centering
    \includegraphics[width=\linewidth]{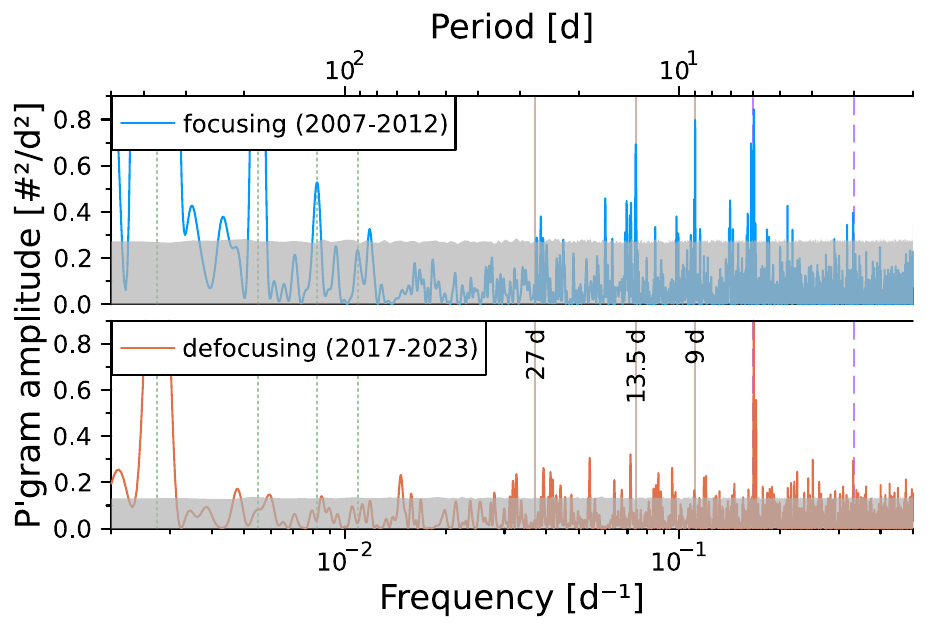}
    \caption{Periodograms of the daily number of dust impacts with morphological types A and B during six years in the focusing phase (2007-2012, blue curve, top panel) and six years in the defocusing phase (September 2017 to August 2023, red curve, bottom panel). The two panels are scaled identically; the estimated $95\%$ significance thresholds are indicated by grey shaded areas. Frequencies of interest are marked by vertical lines as in Fig.~\ref{fig:spectrum}.}
    \label{fig:focusdefocus}
\end{figure}

Figure~\ref{fig:focusdefocus} shows a comparison of the power spectra of the daily number of dust impacts with morphological types A and B during the focusing phase (2007 to 2012) and during the defocusing phase (September 2017 to August 2023), respectively. The seasonal variation ($365\,\si{d}$-peak) is more powerful by roughly an order of magnitude during the focusing phase, indicating that more ISD is measured in addition to IDPs during the focusing phase. The instrumental $6\,\si{d}$-peak is more powerful during the defocusing phase because it affected the instrument only since April 2011 (Sect.~\ref{sec:windwaves}). 

During the focusing phase, clear and powerful spectral peaks can be observed at $13.5\,\si{d}$ and $9\,\si{d}$. Another peak at ${\sim}27\,\si{d}$ is less powerful and more diffuse. 

During the defocusing phase, these solar rotation signatures are less prominent and may vanish. For example, the first harmonic is diminished by a factor of about two, and the second harmonic is not perceptible above neighbouring noise peaks. Furthermore, the spectral peaks that are perceptible do not align exactly with the periods of interest but appear slightly shifted. 

In conclusion, the solar rotation signatures are clearer and more powerful during the focusing phase than during the defocusing phase, indicating that they are caused, at least in large part, by ISD. A caveat to this analysis is that not only ISD but also low-mass IDPs could be affected by the solar magnetic cycle, for example \textbeta-meteoroids or nanodust.

\subsection{Influence of co-rotating interaction regions}\label{sec:ana_cirs}

Because the spacecraft encounters CIRs at the solar rotation period of about $27\,\si{d}$, CIRs are the primary contender to explain the origin of the solar rotation signatures. The higher solar wind speed, stronger IMF, and charging effects due to higher plasma temperatures during CIRs compared to the average solar wind \citep{hajra+2022} result in a stronger Lorentz force that may deflect submicron-sized dust; these dust particles may be charged even more due to their fluffiness \citep{ma+2013}.

\citet{flandes+2011} and \citet{hsu+2010} found that CIRs can periodically affect dust in the Jovian and Saturnian environment, respectively. The particles of these dust streams were identified to be around ${\sim}10\,\si{nm}$ in radius \citep{zook+1996,hsu+2011}. However, \citet{malaspina+2014} estimate from the observed IDP flux that Wind is sensitive to impacts of approximately submicron-sized dust particles; \citet{kellogg+2016} conclude that Wind is insensitive to impacts of highly accelerated nanodust. Thus, particles that are comparable in size to Jovian and Saturnian dust streams are unlikely to be detectable by Wind. Moreover, the mechanism that causes the Jovian and Saturnian dust streams can act on particles with radii of ${\sim}10\,\si{nm}$ but is not expected to extend to submicron-sized dust \citep{hsu+2010}. 

\citet{stcyr+2017} report a reduction of dust impact detections by Wind during CMEs. Calculations by \citet{ragot+2003} find that dust particles with radii of between ca.\ $0.1\,\si{\micro m}$ and $3\,\si{\micro m}$ can be depleted by CMEs in the solar corona.\footnote{We note that the magnetic field of a CME is considerably stronger close to the Sun than at $1\,\si{AU}$ \citep[e.g.][]{patsourakos+2016}.} Simulations by \citet{wagner+2009} indicate that dust particles with radii up to $1\,\si{\micro m}$ may be perturbed by CMEs as they get closer to the Sun than $1\,\si{AU}$ by Poynting-Robertson drag. Similar perturbation effects on dust particles in this size range may also be induced by SIRs. 
This motivates a superposed epoch analysis of dust impact detections during SIRs.

\subsubsection{Superposed epoch analysis of stream interaction regions that are associated with co-rotating interaction regions}
\label{sec:depletion} 

\begin{figure}
    \centering
    \includegraphics[width=\linewidth]{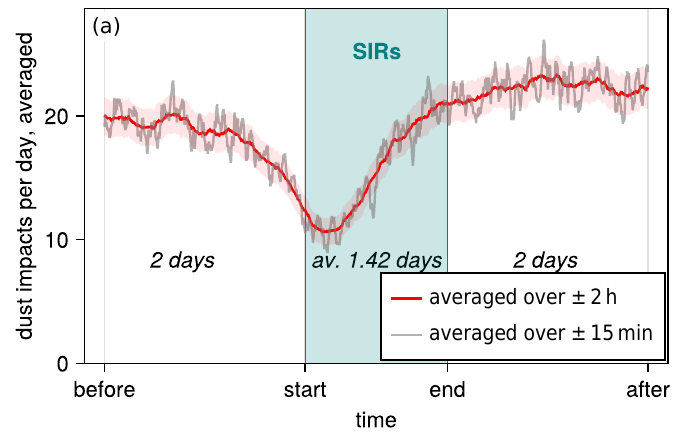}
    \includegraphics[width=\linewidth]{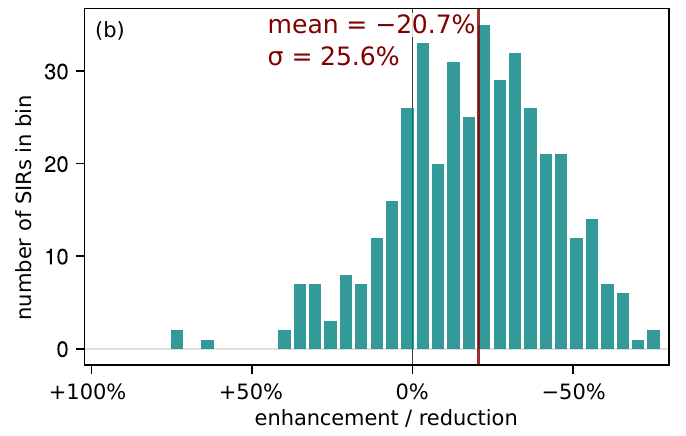}
    \caption{Reduction and enhancement of dust impact detections made by Wind during SIRs. \textit{Top panel:} Scaled superposed epoch analysis of the daily number of dust impacts during SIRs as a moving average within a $30\,\si{min}$ interval (grey) or a $4\,\si{h}$ interval (red). Only the SIRs that are associated with CIRs were taken into account. The shaded red area indicates $95\%$ confidence intervals for a Poissonian distribution.\\ 
    \textit{Bottom panel:} Histogram of the enhancement or reduction factor of dust impacts for the SIRs that are associated with CIRs. The dashed blue line denotes the average reduction of $20.7\%$. The five SIRs that begin on 1 January 2005, 23 November 2006, 15 January 2007, 24 November 2008, and 30 August 2009 were excluded due to low number statistics.}
    \label{fig:depletion_histoscaled}
\end{figure}

Figure~\ref{fig:depletion_histoscaled}a shows a superposed epoch analysis \citep{chree1913} of the daily number of dust impacts for all SIRs that are associated with CIRs from the datasets by \citet{jian+2006,jian2009} and \citet{hajra+2022} since 2005. SIRs that are not associated with CIRs were not taken into account because non co-rotating SIRs are unrelated to the solar rotation period. The time intervals of the SIRs were rescaled to the duration of the respective SIR before superposition. More details of the methodology are presented in Appendix~\ref{app:depletion}. 

As Fig.~\ref{fig:depletion_histoscaled}a indicates, on average, a reduction of dust impacts is observed during these SIRs compared to the preceding and following time periods: dust impact detections are reduced, on average, by $23.8\%\pm2.4\%$ during these SIRs compared to the preceding and following time periods. This reduction appears to begin a few hours before the start of the superposed SIR and is more pronounced during the first half of the superposed SIR compared to the second half.

Figure~\ref{fig:depletion_histoscaled}b shows that SIRs can feature either a reduction or an enhancement of dust impact detections compared to the preceding and following time intervals. However, this distribution is shifted towards a reduction of dust impact measurements during SIRs: only $82$ SIRs of the dataset feature an enhancement, whereas $398$ SIRs feature a reduction. On average, the SIRs reduce the number of dust impact detections by $21\%$ with a standard deviation of $26\%$.

The difference in the average strength of dust impact reduction, $23.8\%$ in the superposed epoch analysis and $21\%$ in the histogram, stems from the difference in methodology: the superposed epoch analysis sums the number of dust impacts over all SIRs, i.e.\ the sum is naturally weighted by the number of dust impacts during the respective SIR. In the histogram, each SIR is assigned the same weight.

The effect of the $82$ SIRs that cause an enhancement of dust impact detections was investigated: artificial gaps were induced in the dust impact time series whenever one of these \enquote*{enhancing SIRs} occurred, beginning three days before the start of the respective SIR and ending three days after the end of each SIR, i.e.\ about a week of data surrounding each enhancing SIR was removed. The periodogram of this modified time series does not significantly differ from the periodogram of the original time series (Fig.~\ref{fig:spectrum}). This indicates that the solar rotation signatures are not associated with only the SIRs that enhance the number of detections of dust impacts. This is important for rejecting the hypothesis that the solar rotation signatures are caused by instrumental effects (see Sect.~\ref{sec:ana_potential}).

When performing the same investigation for the $398$ SIRs that reduce the dust impact observations, the solar rotation signatures were strongly diminished. This indicates that the solar rotation signatures are associated with these \enquote*{reducing SIRs}. Most likely, both the enhancing and the reducing SIRs cause the solar rotation signatures; however, because there are far fewer enhancing SIRs than reducing SIRs, the effect of their removal on the spectrum is comparably minor.

Péronne et al.\ (in prep.)\ investigate how the reduction or enhancement of dust impact detections varies with the properties of the respective SIRs or CMEs, for example with the IMF strength or the presence or absence of interplanetary shocks. 
Further modelling efforts are required to probe the physical mechanism of this reduction of dust impacts during SIRs. Observations with a dust detector that can determine the particles' masses, such as an impact ionisation detector, would be essential in confirming the size range of the affected dust particles.

\subsubsection{Dust spectra for time intervals that are rich and poor in co-rotating interaction regions}\label{sec:casestudy}

To investigate whether the solar rotation signatures are related to CIRs, time intervals with different rates of occurrence and CIR durations were analysed. If the solar rotation signatures are related to CIRs, they should be more powerful during time intervals where multiple long-lasting CIRs occurred and less powerful during time intervals where no or only short-lasting CIRs occurred. 

The datasets by \citet{jian+2006,jian2009} and \citet{hajra+2022} list individual SIRs. While the dataset by \citet{jian+2006,jian2009} flags whether a SIR is associated with CIRs in general, it does not indicate which individual SIRs are recurrences of the same co-rotating structure. It is therefore necessary to identify CIRs in these datasets. The method of CIR identification is detailed in Appendix~\ref{app:cir_ident} and assesses whether two SIRs that could be part of a CIR are separated in time by the solar rotation period. 

\begin{figure}
    \centering
    \includegraphics[width=\linewidth]{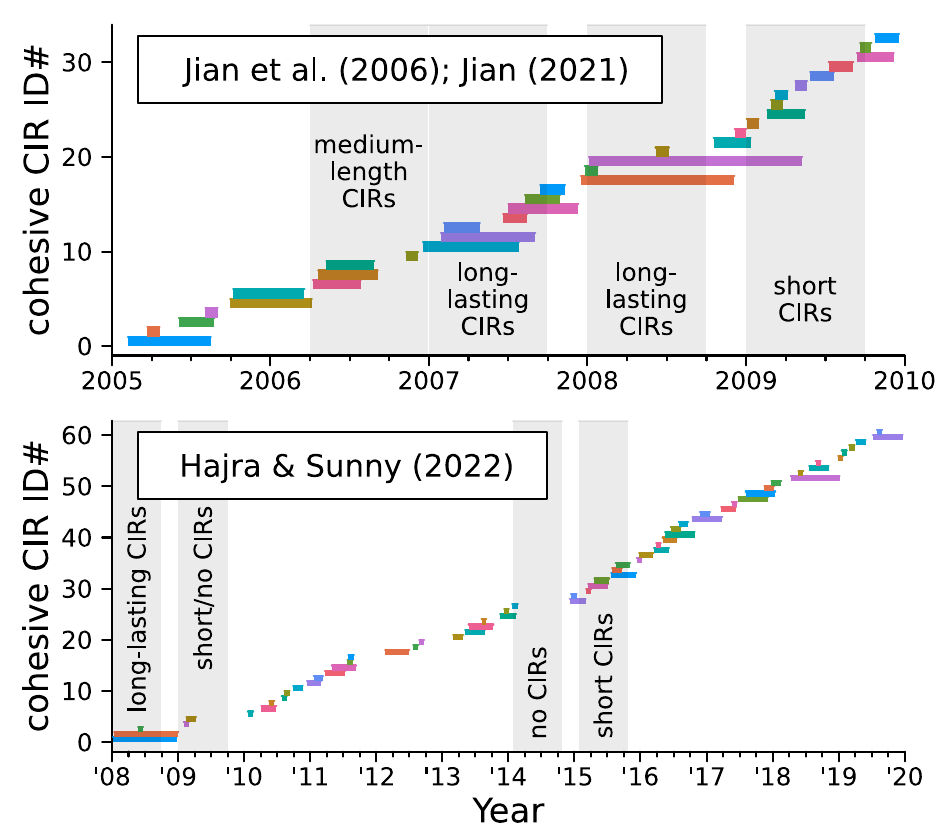}
    \caption{CIRs identified by the method proposed in Sect.~\ref{sec:casestudy} for the SIR datasets of \citet{jian+2006,jian2009} since 2005 (top panel) and \citet{hajra+2022} (bottom panel). Each horizontal bar corresponds to one CIR. Time intervals of interest have been highlighted by shaded areas. We note that the two datasets overlap only for the years 2008 and 2009.}
    \label{fig:unique_cirs}
\end{figure}

The resulting list of CIRs is displayed in Fig.~\ref{fig:unique_cirs}, showing the time interval that is covered by each CIR. \citet{dumbovic+2022} found a CIR that lasted for 27 Carrington rotations from June 2007 to May 2009. The CIRs displayed in Fig.~\ref{fig:unique_cirs} do not include this single long-lasting CIR, but do include two similarly long-lasting CIRs in 2008. 

Six time intervals of interest were selected on the basis of Fig.~\ref{fig:unique_cirs}. For easier comparison, each time interval was selected to be nine months long. These time intervals of interest, given by the day-of-year (doy) are:
\begin{enumerate}[(a)]
    \item 2006 doy 91-365, when multiple CIRs that each last a few months were identified; 
    \item 2007 doy 1-273, when at least two long-lasting CIRs were observed; 
    \item 2008 doy 1-273, when two extremely long-lasting CIRs were identified, coincident with the long-lasting CIR observed by \citet{dumbovic+2022}; 
    \item 2009 doy 1-273, when many short-lasting CIRs were identified in the dataset by \citet{jian+2006,jian2009} and very few CIRs were identified in the dataset by \citet{hajra+2022}; 
    \item 2014 doy 23-295, when no CIRs were identified; and
    \item 2015 doy 23-295, when many short-lasting CIRs were identified.
\end{enumerate}

\begin{figure*}
    \centering
    ~\hfill \textbf{\large Daily number of dust impacts} \hfill~\hfill \textbf{\large Synthetic data} \hfill ~\\
    \includegraphics[width=0.49\linewidth]{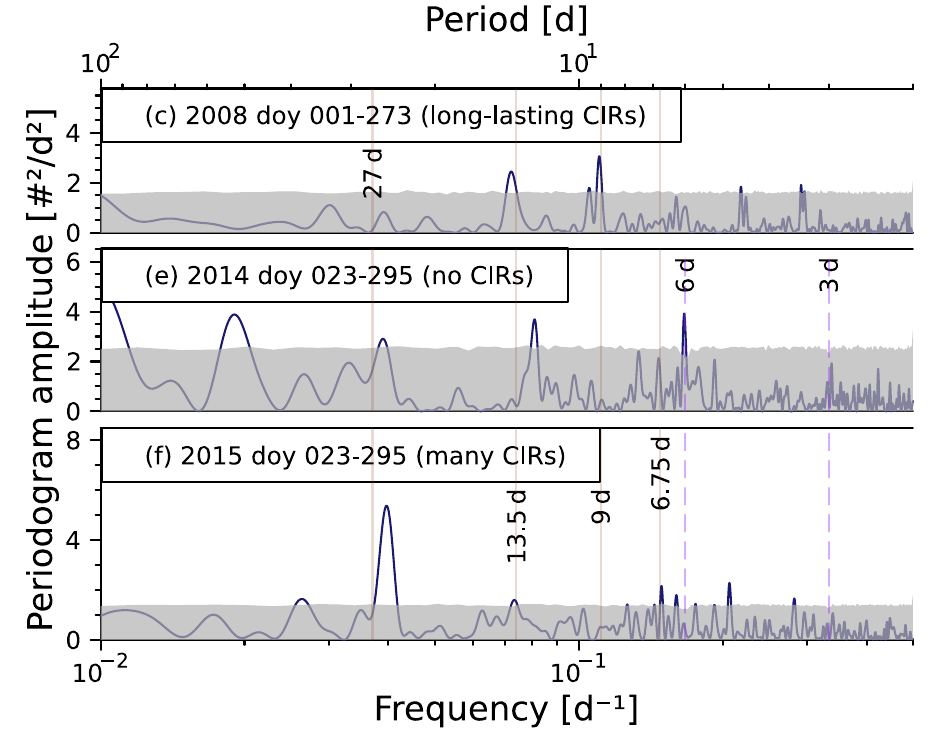}
    \includegraphics[width=0.49\linewidth]{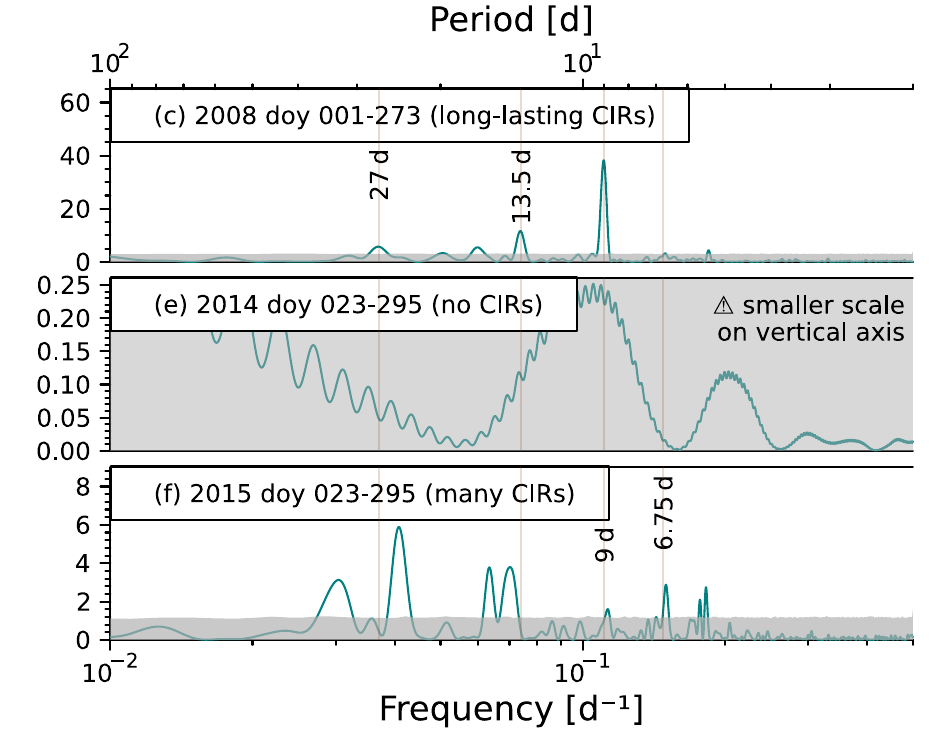}
    \caption{Periodograms for certain time periods corresponding to different systematic rates of occurrence of CIRs for the daily number of dust impacts with morphological types A and B (left, indigo curves) and of a synthetic time series of dust impacts with an absolute dust depletion by SIRs (right, teal curves; see Sect.~\ref{sec:cir-model}). All of the selected time periods are nine months long; their initial and final dates are given as doy. All panels in the left column share the same axis range and scale for easier comparison. The panels in the right column are scaled differently. Frequencies of interest are marked by vertical red lines as in Fig.~\ref{fig:spectrum}, including the third solar rotation harmonic at $6.75\,\si{d}$; the $6\,\si{d}$-periodicity has only been marked in the dust impact spectra since 2011. Estimated $95\%$ significance thresholds are indicated by grey shaded areas.}
    \label{fig:casestudies}
\end{figure*}

Periodograms have been generated for these particular time periods. Figure~\ref{fig:casestudies} (left column) presents three of these time periods: (c) 2008, when two long-lasting CIRs were identified; (e) 2014, when no CIRs were identified; and (f) 2015, when many short-lasting CIRs were identified.

The solar rotation signatures were most easily identifiable when long-lasting CIRs occurred in 2008 (c) and in 2007 (b, not depicted). At times where no CIRs (e, 2014) or multiple short-lasting CIRs (f, 2015; also a and d, 2006 and 2009, not depicted) were identified, the solar rotation signatures were weaker. 

This suggests that the detected solar rotation signatures are connected to CIRs. That long-lasting CIRs cause more powerful solar rotation signatures is reasonable: a long-lasting periodic signal, i.e.\ one long-lasting CIR, causes a stronger spectral peak at the relevant period than multiple short-lasting periodic signals that are individually time-offset, i.e.\ multiple short-lasting CIRs. A periodic signal that consists of multiple peaks in time can generate a spectrum that is more powerful at the harmonics than at the primary (see Appendix~\ref{app:test-spikes}), which explains why different spectra exhibit different relative amounts of periodogram amplitude in the harmonics.

\subsubsection{Synthetic data of dust impacts depleted by co-rotating interaction regions}\label{sec:cir-model}

\begin{figure}
    \centering
    \includegraphics[width=\linewidth]{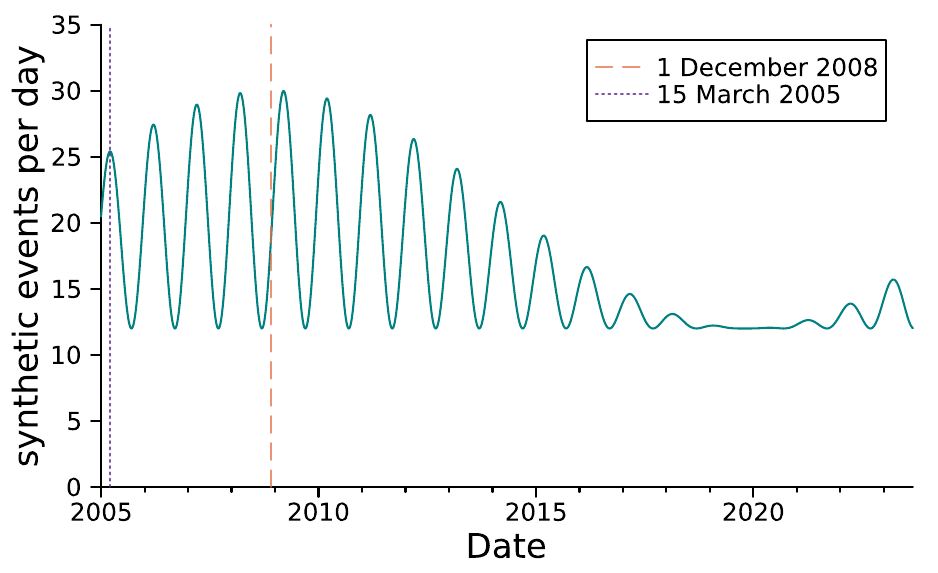}
    \caption{Synthetic time series of IDP and ISD impacts on Wind following Eq.~(\ref{eq:model}) to reproduce the observed number of impacts in Fig.~\ref{fig:overview}a. The times at which the oscillations due to the solar magnetic cycle and the annual variation are at their maxima are indicated by a vertical dashed red line and a vertical dotted purple line, respectively.}
    \label{fig:model}
\end{figure}

To further evaluate the hypothesis that the solar rotation signatures are caused by CIRs a synthetic time series of observed dust impacts on Wind was constructed. This synthetic time series describes the daily number of dust impacts, $N$, as a function of time, $t$, by
\begin{equation}\label{eq:model}
    N(t) = N_{\mathrm{IDP}} + N_{\mathrm{ISD}} \cdot \cos^2 \qty(\frac{\pi\, (t-t_{\min})}{22\,\si{yr}})\cdot \cos^2\qty(\frac{\pi\, (t-t_{\mathrm{ag}})}{1\,\si{yr}}) \ 
\end{equation}
and is displayed in Fig.~\ref{fig:model}. It consists of a constant rate of $N_{\mathrm{IDP}}=12$ IDP impacts per day; the ISD rate periodically changes with the solar magnetic cycle of $22\,\si{yr}$ and the spacecraft's orbital period of $1\,\si{yr}$, with a maximum of $N_{\mathrm{ISD}}=18$ ISD impacts per day (see Sect.~\ref{sec:overview}). The effect of the solar magnetic cycle on the dust has its maximum during solar minimum \citep{landgraf2000,sterken+2012}, here taken as $t_{\min}=\text{1 December 2012}$; and the seasonal cycle has its maximum when the spacecraft moves against the ISD inflow direction, here taken as $t_{\mathrm{ag}}=\text{15 March 2015}$. The synthetic time series was constructed to reproduce the large-scale variations indicated in Fig.~\ref{fig:overview}a and is not physically motivated.

The effect of SIRs was tested by setting the synthetic time series to zero during a SIR, $N(t_i) = 0\ \forall t_i\in \text{SIR}$. This was carried out for all SIRs, not only those that were identified as part of a CIR. This absolute local depletion of dust by a SIR is not intended to be physical; this exaggeration of the effect was chosen to better investigate how the spectrum is influenced. 

In analogy to Sect.~\ref{sec:casestudy}, the case studies of time periods that are of interest with respect to the occurrence of CIRs were repeated for this synthetic time series with dust depletion by SIRs. A selection of the resulting spectra is shown in Fig.~\ref{fig:casestudies} (right column). 

The peaks in the synthetic data spectra are not identical to those found in the genuine spectra but correspond reasonably well. In particular, for the time period in 2008 when long-lasting CIRs were identified (c), the spectral features are similar for both spectra, and the solar rotation signatures are evident in both the genuine dust spectrum and in the synthetic data spectrum. The spectra for the fairly short-lasting CIRs in 2006 and 2009 (a and d; not depicted) match similarly.

In the 2014 time period where no CIRs were observed (e), neither the spectrum of the genuine dust impacts nor that of the synthetic data show strong solar rotation signatures. The weak spectral peak in the synthetic data periodogram at approximately $9\,\si{d}$ is caused by two non-CIR SIRs that are separated by $9\,\si{d}$; it lies below the estimated $95\%$ significance threshold and is not a signature of the solar rotation.
For the many short-lasting CIRs of 2015 (f), both the genuine data and the synthetic data feature only one strong peak adjacent to the solar rotation period and minor peaks at the first and third harmonics.

We note that the genuine and the synthetic spectra do not always fully agree, which may be due to, for example, variations of the CIR parameters, such as the IMF strength, that cause variations in the strength of the observed reduction of dust (see Fig.~\ref{fig:depletion_histoscaled}b). 
Nevertheless, the similarities between the spectra of the synthetic data and of the genuine dust impact data suggest a close relation between CIRs and solar rotation signatures in the dust: it appears that most CIRs locally reduce the number of dust impacts that are observed by the spacecraft at $1\,\si{AU}$ as it is passed by the individual CIR.

\subsection{Influence of the interplanetary magnetic field sector structure}\label{sec:ana_hcs}

An individual crossing of the HCS by the spacecraft is unlikely to have much of an effect due to the time scale: \citet{liou+2021} report an average thickness of the HCS of about $10^5\,\si{km}$ with a maximum thickness of about $10^6\,\si{km}$. The orbital speed of Earth -- and, thus, of the spacecraft -- around the Sun is about $30\,\si{km/s}$. Thus, it would take the spacecraft a few hours to cross a highly inclined HCS. This is a considerably shorter time scale than the duration of a passing SIR (${\sim}36\,\si{h}$; \citealp{jian+2006}). Similarly, the strength of the IMF varies by perhaps $20\%$ as the spacecraft crosses the HCS \citep{liou+2021}, whereas a SIR enhances the magnetic field much more strongly \citep[on average by a factor of two to three; e.g.][]{geyer+2021,hajra+2022}. Thus, the crossings of the HCS are unlikely to have a strong effect on dust particles compared to SIRs. However, the alternating sector structure of the IMF may play a much larger role than individual spacecraft crossings of the HCS.

The alternating polarity of the magnetic field in adjacent sectors of the IMF was proposed as the cause of the periodic appearance of Jovian dust streams \citep{hamilton+1993}. However, \citet{flandes+2011} find a strong correlation between Jovian dust streams and the compression regions of, for example, CIRs. Similarly, \citet{hsu+2010} find that Saturnian dust streams result from the combination of both CIRs and the alternating sector structure. 

It is therefore unlikely that the sector structure alone is the sole cause of the solar rotation signatures. Case studies similar to Sect.~\ref{sec:casestudy} were performed to confirm or refute this hypothesis, making use of the dust impact data and of the daily IMF polarity data given by \citet{svalgaard2023_struc}. This is presented in detail in Appendix~\ref{app:hcs-casestudies}. 

The spectra of the IMF polarity data (see Fig.~\ref{fig:casestudy-hcs}, right column) are highly correlated to the IMF sector structure stability: they feature strong $27\,\si{d}$-peaks for a two-sector structure and strong $13.5\,\si{d}$-peaks for a four-sector structure, and the solar rotation signatures in the IMF polarity data are weakened during times of a chaotic sector structure. 
In contrast, the dust impact spectra (see Fig.~\ref{fig:casestudy-hcs}, left column) do not show this correlation: for example, during a stable four-sector structure, a powerful $27\,\si{d}$-peak and a weak $13.5\,\si{d}$-peak are observed; and the solar rotation signatures can be powerful during a chaotic IMF sector structure and weak during a stable IMF sector structure.

This indicates that the alternating polarities of the IMF sector structure are not the sole cause of the solar rotation signatures. However, this only holds true for the local sector structure at the spacecraft; ISD travelling through the heliosphere may nevertheless experience alternating phases of focusing and defocusing with respect to the ecliptic plane as it encounters the alternating polarities of the IMF sector structure. Similarly, dust may be affected differently when CIRs and HCS crossings coincide, as is the case farther away from the Sun, for example in Jupiter's or Saturn's orbits \citep{gosling+1999}. 
Nevertheless, while the local IMF sector structure is expected to affect cosmic dust, it is unlikely to be the sole cause of the observed solar rotation signatures.

\subsection{Influence of external effects}\label{sec:ana_instrumental}

It may be possible that the solar rotation signatures in the dust impact data do not reflect a genuine pattern in the dust environment but are artificially caused by external effects. One of these external effects is a periodic change of the spacecraft's floating potential when encountering CIRs (Sect.~\ref{sec:ana_potential}); another external effect is a periodic insensitivity to dust impacts upstream of interplanetary shocks associated with CIRs (Sect.~\ref{sec:ana_neuralnet}). The influence of these two potential artificial origins of the solar rotation signatures are investigated below.

\subsubsection{Influence of the spacecraft's floating potential}\label{sec:ana_potential}
 
\citet{wilson+2023} report evidence of solar rotation signatures in the floating potential of the Wind spacecraft and hypothesise that CIRs are the origin. Thus, one must consider that the solar rotation signatures in the dust impact spectra may not stem from a genuine reduction of dust impacts as suggested in Sect.~\ref{sec:depletion}; instead, periodic changes of the spacecraft floating potential could artificially induce these signatures by periodically reducing the instrument's sensitivity to dust impacts.

\begin{figure}
    \centering
    \includegraphics[width=\linewidth]{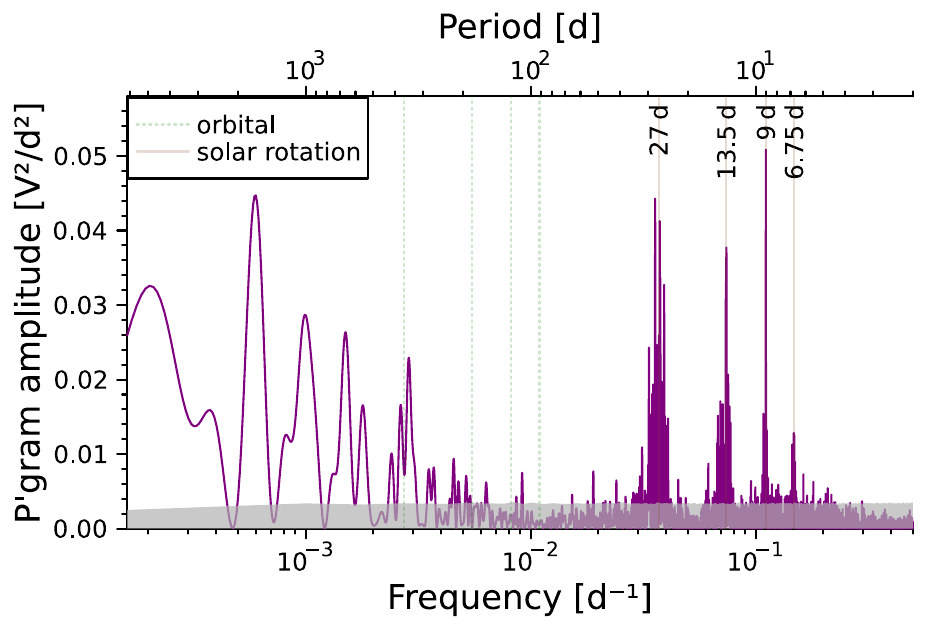}
    \caption{Periodogram of Wind's floating potential from 2005 to 2021 using Method 4 of \citet{wilson+2023} for parallel pitch angles (cf.\ \citealp[][Fig.~6]{wilson+2023}). The data were averaged over each day before generating the spectrum. The vertical lines are as in Fig.~\ref{fig:spectrum}. The estimated $95\%$ significance thresholds are indicated by a grey shaded area.}
    \label{fig:pgram-scpot}
\end{figure}

The dataset of \citet{wilson+2023_data}, using the spacecraft potential derived by the local minimum of the electron energy distribution function with parallel pitch-angles as in \citet[][Fig.~6]{wilson+2023}, yields the power spectrum displayed in Fig.~\ref{fig:pgram-scpot}. Spectra for the spacecraft potential derived with different methods of \citet{wilson+2023} are quantitatively but not qualitatively different. High periodogram amplitudes are observed at the primary and the first two harmonics of the solar rotation period; the third harmonic is less powerful but still notable. These solar rotation signatures are much more prominent in the spacecraft potential spectrum than in the dust impact spectrum (see Fig.~\ref{fig:spectrum}).

The question remains whether the change of the spacecraft floating potential during a CIR decreases the instrument's sensitivity to dust impacts, artificially inducing a reduction of observed dust impacts, or if it increases the instrument's sensitivity, counteracting a genuine reduction of dust impacts.
Within a CIR the plasma density is increased compared to the surrounding interplanetary medium (Sect.~\ref{sec:cir}). This density enhancement increases the electron thermal current Wind experiences, decreasing the spacecraft's floating potential to lower positive values \citep{garrett1981}. The spacecraft's charging timescales of tens of milliseconds are negligible compared to the duration of the CIRs of a few days \citep{chen+2013,wang+2014}. 

\begin{figure}
    \centering
    \includegraphics[width=\linewidth]{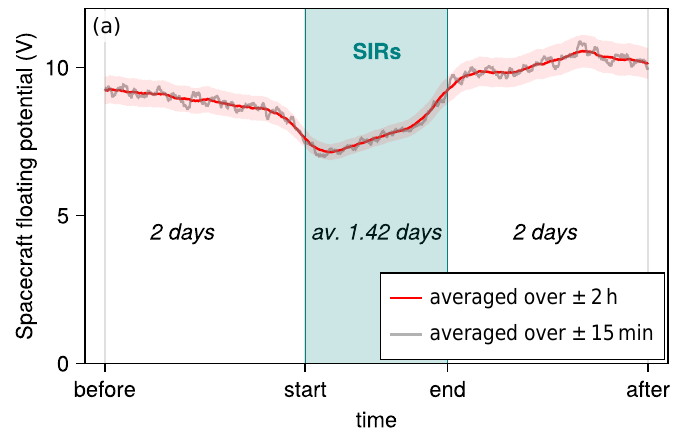}
    \includegraphics[width=\linewidth]{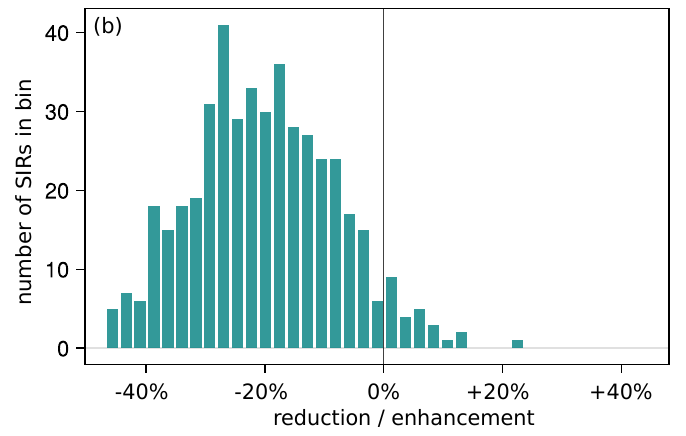}
    \caption{Scaled superposed epoch analysis (top panel) and histogram of the reduction or enhancement during each event (bottom panel) of Wind's floating potential during all SIRs that are associated with CIRs since 2005, in analogy to Fig.~\ref{fig:depletion_histoscaled}. The shaded red confidence band in the top panel corresponds to the standard deviation and a $1\,\si{V}$ measurement uncertainty for individual measurements of the floating potential. The first fourteen SIRs of 2005 were not taken into account because no spacecraft floating potential data was available. }
    \label{fig:depletion_potential}
\end{figure}

The results of a superposed epoch analysis (see Appendix~\ref{app:depletion}) of the floating potential data of Wind during CIRs are displayed in Fig.~\ref{fig:depletion_potential}a: the spacecraft floating potential is, on average, reduced during CIRs; only 30 out of 456 events feature an enhancement of the floating potential (Fig.~\ref{fig:depletion_potential}b). A lower positive floating potential increases the signal amplitude of a dust impact measured by a dipole antenna and, thus, increases the instrument's sensitivity to dust impacts \citep{shen+2023}. Therefore, the passage of a CIR should increase Wind's sensitivity to dust impacts.

However, Sect.~\ref{sec:depletion} showed that the observed dust impacts are on average reduced during CIRs, contrary to the effect that is imposed by the change of the floating potential. While some CIRs appear to enhance the number of dust impact detections, it was shown in Sect.~\ref{sec:depletion} that these CIRs cannot solely be responsible for the solar rotation signatures. Thus, the observed solar rotation signatures cannot be artificially created through CIR-induced changes of the floating potential but must come from a genuine reduction of dust impacts during CIRs. 
This indicates that the solar rotation signatures are not caused by periodic changes of the spacecraft's floating potential but genuine patterns in the dust environment.\footnote{The dust impact data of the STEREO spacecraft were measured with monopole antennas, which are only weakly affected by changes to the spacecraft floating potential \citep{shen+2023}; thus, the presence of the solar rotation signatures in the STEREO data by Chadda et al.\ (in prep.)\ corroborate our interpretation as genuine patterns in the dust.}

\subsubsection{Influence of the event selection by the fast time domain sampler}\label{sec:ana_neuralnet}

The data acquisition rate of the TDSF instrument is much higher than its data transmission rate. Therefore, the events measured by the TDSF are ranked by an algorithm, and only the highest-quality events are transmitted to Earth \citep{bougeret+1995}. Although the primary criterion by which the algorithm assesses an event's quality is the event's timestamp -- a new event would replace the oldest event in the buffer --  there were times when instead the events' signal amplitudes were used as the primary criterion \citep{goetz2024}. 

Ordinarily, the signals of dust impacts have higher amplitudes than those of most plasma waves. However, upstream of an interplanetary shock, bursts of high-amplitude plasma waves can occur \citep{wilson+2010}. These signals could fill the TDSF buffer and essentially make the instrument insensitive to dust impacts. If this were to systematically happen during CIRs, an artificial reduction of dust impacts by CIRs would be observed, resulting in artificial solar rotation signatures.

However, interplanetary shocks only rarely fill the TDSF buffer with high-amplitude events. These bursts of high-amplitude plasma waves following a shock last only for a few minutes \citep{cohen+2020}, and the TDSF buffer has room for only about twenty events that are continuously written onto the spacecraft's tape recorder \citep{goetz2024}. In contrast, the reduction of dust impact detections associated with CIRs is observed on timescales of one to two days (Sect.~\ref{sec:depletion}). It is unlikely that the day-long observed reduction of dust impacts could have been caused by a possible minutes-long insensitivity to dust impacts.

Furthermore, less than a third of all CIRs recorded by \citet{jian+2006} are associated with interplanetary shocks, whereas most CIR passages are associated with a reduction of dust impact detections (see Sect.~\ref{sec:depletion}); it is, thus, unlikely that the observed reductions of dust impacts are caused by an instrumental insensitivity to dust impacts upstream of interplanetary shocks. 
This indicates that it is unlikely that the solar rotation signatures are artificially caused by selection biases of the TDSF ranking algorithm, but that the solar rotation signatures are genuine patterns in the dust environment.\footnote{Unlike the TDSF of Wind/WAVES, the dust impact data measured with S/WAVES onboard the STEREO spacecraft is not affected by this ranking algorithm's selection issue, reaffirming that the solar rotation signatures are not caused by external influences but are genuine features of the dust environment.}

\section{Summary and conclusions}\label{sec:summary}

Signatures of the solar rotation were discovered in the observations of cosmic dust impacts on the Wind spacecraft. The time series of these dust impact data \citep{malaspina+wilson2016}, measured by Wind's plasma wave instrument, was investigated in detail. A frequency analysis of the daily number of dust impacts measured at $\mathrm{L}_1$ between 1 January 2005  and 31 August 2023 yielded the following conclusion (Sect.~\ref{sec:spectrum}): Solar rotation signatures are evident in the dust impact data as spectral peaks at ${\sim}27\,\si{d}$, $13.5\,\si{d}$, and $9\si{d}$.

Further, we investigated whether these solar rotation signatures are evident in the interplanetary or the interstellar component of dust impacts measured on Wind. More ISD impacts were observed when the spacecraft moved against the ISD inflow direction (February to April) compared to when it moved with the ISD inflow direction (August to October). Seasonal variations of the measured IDP impacts were assumed to be negligible. The solar rotation signatures were found in spectra of the daily number of dust impacts for both orbital configurations but were considerably stronger when the spacecraft moved against the ISD inflow direction, that is, when more ISD impacts were measured (Sect.~\ref{sec:withagainst}).

Measurable dust impacts from ISD occur more frequently during the focusing phase of the solar magnetic cycle compared to the defocusing phase of the solar magnetic cycle. The solar rotation signatures were found in spectra of the daily number of dust impacts during both the focusing phase and the defocusing phase of the solar magnetic cycle but were considerably stronger during the focusing phase, that is, when more ISD impacts were measured (Sect.~\ref{sec:focusdefocus}). This yielded the following conclusion: The solar rotation signatures are not caused exclusively by IDPs but are, at least partially, caused by ISD.

It is known that Jovian and Saturnian dust streams are affected by CIRs \citep{hsu+2010,flandes+2011}, and we therefore hypothesised that CIRs cause the solar rotation signatures. For this reason, a superposed epoch analysis for all CIRs was performed, and we found that dust impact detections are, on average, reduced during CIRs compared to the preceding and following time intervals (Sect.~\ref{sec:depletion}). Additionally, case studies of time periods of particular rates of occurrence of CIRs were investigated. We found that during time periods of long-lasting CIRs, the solar rotation signatures are powerful and that during time periods where no CIRs were observed, the solar rotation signatures vanish (Sect.~\ref{sec:casestudy}). A synthetic time series of dust impacts that are completely depleted during CIRs generated comparable spectra for these case studies (Sect.~\ref{sec:cir-model}). This yielded the following conclusions: The solar rotation signatures are correlated with and most likely caused by CIRs, and the number of dust impact detections is, on average, reduced during CIRs compared to the preceding and following time periods.

A competing hypothesis for the origin of the solar rotation signatures lies in the sector structure of the IMF, which is known to affect Jovian and Saturnian dust streams \citep{hamilton+1993,hsu+2010}. Therefore, case studies of time periods when the IMF sector structure was stable or chaotic were investigated. We found that the solar rotation signatures for spectra of the daily number of dust impacts are not correlated with the solar rotation signatures of the IMF polarity. Powerful solar rotation signatures were observed in the dust impacts even at times when the sector structure was chaotic, and weak solar rotation signatures were observed at times when the sector structure was stable (Sect.~\ref{sec:ana_hcs}). This yielded the following conclusion: The IMF sector structure local to the spacecraft cannot be the sole cause of the solar rotation signatures.

Solar rotation signatures have also been observed in spectra of Wind's floating potential \citep{wilson+2023}. It is known that changes in the spacecraft's floating potential affect the signal amplitudes of a dust impact measured by dipole antennas \citep{shen+2023}. Therefore, it may be possible that the solar rotation signatures in the dust impact spectra do not stem from a genuine modulation of dust impacts but are artificially induced by changing the instrument's sensitivity to dust impacts whenever a CIR passes. However, passing CIRs enhance the instrument's sensitivity to dust impacts, partially counteracting the observed reduction of dust impacts during CIRs instead of artificially inducing it (Sect.~\ref{sec:ana_potential}). Thus, we made the following conclusion: Periodic changes of the spacecraft's floating potential alone are an unlikely explanation of the solar rotation signatures in the dust impact data.

A similarly artificial origin of the solar rotation signatures could come from the fact that the TDS becomes temporarily insensitive to dust impacts upstream of interplanetary shocks due to a selection bias by the instrument. Interplanetary shocks can be associated with CIRs \citep{jian+2006} and can thus artificially cause the solar rotation signatures. However, the timescale on which the instrument would be biased is much shorter than the timescale of the observed reduction of dust impact measurements during CIRs. Furthermore, only a fraction of CIRs are associated with interplanetary shocks (see Sect.~\ref{sec:ana_neuralnet}). 
This yielded the following conclusion: A selection bias of the instrument against dust impact signals upstream of interplanetary shocks is unlikely to be the cause of the solar rotation signatures in the dust impact data.

In total, solar rotation signatures were found in the frequency spectra of the daily number of dust impacts recorded by Wind at periods of ${\sim}27\,\si{d}$ and its harmonics. Known instrumental effects were ruled out as the possible origin of these signatures, indicating that the signatures are genuine features of the dust environment. The most likely cause of these signatures is CIRs that reduce the number of dust impact detections as they pass the spacecraft. The solar rotation signatures showcase the close link between heliospheric and dust sciences, highlighting the need for cooperation and utilising the synergies between the two \citep[cf., e.g.][]{sterken+2023}.

The physical mechanism that causes the observed reduction of dust impacts during CIRs is not well understood. While it may be similar to the mechanisms proposed by \citet{ragot+2003} and \citet{wagner+2009} for CMEs, further modelling efforts are required to investigate this possibility.

In order to examine the influence of the solar rotation on dust in greater detail, it is essential to not only know the signal amplitude of a particle impact but to be able to infer the impact speed and impactor mass and size from it. Likewise, being able to constrain the direction from which an impacting dust particle comes would facilitate clearer distinctions between interstellar and interplanetary origins of the particles. 
This requires space-based missions with dedicated dust detectors. In the near future the Interstellar Mapping and Acceleration Probe at $\mathrm{L}_1$ will carry the Interstellar Dust Experiment; however, this dust detector infers the impact speed of a dust particle from orbital dynamics and from the particle's composition and does not determine it from measurements \citep{mccomas+2018}. A complementary (large-area) \enquote*{dust telescope} (i.e.\ a trajectory grid combined with a time of flight mass spectrometer) that can directly determine the impact speed would be essential to have, for example, on board the Lunar Gateway (\citealp{wozniakiewicz+2021,arnet2023}; Sterken et al., in prep.) or on board the proposed SunCHASER mission at $\mathrm{L}_4$ \citep{posner+2021,cho+2023}.

\begin{acknowledgements}
We would like to thank the anonymous referee for excellent and thought-provoking comments, and Keith Goetz, Ming-Hsueh \enquote*{Mitchell} Shen, O.~Chris St.~Cyr, Felix Dannert, Markus Bonse, Thomas Birbacher, Ingo Waldmann, and Kai Hou \enquote*{Gordon} Yip for their suggestions and advice. V.~J.\ Sterken, S.\ Hunziker, L.~R.\ Baalmann, and A.\ Péronne received funding from the European Union’s Horizon 2020 research and innovation programme under grant agreement N$^\circ$~851544 \-- ASTRODUST. S.\ Chadda and D.~M.\ Malaspina acknowledge support from STEM Routes Uplift program (Packard Foundation 2021-72553), the Undergraduate Research Opportunities Program (UROP) at the University of Colorado, Boulder, NASA grant \#80NSSC23K0286, and NASA grant \#80NSSC22K0753. This work makes use of the Julia programming language, v1.10.2 \citep{bezanson+2017}, and its packages LsqFit.jl, Plots.jl \citep{Plots.jl}, and SPICE.jl \citep{SPICE}.\\
\textit{Author contributions:} L.R.B.\ curated the data, performed the data analyses, and wrote and revised the manuscript. S.H.\ performed a preliminary data analysis and discovered the solar rotation signatures. A.P.\ performed the superposed epoch analyses and supplementary data analyses (cf.\ Péronne et al., in prep.) and revised the manuscript. J.W.K.\ devised the frequency analysis method and supported the data analyses; J.W.K.\ and K.-H.G.\ advised on the data analyses and the physical interpretation of their results, and revised the manuscript. D.M.M.\ and L.B.W.III acquired and provided the data and revised the manuscript; D.M.M.\ advised on the physical interpretation, and L.B.W.III advised on the instrumentation. C.S.\ performed a preliminary data analysis. S.C.\ enabled comparisons to spectra of the STEREO data (Chadda et al., in prep.). V.J.S.\ initiated and supervised the project, contributed to the original discovery, advised on the data analyses and the physical interpretation of their results, revised the manuscript, and acquired funding.
\end{acknowledgements}

\bibliographystyle{aa}
\bibliography{wind}

\appendix

\section{Data reduction}\label{app:reduction}

\begin{figure*}
    \centering
    \includegraphics[width=\linewidth]{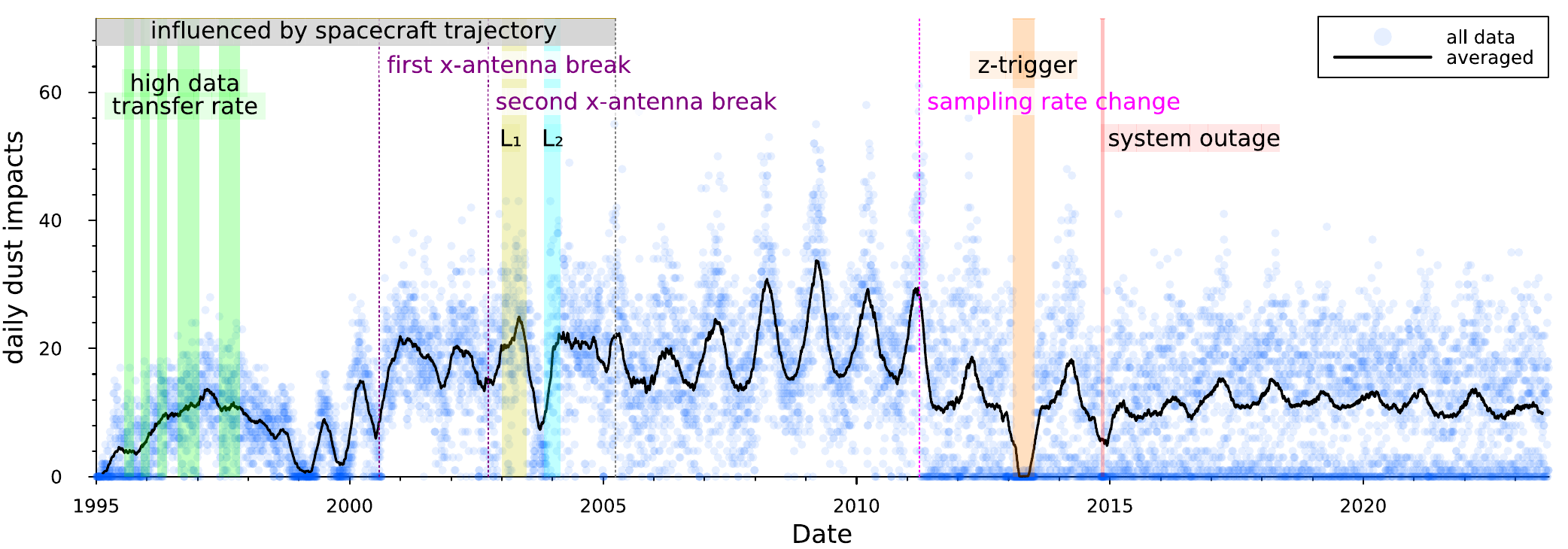}
    \caption{Unprocessed data presented as the daily number of impacts. Each pale blue disc corresponds to the number of impacts during one calendar day; a moving average within a centred window of $91\,\si{d}$ is given by the black line. The many external influences (see Appendix~\ref{app:lims_summary}), marked by vertical lines and shaded areas, are not taken into account in this representation of the unprocessed data.}
    \label{fig:overview_alltime}
\end{figure*}

The daily number of dust impacts contained in the dataset is depicted in Fig.~\ref{fig:overview_alltime} without applying the corrections of Sect.~\ref{sec:database}; for convenience, a moving average with a centred window of $91\,\si{d}$, corresponding to roughly a quarter of the spacecraft's orbital period, is graphed as well. 
Many external influences affected the recordings of dust impacts on Wind in its almost thirty years of service (Appendix~\ref{app:lims_summary}). These influences are not taken into account in Fig.~\ref{fig:overview_alltime}, but the times when they occurred are marked by vertical lines and shaded areas in the figure. 

These influences include the times when a twice-as-high data transfer rate was used before 1998, the two incidents on 3 August 2000 and 25 September 2002 in which the $x$-antenna was shortened, the change of the sampling rate on 1 April 2011, and the two time intervals when no dust impacts could be recorded due to an incorrect trigger setting in 2013 and due to a system outage in late 2014. For convenience, the times when Wind briefly visited $\mathrm{L}_1$ in 2003 and $\mathrm{L}_2$ in 2003-2004 are indicated as well. Before the spacecraft began permanently orbiting $\mathrm{L}_1$ in 2004-2005, the number of daily impacts appears to be strongly influenced by the distance to Earth (Appendix~\ref{app:data_dist_dep}); this time interval is marked as well.

These external influences must be taken into account when analysing the daily number of dust impacts. Changes to the instrument that influenced the general availability of the data, such as the use of the high data transfer rate before 1998, can be approximately corrected for by normalising the daily number of dust impacts by the total number of events the TDSF registered (Appendix~\ref{app:tdsf}). However, this introduces biases and should only be used if no other means of correction are available.

Another external influence that must be taken into account is the change of the sampling rate in 1 April 2011 (Appendix~\ref{app:data_ratechange}), which is especially evident in Fig.~\ref{fig:overview_alltime}: since April 2011, the moving average of the daily number of dust impacts is systematically shifted to lower values, and days with no impacts occur frequently and periodically. 

Some influences cannot be corrected for directly, such as the distance dependence of the daily number of dust impacts (Appendix~\ref{app:data_dist_dep}) or the antenna cuts. These must be taken into account otherwise, for example by removing the affected data.

\subsection{Summary of known influences}\label{app:lims_summary}

To summarise, the dust impact dataset is externally influenced by instrumental effects. Chronologically ordered, the known influences are:
\begin{itemize}
    \item The quality of a TDSF event is assessed by a ranking algorithm, referred to as a neural network. If the memory buffer of the time domain sampler (TDS) subsystem is full, the lowest-quality event is discarded. The best-quality event from the TDS memory buffer is written onto the spacecraft's tape recorder and later transmitted to Earth whenever the WAVES telemetry stream has sufficient idle capacity \citep{bougeret+1995}. The effect of a possible selection bias of the ranking algorithm is investigated in Sect.~\ref{sec:ana_neuralnet}.
    \item The TDS subsystem has the lowest telemetry priority of all WAVES subsystems \citep{goetz2022}. If other subsystems deliver more data, less telemetry bandwidth is available for the TDS. 
    \item The sampling rate of the TDSF is changed from $120\,\si{kS/s}$ to a lower setting of $30\,\si{kS/s}$ for one hour at 21:00 Terrestrial Time (TT) every two days. During this time, no dust impacts can be measured.
    \item Until the end of 1998, a twice-as-high telemetry acquisition bit rate was used whenever the spacecraft was close to Earth (Appendix~\ref{app:tdsf}). During these time periods, more detections of dust impacts could be sent to Earth.
    \item For unknown reasons the TDSF measured a low number of daily events from 1998 to 2000 (Appendix~\ref{app:tdsf}). This also decreased the number of observed dust impacts.
    \item The $X_+$-antenna arm was cut on 3 August 2000 and on 25 September 2002. This has altered the amplitude distribution measured by the $x$-dipole \citep{kellogg+2016}; see Appendix~\ref{app:data_removal}.
    \item For unknown reasons the dust impacts are correlated with the distance to Earth until the spacecraft began continuously orbiting $\mathrm{L}_1$ in 2005 (Appendix~\ref{app:data_dist_dep}). 
    \item Since 1 April 2011 the TDSF sampling rate is periodically changed from $120\,\si{kS/s}$ to $30\,\si{kS/s}$. For $45\,\si{h}\,36\,\si{min}$ out of every six days no dust impacts can be measured (cf. Appendix~\ref{app:data_ratechange}).
    \item From 1 February 2013 to 8 July 2013 the $z$-dipole was used as a trigger for the TDS. Only spurious events were measured during this time.
    \item From 25 October 2014 to 24 November 2014 the Wind spacecraft was unable to measure any dust impacts due to a double single event upset, likely caused by a cosmic ray particle impact.
\end{itemize}

For long-term analyses it is expedient to only use the dataset since 1 January 2005, when Wind continuously orbited $\mathrm{L}_1$. The periodically occurring measurement gaps and the general measurement gaps of 2013 and 2014 must be accounted for. Only dust impacts with morphological types A and B were taken into account (see Sect.~\ref{sec:database}; Appendix~\ref{app:type_cd}). When investigating the signal amplitudes generated by the dust impacts, only the signal amplitudes measured by the $y$-antenna (Channel 2) should be used (see Appendix~\ref{app:data_removal}).

\subsection{Removal of events}\label{app:data_removal}

The dataset contains a total of $\num{141438}$ dust impacts. According to the location flag given in the database, $\num{843}$ ($\approx 0.60\%$) were measured when Wind was located in the Earth's magnetosphere, and one was measured in the lunar wake; $\num{1417}$ ($\approx 1.00\%$) entries are assigned the location flag \enquote*{N/A}. These entries were removed from the dataset, resulting in $\num{139177}$ valid entries, $98.4\%$ of the whole dataset. Reducing the dataset to the time period since 2005 further decreased the total number of impacts to $\num{95895}$, and taking into account only impact signals with morphological types A and B yielded a total of only $\num{83203}$ dust impacts. 

From 1 February 2013 to 8 July 2013, the $z$-antenna was used as the trigger for TDSF recordings for $157\,\si{d}$ \citep{malaspina+wilson2016}. 30 spurious events were measured during this period; these events were removed from the dataset. 

\begin{figure*}
    \centering
    \includegraphics[width=\linewidth]{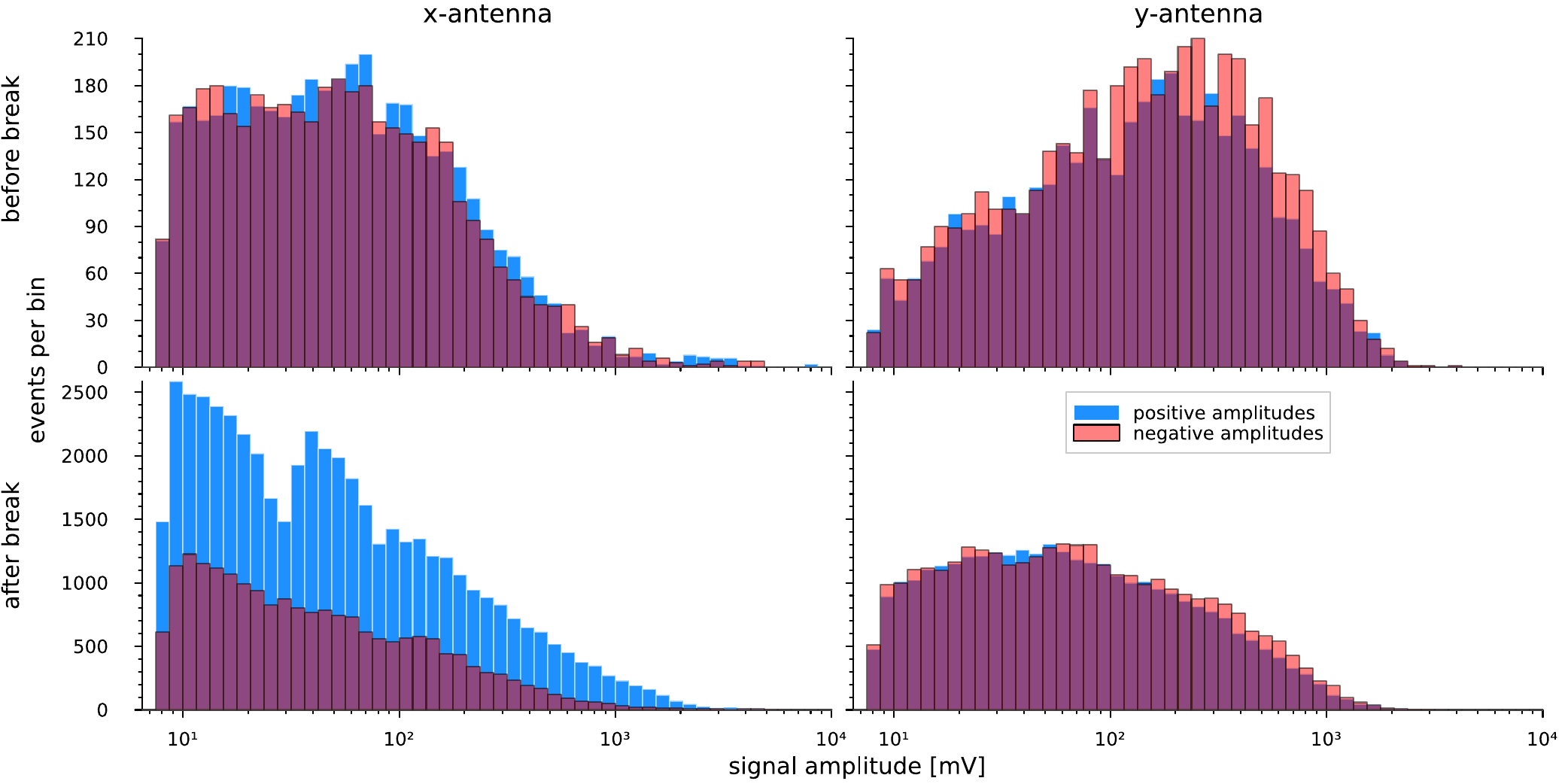}
    \caption{Amplitude distribution of type A and B dust impact signals before the first antenna break on 8 August 2000 (top row) and after the second antenna break on 9 September 2002 (bottom row) for the $x$-antenna (Channel~1; left column) and for the $y$-antenna (Channel~2; right column), separated into positive (blue) and negative (red) amplitudes.}
    \label{fig:distr-breaks}
\end{figure*}

Figure~\ref{fig:distr-breaks} displays the amplitude distributions for both the $x$- and the $y$-antenna before the first antenna break and after the second break of the $x$-antenna. After the $x$-antenna was broken, its amplitude distribution was affected by the asymmetry of its two arms. For example, the $x$-antenna measured many more impacts with positive amplitudes than with negative amplitudes. For this reason, only the amplitude values recorded by the $y$-antenna were used (see Sect.~\ref{sec:amplitude}). This reduced the number of available dust impact signals to $\num{58053}$, about two thirds of all impacts of morphological types A and B. 

If the signal amplitudes are not of interest, as was the case for the frequency analyses in Sect.~\ref{sec:fourier}, all $\num{83203}$ dust impacts with morphological types A and B can be taken into account. Because dust impacts signals of types C and D are rare compared to those of types A and B, the power spectra of dust impacts with only morphological types A and B do not systematically differ from the spectra of dust impacts with all morphological types, except that the slightly reduced number of dust impacts results in slightly lower periodogram amplitudes. Dust impacts with morphological types C and D are briefly investigated in Appendix~\ref{app:type_cd}.

\subsection{Normalisation of the dust impact rate by the total event rate measured by the fast time domain sampler}\label{app:tdsf}

The daily number of dust impacts in Fig.~\ref{fig:overview_alltime} shows many unexpected features. One of these is the bell-shaped feature spanning the time interval from 1995 to 1999, during which the data transfer rate repeatedly changed: until the end of 1998, a twice-as-high telemetry acquisition bit rate was in use whenever the spacecraft was closer than ${\sim}100\,R_{\otimes}\approx 6.4\times 10^{5}\,\si{km}$ to Earth; after 1998 this mode was no longer used due to concerns regarding the tape recorders on Wind \citep{goetz2022}. Of all WAVES subsystems the TDS benefited most from the twice-as-high bit rate; it was allocated three times as much bandwidth compared to the low bit rate \citep{bougeret+1995}. This strongly affects the rate of measured TDSF events. 
Another unexpected feature in Fig.~\ref{fig:overview_alltime} is the low number of dust impacts around the year 1999, which is caused by unspecified instrumental effects. 

\begin{figure}
    \centering
    \includegraphics[width=\linewidth]{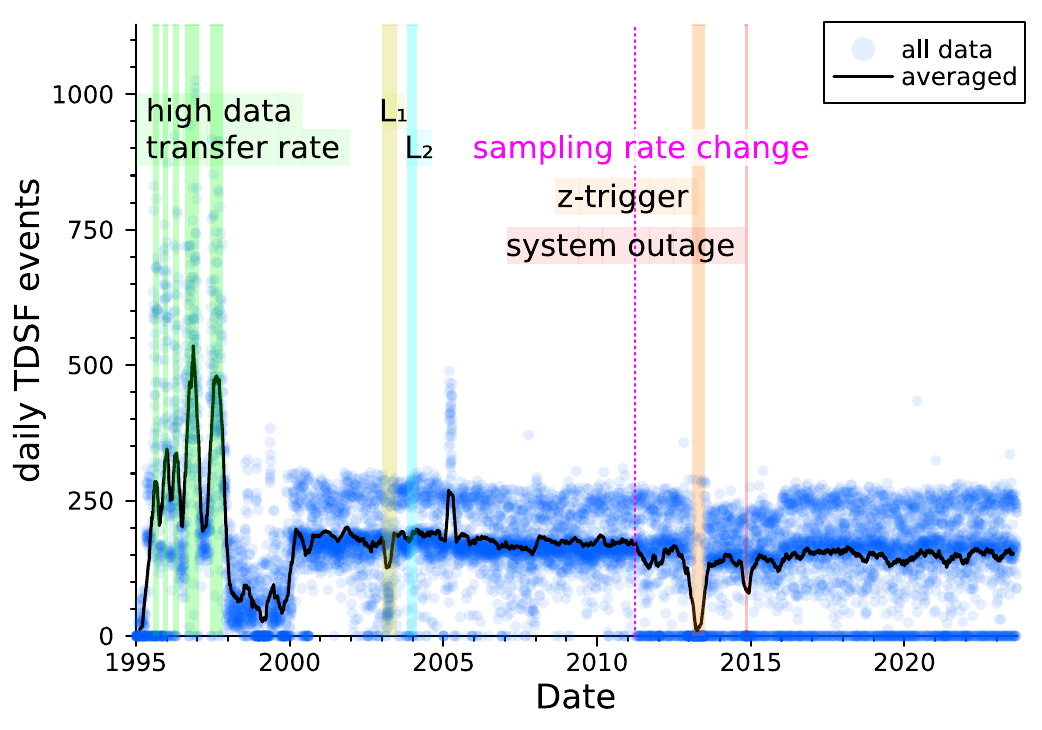}
    \caption{Daily number of TDSF events. The figure is displayed in the same manner as Fig.~\ref{fig:overview_alltime}.}
    \label{fig:tds_overview}
\end{figure}

These two features do not reflect physical changes in the dust environment but were caused by changes of the instrumentation. It can therefore be expedient to normalise the daily number of dust impacts by the daily number of all TDSF events, which is available in the dust impact dataset \citep{malaspina+wilson2016} and plotted in Fig.~\ref{fig:tds_overview}.

These data show the effect of the high data transfer rate until 1998 as the globally highest peaks, and that the overall rate of TDSF events between 1998 and 2000 is much lower compared to other time intervals. It is, at this point, not known why the daily number of TDSF events is much lower from 1998 to 2000 compared to other time intervals. The sudden increase in 2000 from, on average, well below 100 events per day to almost 200 events per day aligns with the shutdown of the Transient Gamma-Ray Spectrometer (TGRS) in January 2000 \citep{wilson+2021}; however, the WAVES instrument was always allocated the same portion of the bandwidth. 

The normalisation of the dust impact rate by the TDSF event rate does not correct for all external influences and introduces biases on the dust impact rate. This normalisation is only appropriate under the assumption that the TDSF rate is a useful measure for how many dust impacts could have been recorded on a given day. For example, when the twice-as-high telemetry rate was in use, the TDS instrument was allocated three times as much bandwidth. The normalisation implies that this telemetry rate would similarly allow three times as many dust impacts to be transmitted to Earth. 

This is, however, not necessarily true. Because the TDSF data acquisition rate of more than $2\,\si{Mb/s}$ is orders of magnitude higher than the telemetry rate of the WAVES instrument of $944\,\si{b/s}$, the events to be transmitted are selected by a ranking algorithm, referred to as a neural network \citep{bougeret+1995}. There is no fundamental reason to assume that this ranking algorithm is unbiased with respect to dust impact signals compared to other plasma wave signals. The effect of a selection bias against dust impact signals when high-amplitude plasma waves are registered upstream of interplanetary shocks is investigated in Sect.~\ref{sec:ana_neuralnet}.

Another bias is evident in the data: the daily number of TDSF events shows two distinct levels in Fig.~\ref{fig:tds_overview}: one slightly below 200 events per day, and one at approximately 250 events per day. The TDSF rate varies between these two levels regularly with a period of $6\,\si{d}$ for unknown reasons. Performing spectral analysis on the daily number of TDSF events reveals this periodicity as the most powerful peak of the periodogram. For the daily number of dust impacts before April 2011, spectral analysis finds only a weak $6\,\si{d}$-peak; after April 2011 this spectral peak becomes more powerful due to the periodic sampling rate changes. 

If the $6\,\si{d}$-periodicity of the TDSF events had directly caused a similarly powerful $6\,\si{d}$-periodicity of the dust impact data, normalising the daily number of dust impacts by the daily number of TDSF events should decrease the periodogram amplitude of this peak. However, performing this normalisation artificially increases the power of the $6\,\si{d}$-peak in the dust data, indicating that normalising the dust impact data by the TDSF data artificially induces instrumental biases.

To summarise, normalising the dust impact data by the TDSF data has the advantage of accounting for the instrumental effects that predominantly occurred before the year 2000; its disadvantage is that the normalisation induces patterns of the TDSF data that may not be present in the dust impact data, such as a powerful $6\,\si{d}$-periodicity. 
Therefore, it is essential to normalise the dust impact data by the TDSF data when investigating the time period before 2000, especially if short-term variations on timescales similar to $6\,\si{d}$ are not of particular interest. This is the case in Appendix~\ref{app:data_dist_dep}, which investigates the time period before Wind reached $L_{\mathrm{1}}$ in 2004. 

When short-term variations are of interest, especially if only the dust impacts observed at $\mathrm{L}_1$ post-2004 are investigated, the main advantage of the normalisation no longer applies and its disadvantage becomes relevant. Thus, with the exception of Appendix~\ref{app:data_dist_dep}, the dust impact data were not normalised by the TDSF data within the scope of this study.

\subsection{Correction for the sampling rate change in April 2011}\label{app:data_ratechange}

Since April 2011, due to a periodic change of the sampling rate, the TDSF is insensitive to dust impacts for $45\,\si{h}~36\,\si{min}=1.9\,\si{d}$ out of every six days (see Sect.~\ref{sec:windwaves}). Therefore, when investigating long-term patterns of the daily number of dust impacts, these periodic measurement gaps must be accounted for. When calculating the daily number of dust impacts it is expedient to delete the two days during which the $1.9\,\si{d}$-long gap occurred, even though this discards an additional $2\,\si{h}~24\,\si{min}$ of dust impacts every $6\,\si{d}$. 

The daily number of TDSF events is also affected by the sampling rate change, but its influence is not so easily quantified. In analogy to the treatment of the dust impact dataset, it is expedient to delete the same two full days of TDSF data whenever the sampling rate was changed.

\begin{figure}
    \centering
    \includegraphics[width=0.49\linewidth]{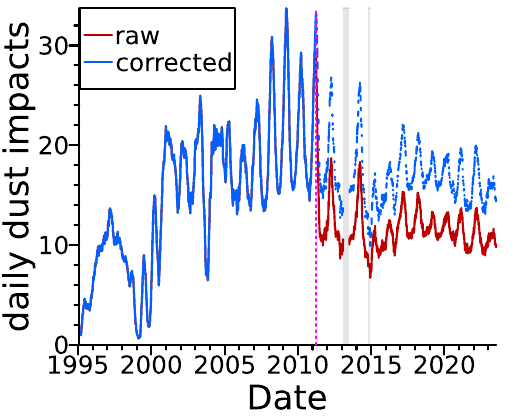}\hfill\includegraphics[width=0.49\linewidth]{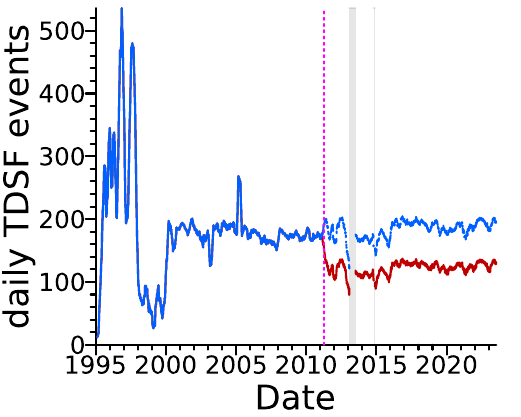}
    \includegraphics[width=\linewidth]{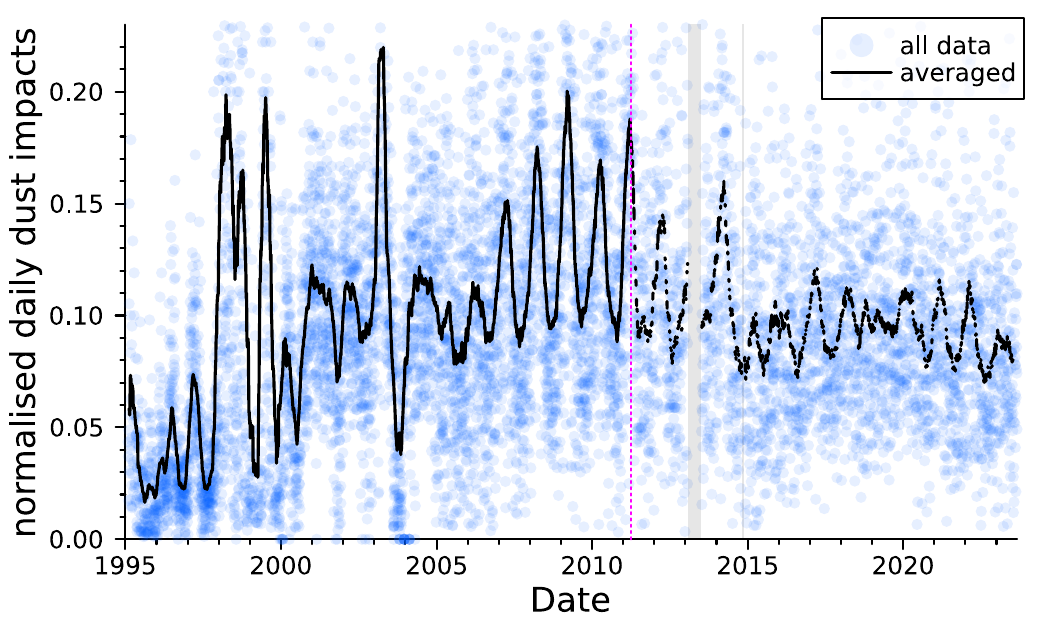}
    \caption{Daily number of dust impacts and TDSF events, corrected for the sampling rate change in April 2004. \textit{Top:} Daily number of dust impacts (left) and daily number of TDSF events (right) as moving averages with centred windows of $91\,\si{d}$. The red curves depict the \enquote*{raw} data -- same as in Figs.~\ref{fig:overview_alltime}~and~\ref{fig:tds_overview} but additionally taking into account the data gaps of 2013 and 2014, which are marked by grey bars. The blue curves depict the same data but corrected for the change of the sampling rate in April 2011, which is marked by a vertical dotted magenta line.\\ 
    \textit{Bottom:} Daily number of dust impacts normalised by the daily number of TDSF events, graphed for each individual day (blue discs) and as a moving average with a centred window of $91\,\si{d}$ (black curve). The data have been corrected for the sampling rate change and the data gaps.}
    \label{fig:ratefix}
\end{figure}

This is depicted in Fig.~\ref{fig:ratefix}, where the top left panel shows a $91\,\si{d}$ centred moving average of the daily number of dust impacts before and after correcting for the sampling rate change by deleting two full days of data; the top right panel shows the same for the daily number of TDSF events. As the figure indicates, the corrected data shows a similar pattern post-2011 and pre-2011: for the daily number of dust impacts, the yearly minimum before 2011 ranged around $15$ impacts per day; for the uncorrected (\enquote*{raw}) data, it dropped to about $10$ impacts per day, whereas for the corrected data the minima remain at about $15$ impacts per day. Similarly, the TDSF rate before April 2011 stands at about $180$ events per day; the uncorrected (\enquote*{raw}) data dropped to about $150$ events per day after April 2011, and the corrected data again match the pre-2011 level.

The bottom panel of Fig.~\ref{fig:ratefix} depicts the TDSF-normalised daily number of dust impacts. The figure shows that the bell-shaped feature before 1998 now resembles the yearly modulation expected from Wind's orbit around the Sun. The high values in 1998 and 1999 are most likely unphysical; during this time, the TDSF rate was unusually low for unknown reasons. The unusual feature in 2003-2004, which is present in both the dust impacts, the TDSF events, and the normalised dust impacts, is assumed to be unphysical; its origin is unknown. After 2005, the normalised dust impact rate is overall similar to the unnormalised dust impact rate because the TDSF rate remains, on average, reasonably constant.
The time interval before 2005, when Wind's orbit frequently changed, is investigated in Appendix~\ref{app:data_dist_dep}.

\subsection{Distance dependence pre 2005}\label{app:data_dist_dep}

\begin{figure}
    \centering
    \includegraphics[width=\linewidth]{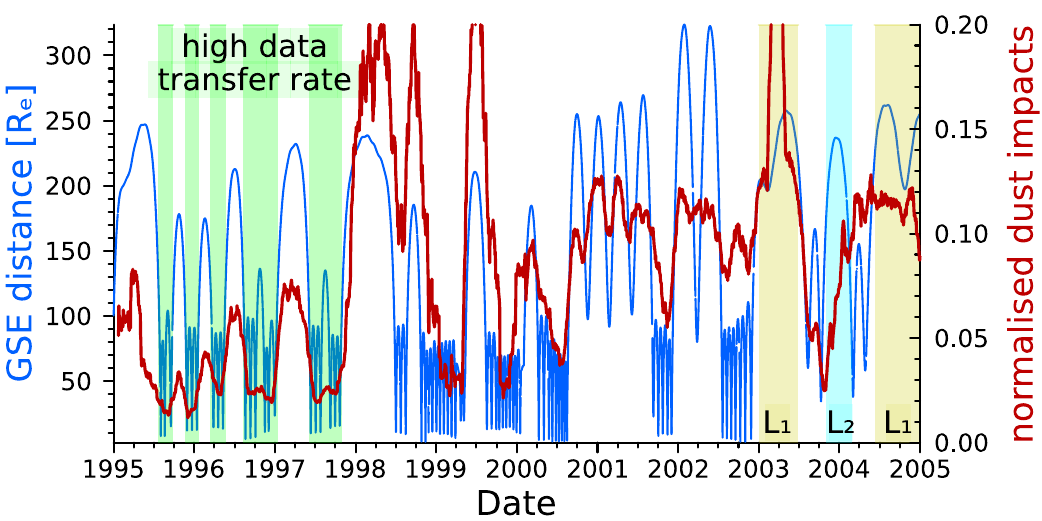}
    \caption{Distance of the spacecraft to Earth (smooth blue curve; left vertical axis) and normalised daily dust impacts (jagged red curve; right vertical axis) with a centred moving average of $45\,\si{d}$. The time intervals when the high data transfer rate was in use and the times spent at the Lagrange points are indicated by shaded areas. The distance to Earth is given in Earth radii, $1\,R_{\mathrm{e}}\approx6378\,\si{km}$.}
    \label{fig:dist_vs_relcounts}
\end{figure}

Figure~\ref{fig:dist_vs_relcounts} shows both the normalised daily number of dust impacts and the spacecraft's distance to Earth in the time interval of 1995 to 2005. During this time, Wind frequently changed orbits \citep[cf.][Fig.~1]{wilson+2021}. For convenience, the normalised number of dust impacts has been graphed in Fig.~\ref{fig:dist_vs_relcounts} as a centred moving average with a window width of $45\,\si{d}$. As the figure shows, the two displayed curves appear to be correlated with each other; a longer distance to Earth coincides with a higher normalised number of dust impacts. This correlation is not perfect: in the time interval between 1998 and 2000, where fewer TDSF events were measured, the normalised number of dust impacts is much higher than for other time intervals, excepting the brief visit to $\mathrm{L}_1$ in 2003; and during the visit to $\mathrm{L}_2$ in 2003-2004, no such correlation between the normalised number of dust impacts and the distance to Earth seems apparent. However, over the entire time interval between 1995 to 2005, the two curves are correlated; Pearson's correlation coefficient gives a value of $0.5$ compared to $-0.03$ for the time period between 2005 and 2021, when Wind continuously orbited $\mathrm{L}_1$.

This apparent distance dependence of the rate of dust impacts on the spacecraft is unexpected. Its cause is unknown and requires further investigation. To avoid unduly affecting the data analysis, all data before 2005 were excluded for the analyses performed in this work.

\FloatBarrier
\section{Details of the frequency analysis}\label{app:methodeval}

The method for frequency analysis proposed by \citet[][Suppl.~Mat.]{kirchner+2013} resembles the Lomb-Scargle transform \citep{lomb1976,scargle1982} with the time-shift factor, $\tau$, modified to produce mutually orthogonal cosine and sine terms and thus eliminate the phase and amplitude artefacts of the Lomb-Scargle method; see Eqs. (S5-S9) of \citet{kirchner+2013} for a formal description. 

Within the scope of this publication, spectra were evaluated on a frequency grid that ranged from the fundamental frequency, $f_{\min}=1/(t_{\max}-t_{\min})$, where $t_{\max}-t_{\min}$ is the total time interval covered by the dataset, to $f_{\max}=f_{\mathrm{Ny}}=0.5f_{\mathrm{S}}$, where $f_{\mathrm{Ny}}$ is the Nyquist frequency of evenly sampled data, which is half the sampling frequency, $f_{\mathrm{S}}$. This grid was divided into $n/2$ equidistant points, where $n$ is the number of data points. If the dataset contained no gaps, this frequency grid would reduce to the natural frequencies of the DFT. The frequency grid was oversampled by a factor of $k$, which increased the sampling density by that factor $k$. 

As an example, the general dataset at $\mathrm{L}_1$ covers the time range from 7 January 2005 to 31 August 2023, which is a total time span of $6810\,\si{d}$. Thus, the fundamental frequency is $f_{\min}=1/(6810\,\si{d})$. Because the time series of daily numbers of impacts are analysed, the sampling period is $1\,\si{d}$ and, thus, the Nyquist frequency is $f_{\max}=1/(2\,\si{d})$. If the dataset were to contain no gaps, i.e.\ if it were evenly sampled, the frequency grid would consist of $n/2=3405$ equidistant frequencies ranging from $f_{\min}=1/(6810\,\si{d})$ to $f_{\max}=1/(2\,\si{d})$.

However, in this dataset, only $n=5161$ data points contain evaluable data and, thus, the frequency grid is coarser than the frequency grid of evenly sampled data. This is ameliorated by oversampling the frequency grid by a factor $k$, i.e.\ the frequency grid ranging from the same $f_{\min}$ to $f_{\max}$ is divided into $nk/2$ equidistant frequencies. Typical values are for $k$ are $k\in\{2,5,10,15,20\}$, depending on the desired sampling density and computational cost. While oversampling doesn't increase the resolution of the spectrum, i.e.\ neighbouring peaks cannot be resolved any better on an oversampled frequency grid, it does increase the precision of the frequency corresponding to a peak. 
For evenly sampled data, the method proposed by \citet{kirchner+2013} perfectly reproduces the results of the DFT.

The spectra are given as periodograms; the periodogram amplitude is the squared amplitude of the fitted sinusoids, here in units of $(\text{amplitude})^2=(\text{number of dust impacts per day})^2$. In the literature, spectra are also often given in terms of the power spectral density (PSD), which has the dimension $(\text{amplitude})^2/\text{frequency}$; here the frequency is measured in $\si{d}^{-1}$. The periodogram amplitude can be converted to the PSD by multiplying with the total time interval covered by the data. The advantage of the periodogram compared to the PSD is that the periodogram amplitudes of peaks that correspond to periodic signals are independent from the length of the dataset; the broadband noise level decreases for longer datasets. Conversely, in a PSD the broadband noise level is independent of the length of the dataset, whereas the spectral power of peaks that correspond to periodic signals increase with the dataset length. In this work, spectra are represented by periodograms, facilitating a comparison of the peaks of periodic signals between data subsets of different lengths. 

Before computing a spectrum, the data were linearly detrended by subtracting from the data a linear fit to the data. The linear fitting was performed by the LsqFit.jl\footnote{\url{https://github.com/JuliaNLSolvers/LsqFit.jl}} package for the Julia programming language. 

\begin{figure}
    \centering
    \includegraphics[width=\linewidth]{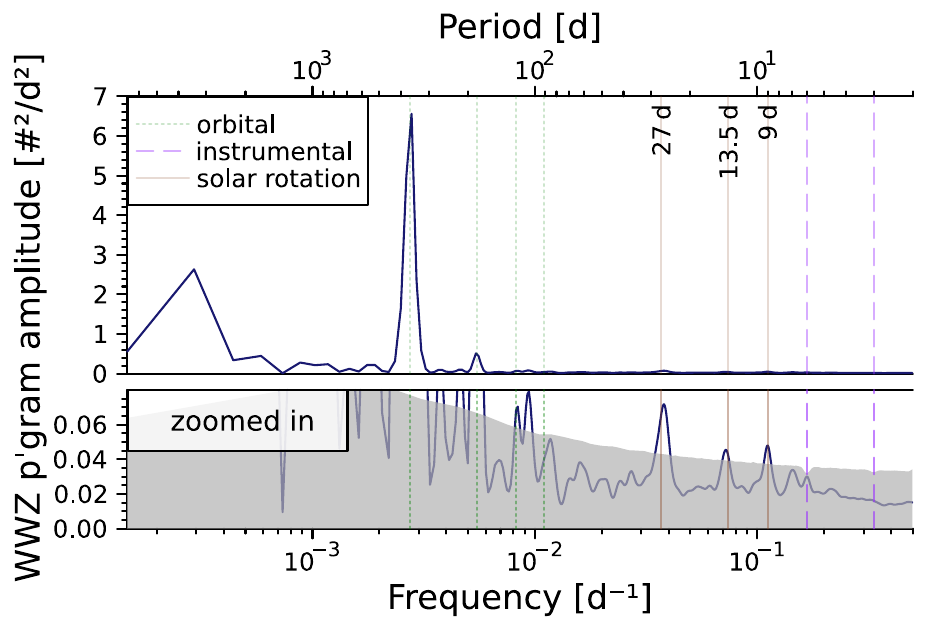}
    \caption{WWZ periodogram of the daily number of dust impacts recorded at $\mathrm{L}_1$, in analogy to Fig.~\ref{fig:spectrum}. The bottom panel shows a zoomed-in view of the full spectrum, with the estimated $95\%$ significance thresholds indicated by a grey shaded area.}
    \label{fig:wwz}
\end{figure}

In addition to the phase and amplitude artefacts mentioned above, uneven sampling can also result in spectral aliasing, in which strong signals at one frequency appear as spurious signals at other frequencies. These aliases can be effectively suppressed using Kirchner and Neal's variant of Foster's Weighted Wavelet Z approach \citep[WWZ;][]{foster1996,kirchner+2013}. Figure~\ref{fig:wwz} shows this approach applied to the same data shown in Fig.~\ref{fig:spectrum}. The solar rotation peaks at $27\,\si{d}$, $13.5\,\si{d}$, and $9\,\si{d}$ are clearly visible in both Fig.~\ref{fig:wwz} and Fig.~\ref{fig:spectrum}, indicating that they are not aliases generated by the uneven sampling of the time series. The much stronger peak at $6\,\si{d}$ in Fig.~\ref{fig:spectrum}, which arises as an alias of the $365\,\si{d}$-cycle, is suppressed below the $95\%$ significance threshold in Fig.~\ref{fig:wwz}.

\subsection{Estimation of the $95\%$ significance thresholds}\label{app:noiselevel}

The $95\%$ significance thresholds of a periodogram were estimated through bootstrapping:
\begin{compactenum}[(1)]
    \item The non-NaN values of the underlying time series were reshuffled, i.e.\ randomly reordered. NaN values, i.e.\ known measurement gaps, were not affected by this reshuffling and remained at the same timestamps. The total sum of all values remained the same.
    \item The periodogram of the reshuffled time series was calculated with the previously introduced methodology; the time series was linearly detrended beforehand. To decrease computational costs, the frequency grid was typically subsampled by a factor of $k=0.1$. 
    \item This was repeated for a total of 5000 periodograms of randomly reshuffled time series. 
    \item For each frequency the $95\%$ significance thresholds were estimated as the $95^{\textrm{th}}$ percentile of the periodogram amplitudes of all 5000 periodograms.
\end{compactenum}

Due to the measurement gaps, i.e.\ because the time series is not evenly sampled, the estimated $95\%$ significance thresholds are not expected to be constant over all frequencies.

\subsection{Effect of the $6\,\si{d}$-periodic gaps}\label{app:eval_6dgap}

Every six days, measured by the day-of-year of each year, a change of the instrument's sampling rate made it impossible to detect impacts of dust particles for $45\,\si{h}\,36\,\si{min}$ (Sect.~\ref{sec:windwaves}). Because the daily number of dust impacts are analysed in this publication, this leads to two days of very few dust impacts every six days. To combat this, the entire two days were removed from the dataset, creating periodic gaps (Sect.~\ref{app:data_ratechange}).

Periodic gaps do not, by themselves, create additional peaks in the power spectrum. However, where a sinusoid signal of frequency $f_{\mathrm{sig}}$ will show a spectral peak at only $f_{\mathrm{sig}}$, the same signal with $f_{\mathrm{gap}}$-periodic gaps will show additional peaks. These stem from the convolution of the original signal with the receiver function, which is zero during gaps and one everywhere else. These peaks appear at frequencies $\abs{f_{\mathrm{sig},i} \pm f_{\mathrm{gap},j}}$, where $i$ iterates over every genuine frequency of the signal, and $j$ iterates over the relevant frequency and all harmonics of the receiver function. 

Therefore, a spectrum of data with periodic gaps will be affected. In the case of $f_{\mathrm{sig}}\ll f_{\mathrm{gap}}$ these peaks will appear close to $f_{\mathrm{gap}}$; for example for the seasonal variability ($f_{\mathrm{sig}}^{-1}\approx365\,\si{d}$) and $f_{\mathrm{gap}}^{-1}=6\,\si{d}$, additional peaks appear at $f_{\mathrm{new}}^{-1}\in\{5.9\,\si{d}, 6.1\,\si{d}\}$. For $f_{\mathrm{sig}}\lesssim f_{\mathrm{gap}}$, these gaps are more easily confused with genuine periodicities; for example, for $f_{\mathrm{sig}}^{-1}=9\,\si{d}$ an additional peak would appear at $f_{\mathrm{new}}^{-1}=18\,\si{d}$.

\begin{figure}
    \centering
    \includegraphics[width=\linewidth]{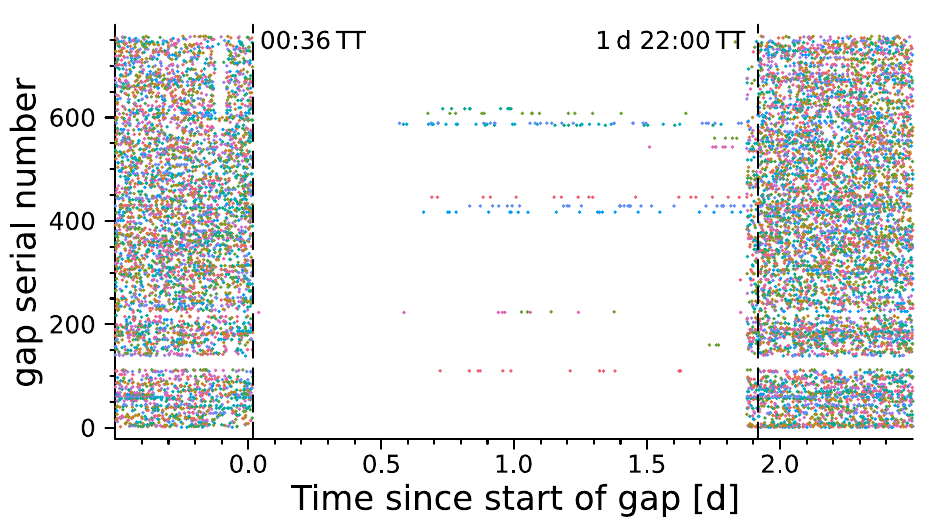}
    \caption{Superposed epoch analysis of all gaps stemming from the $6\,\si{d}$-periodic sampling rate change; serial number of the gaps versus time since the start of the two days of removed data. Each dot corresponds to one observed dust impact; each line of identically coloured dots corresponds to the same epoch, i.e.\ gap serial number. The vertical dashed black lines correspond to the reported timing of the sampling rate change.}
    \label{fig:chree-gaps}
\end{figure}

This effect is one reason why peaks can appear at a period of $6\,\si{d}$ and its $3\,\si{d}$-harmonic in the spectra, for example in Fig.~\ref{fig:spectrum}. It can be suppressed by more elaborate methods such as the WWZ (see Fig.~\ref{fig:wwz}). Another potential source of these spectral peaks could have been that the timing of the measurement gaps was misaligned with the removal of the data. This, however, could be disproved (see Fig.~\ref{fig:chree-gaps}). 

Figure~\ref{fig:chree-gaps} shows a superposed epoch analysis \citep{chree1913} of all gaps corresponding to the $6\,\si{d}$-periodic sampling rate changes. The sampling rate was reported to be reduced from $\text{00:24 TT}=0.01\bar{6}\,\si{d}$ until $1\,\si{d}+\text{22:00 TT}=1.91\bar{6}\,\si{d}$, which has been marked in the figure. While the start time of these sampling rate changes agrees with the observations, many dust impacts were already measured in the hour after $1\,\si{d}+\text{21:00 TT}=1.875\,\si{d}$. Nevertheless, the interval of reduced sampling rate lies fully within the interval $[0, 2]\,\si{d}$ that was removed from the dataset. The observational gaps are not misaligned with the data that was removed. 

Some additional features remain in Fig.~\ref{fig:chree-gaps}: the two horizontal gaps around gap serial numbers 130 and 220 correspond to the long-term gaps of 2013 and 2014 (see Sect.~\ref{app:data_removal}), respectively. For 13 gaps dust impacts were observed during the time where the detector was unable to detect dust impacts; the reason for this is unknown. The void that appears between $[-0.125,-0.08\bar{3}]\,\si{d}$, especially visible at high gap serial numbers, corresponds to a short-term sampling rate change that occurred every $2\,\si{d}$ (see Appendix~\ref{app:lims_summary}).

\begin{figure}
    \centering
    \includegraphics[width=\linewidth]{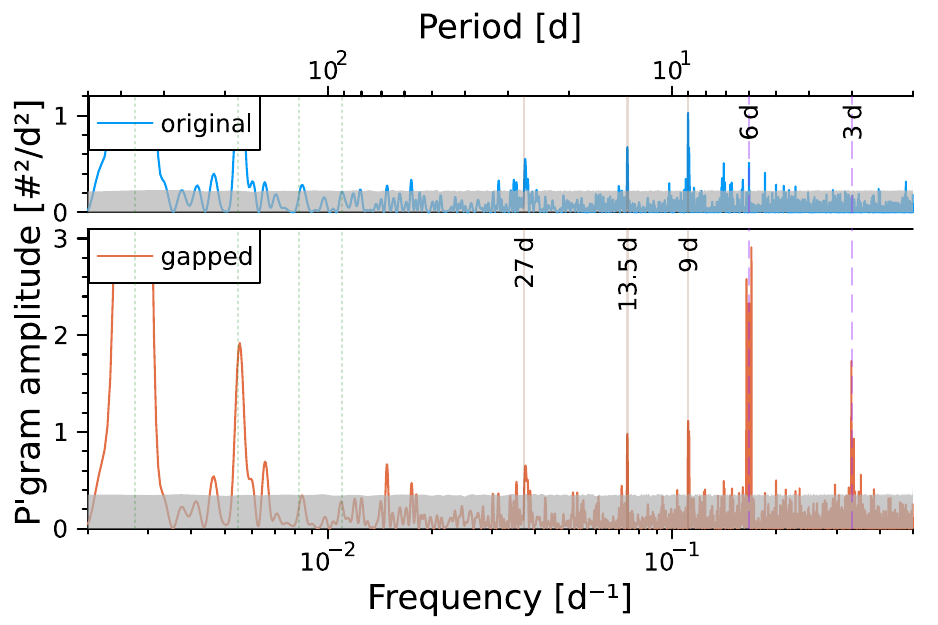}
    \caption{Periodogram of the daily number of dust impacts with morphological types A and B before April 2011, original data (blue curve, top panel) and data with artificially induced $6\,\si{d}$-periodic gaps (orange curve, bottom panel), evaluated at the DFT's natural frequencies and oversampled by a factor of five. The two panels are scaled identically. Frequencies of interest are marked by vertical lines, as in Fig.~\ref{fig:spectrum}. The estimated $95\%$ significance thresholds are indicated by grey shaded areas.}
    \label{fig:artifgap}
\end{figure}

The effect of the $6\,\si{d}$-periodic gaps on the generated spectra is displayed in Fig.~\ref{fig:artifgap}. For the data before the sampling rate was periodically changed in April 2011, artificial gaps following the same pattern at which the instrumental gaps occurred were induced on the originally continuous time series; the spectra for both the original and the gapped time series are shown in the figure. 

The most obvious difference between the two spectra are the additional peaks close to $6\,\si{d}$ and $3\,\si{d}$, which result from the convolution of the original time series with the receiver function, which is zero during gaps and one everywhere else. Some peaks of the original spectrum, such as the $9\,\si{d}$- and the $13.5\,\si{d}$-peaks, appear at the exact same period but have slightly elevated powers. The ${\sim}27\,\si{d}$-peak has been shifted from $26.69\,\si{d}$ to $26.59\,\si{d}$ and likewise increased in power. An additional peak at $67.56\,\si{d}$ has appeared. These are effects that must be taken into account when evaluating the spectra.

\subsection{Periodic signal with multiple peaks}\label{app:test-spikes}

\begin{figure}
    \centering
    \includegraphics[width=\linewidth]{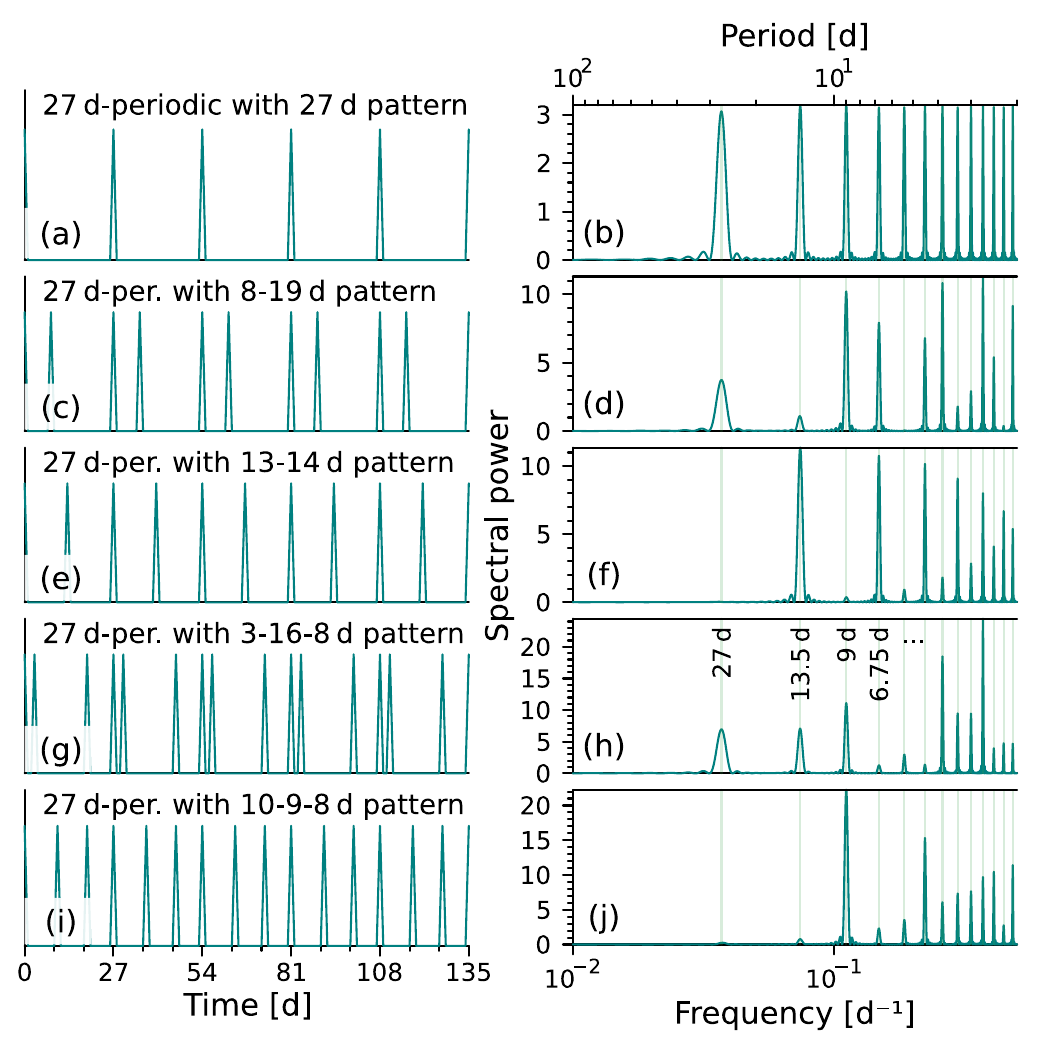}
    \caption{Time series (left column) and corresponding periodograms (right column) for $27\,\si{d}$-periodic test signals with a different pattern of spikes per period. The spectral peaks mark the $27\,\si{d}$-period and its harmonics.}
    \label{fig:test-spikes}
\end{figure}

This appendix illustrates the influence of a spike-train-like signal with multiple spikes per period, corresponding to, for example, multiple CIRs occurring in the same time period (see Sect.~\ref{sec:casestudy}). Figure~\ref{fig:test-spikes} shows five examples for these spiky functions and the corresponding periodograms. 

The first example (a) shows a $27\,\si{d}$-periodic signal consisting of a single spike, akin to a single, long-lasting CIR. The corresponding power spectrum (b) shows peaks at the primary and all harmonics of the $27\,\si{d}$-frequency. All peaks have roughly the same power. This is expected; it is well known that the Fourier spectrum of a Dirac comb is itself a Dirac comb \citep[e.g.][]{vanderplas2018}. 

In the second example (c-d) another spike was added to the periodic signal; the spikes occur in an alternating pattern of eight and nineteen days, abbreviated as an 8-19 pattern. The amplitude of the spikes is the same as in the first example. The periodogram shows peaks at the same frequencies; however, their power is distributed differently. The first harmonic ($13.5\,\si{d}$) has become less powerful and the second and third harmonics have become more powerful than the primary peak, for example. 

The relative power of the peaks depends on the phase shift between the spikes. If the spikes occur close to each other (e.g.\ a 2-25 pattern), the primary frequency remains the most powerful peak and the power decreases for higher harmonics. If the spikes are equidistant, the signal becomes $13.5\,\si{d}$-periodic; for a 13-14 pattern (e-f), the $13.5\,\si{d}$ peak and all other odd harmonics of the $27\,\si{d}$ are powerful, whereas the primary and all even harmonics almost vanish. 

In the fourth and fifth examples (g-j), a third spike per period has been added. Similar to the case of two spikes per period, if the spikes are unevenly distributed among the period, the powers are unevenly distributed: if the spikes occur closely together, the primary frequency is powerful and the power decreases for higher harmonics (g-h); if the spikes are evenly distributed, the signal becomes $9\,\si{d}$-periodic, so only the second harmonic and every successive third harmonic remain, whereas all other harmonics and the primary peak vanish (i-j).

Thus, assuming that CIRs cause spike-like features in the time series, for example by reducing or enhancing detections of dust particle impacts, the number of CIRs and how they are spatially distributed among one solar rotation will affect the powers of the solar rotation signatures. The spectral signature of a reduction or an enhancement are indistinguishable in a periodogram.

\subsection{Identification of CIRs}\label{app:cir_ident}

CIRs are usually not identified directly as cohesive, co-rotating structures but as individual SIRs that pass by a spacecraft (see Sect.~\ref{sec:cir}). The dataset by \citet{jian+2006,jian2009}, for example, used measurements made with Wind and with the Advanced Composition Explorer to identify these individual SIRs. Although the dataset contains a flag that indicates whether the respective SIR is part of a CIR, it does not assign these SIRs to specific CIRs. Therefore, the dataset contains no identification of CIRs nor any indication on their number of co-rotations. Similarly, the dataset by \citet{hajra+2022} only contains a list of individual SIRs without any indication of CIRs. To identify CIRs in these datasets, it is therefore necessary to identify which SIRs are part of CIRs.

If a SIR succeeds another SIR by roughly a Carrington rotation period, it is likely that both SIRs are the same recurring structure that co-rotates with the Sun, which is then identified as a CIR. Using the two datasets of individual SIRs, a set of SIRs were identified as a CIR if all constituent SIRs occurred at multiples of a Carrington rotation period, $T=n\cdot T_{\mathrm{C}}, n\in\mathbb{N}$, to each other, where $T_{\mathrm{C}}=27.2753\,\si{d}$, allowing for some slack, $T\pm t_{\mathrm{sl}}$.

This method will not perfectly identify every CIR correctly: if two unrelated SIRs accidentally occur within $T\pm t_{\mathrm{sl}}$, they will falsely be recognised as belonging to the same CIR. If, for any reason, a single SIR during a long-lasting CIR is not detected, the CIR will be identified as two unrelated CIRs. Different choices of $t_{\mathrm{sl}}$ lead to different sets of CIRs. If a SIR could ambiguously be related to two CIRs that are unrelated amongst themselves, the SIR is assigned to the earlier CIR even if it were naturally part of the second one. 

In practice, a SIR was recognised as part of a potential CIR if it fulfilled two criteria:
\begin{enumerate}[1.]
    \item the individual SIR had to occur within $T_{\mathrm{C}}\pm 3\,\si{d}$ after the previous SIR that is part of the potential CIR, and
    \item the individual SIR had to occur within $n\cdot T_{\mathrm{C}} \pm 5\,\si{d}, n\in\mathbb{N}$ of all other previous SIRs that are part of the same CIR.
\end{enumerate}
Applying this method lead to the list of CIRs that are displayed in Fig.~\ref{fig:unique_cirs}.

\subsection{Details of the interplanetary magnetic field sector structure case studies}\label{app:hcs-casestudies}

This appendix presents case studies for the IMF sector structure (see Sect.~\ref{sec:ana_hcs}). 
The list of sector structure polarities by \citet{svalgaard2023_struc} gives the polarity of the IMF at Earth for each day as either an \enquote*{X} (negative polarity, pointing towards the Sun), as a period (positive polarity, pointing away from the Sun), or an asterisk (mixed polarity, \enquote{zero}). To calculate power spectra of the sector structure, these values were mapped to $-1$, $+1$, and $0$, respectively.

Several time periods of noteworthy sector structures can be observed: a stable four-sector structure in 2006; a chaotic structure in 2012; a stable two-sector structure in 2016; a chaotic structure from June 2020 until June 2021; and a stable two-sector structure in 2022. All of these time periods were taken to be $365\,\si{d}$ long. If the solar rotation signatures were caused by the local IMF sector structure, clear solar rotation signatures should be observable during times of a stable sector structure, and no solar rotation signatures should be observable during times of a chaotic sector structure. 

\begin{figure*}
    \centering
    ~\hfill \textbf{\large Daily number of dust impacts} \hfill~\hfill \textbf{\large Daily IMF polarity} \hfill ~\\
    \includegraphics[width=0.49\linewidth]{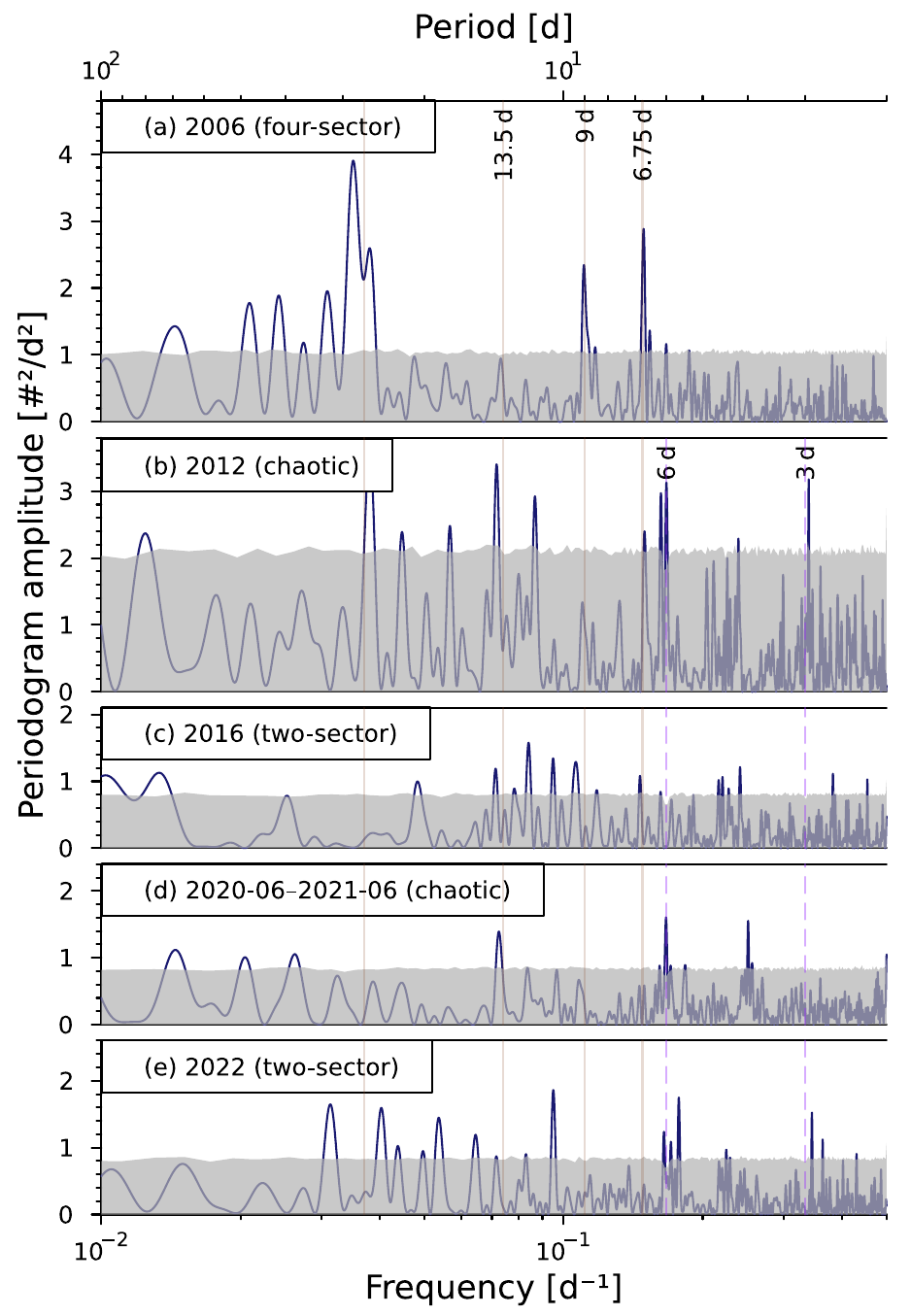}
    \includegraphics[width=0.49\linewidth]{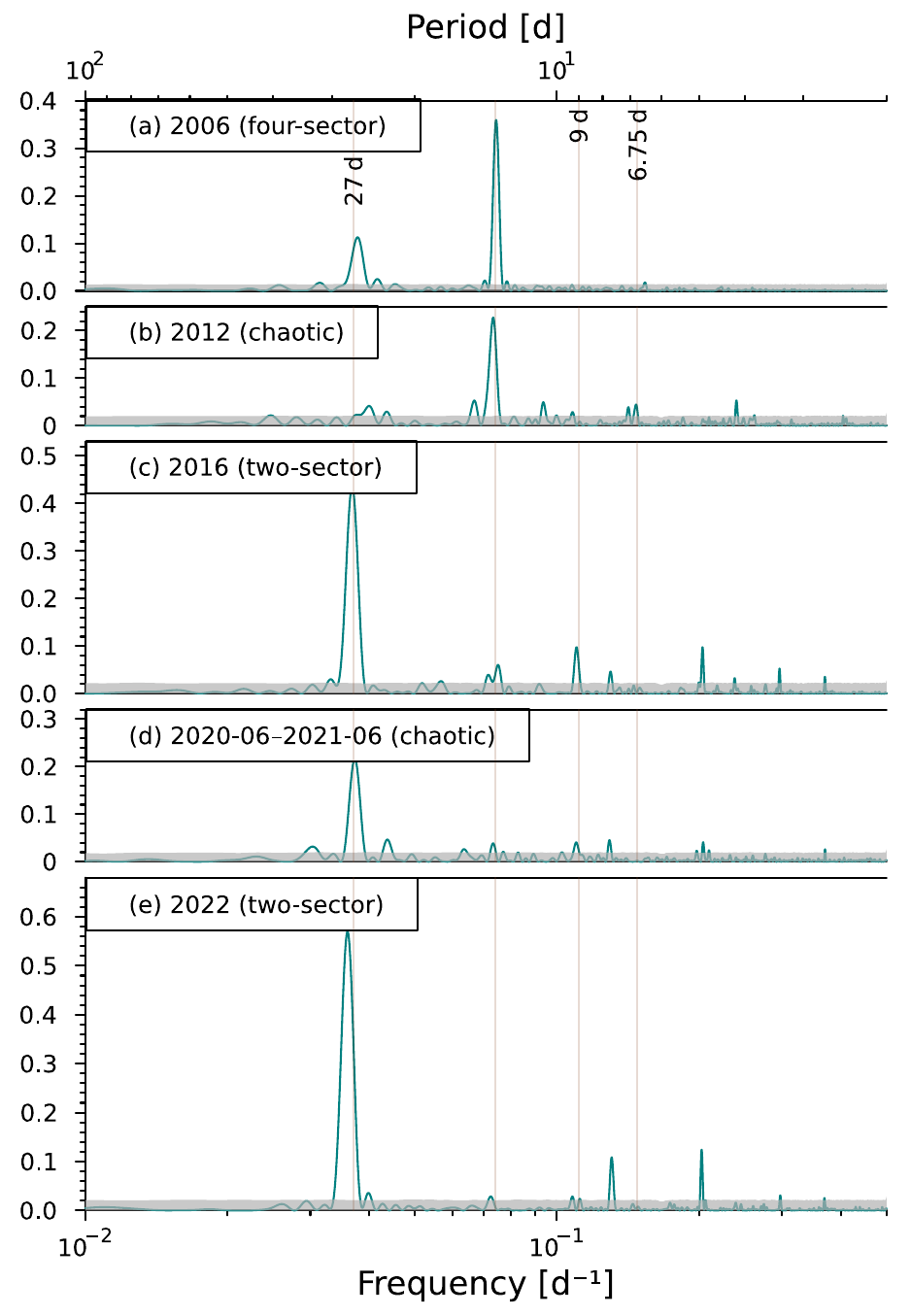}
    \caption{Periodograms for certain time periods corresponding to different sector structures of the IMF, the daily number of dust impacts with morphological types A and B (left, indigo curves), and the sector boundary structure (right, teal curves). All selected time periods last $1\,\si{yr}$. The axes of the periodogram amplitude are scaled identically within each column. The vertical lines are as in Fig.~\ref{fig:casestudies}. The estimated $95\%$ significance thresholds are indicated by grey shaded areas.}
    \label{fig:casestudy-hcs}
\end{figure*}

The periodograms for the respective time period of interest are displayed in Fig.~\ref{fig:casestudy-hcs} for both the daily number of dust impacts (left column) and the daily polarity of the IMF (right column). 
The IMF polarity spectra feature a strong solar rotation signature at the primary peak (about $27\,\si{d}$) when the sector structure is in a stable two-sector configuration (2016 and 2022; c and e, respectively). When the sector structure is in a stable four-sector configuration (2006, a), the first harmonic (about $13.5\,\si{d}$) is most powerful and the primary peak is less powerful. When the sector structure is chaotic (2012, 2020; b and d, respectively), these features are less powerful.

The dust impact spectra show no correlation with the IMF polarity spectra or with the sector structure stability. The solar rotation signatures can be both powerful  (2006, a) or weak (2016, 2022; c and e, respectively) during a stable sector structure, and also powerful (2012, b) or weak (2020-2021, d) during a chaotic structure. 
This indicates that there is no strong correlation between the observed dust impacts and the local IMF polarity.

\subsection{Dust impacts with morphological types C and D}\label{app:type_cd}

As mentioned in Sect.~\ref{sec:database}, the analyses of Sects.~\ref{sec:theory} \&~\ref{sec:fourier} only take into account dust impacts that generated signals with morphological types A and B. This is motivated by the unexplained features of the time series of dust impacts with types C and D before 2005, especially for the impacts observed by the $x$-antenna \citep[cf.][Fig.~5]{malaspina+wilson2016}. However, since 2005, the time series of dust impacts with types C and D does not show comparably exceptional features.

\begin{figure}
    \centering
    \includegraphics[width=\linewidth]{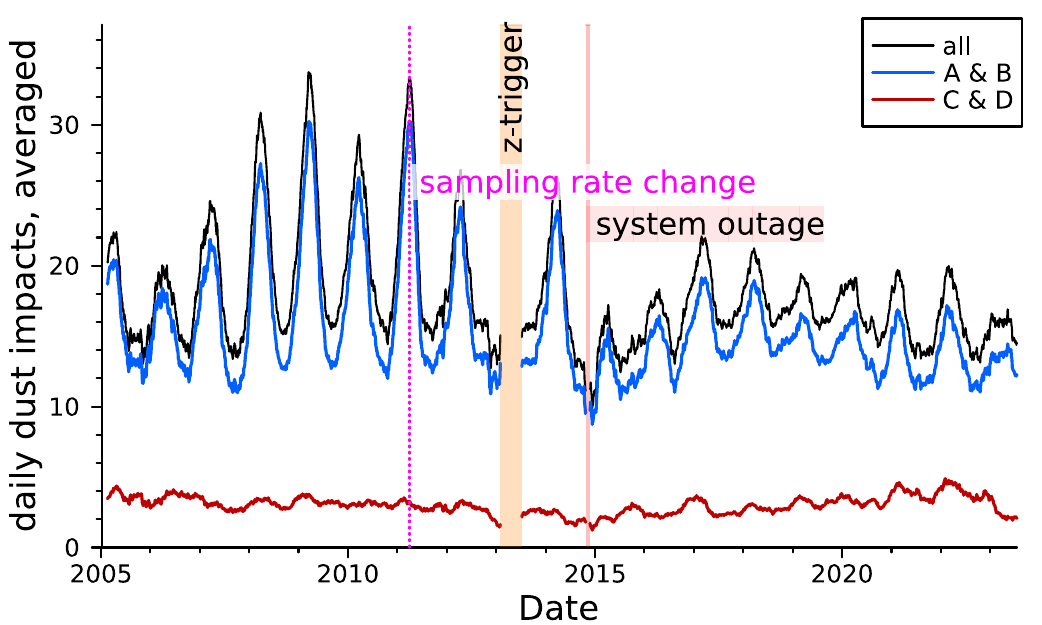}
    \caption{Daily number of dust impacts on Wind at $\mathrm{L}_1$ of all morphological types (black), of only types A and B (blue), and of only types C and (red), given as a centred moving average with a width of $91\,\si{d}$. The events are marked the same as in Fig.~\ref{fig:overview}a; the dataset has been corrected as per Sect.~\ref{sec:database}.}
    \label{fig:overview-cd}
\end{figure}

This is shown in Fig.~\ref{fig:overview-cd}, which compares the time-averaged daily number of dust impacts of types C and D with those of types A and B (see Fig.~\ref{fig:overview}a). Generally, dust impacts with types C and D occur less frequently than those of types A and B. Furthermore, the seasonal variation of dust impacts of types C and D is considerably less apparent than for types A and B, indicating that dust particles that generate signals of types C and D may predominantly be interplanetary in nature.

\begin{figure}
    \centering
    \includegraphics[width=\linewidth]{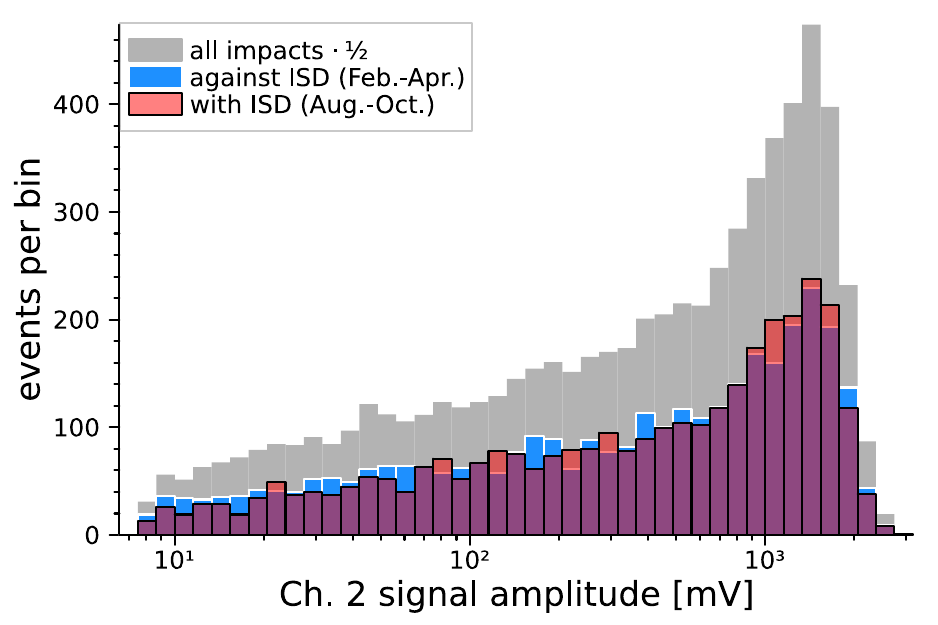}
    \caption{Amplitude distributions of all impacts (grey) with morphological types C and D measured by the $y$-antenna, impacts measured only when the spacecraft moved against the ISD inflow direction (February to April; blue and white), and impacts measured only when the spacecraft moved with the ISD inflow direction (August to October; red and black). The distribution of all impacts has been scaled by a factor of $0.5$ to tighten the histogram's vertical axis (see Fig.~\ref{fig:distr-ch2-againstwith}.)}
    \label{fig:distr-ch2-againstwith-cd}
\end{figure}

This is further supported by the amplitude distribution of the dust impacts with types C and D, which is shown in Fig.~\ref{fig:distr-ch2-againstwith-cd}. As in Fig.~\ref{fig:distr-ch2-againstwith} for types A and B, the amplitude distributions for the three months when the spacecraft moved against or with the ISD inflow direction, respectively, are graphed as well. The distributions when moving against and with the ISD inflow direction are strikingly similar for types C and D, indicating that these dust impacts predominantly stem from IDPs. Furthermore, in contrast to the amplitude distribution of signals of types A and B, which decreases at amplitudes above $A\gtrsim 200\,\si{mV}$, the amplitude distribution of types C and D increases with the amplitude over the entire amplitude range; the sudden decline at the very highest amplitudes ($A>1.5\times 10^3\,\si{mV}$) is most likely caused by a saturation of the instrument and not a physical feature of the amplitude distribution. 

\begin{figure}
    \centering
    \includegraphics[width=\linewidth]{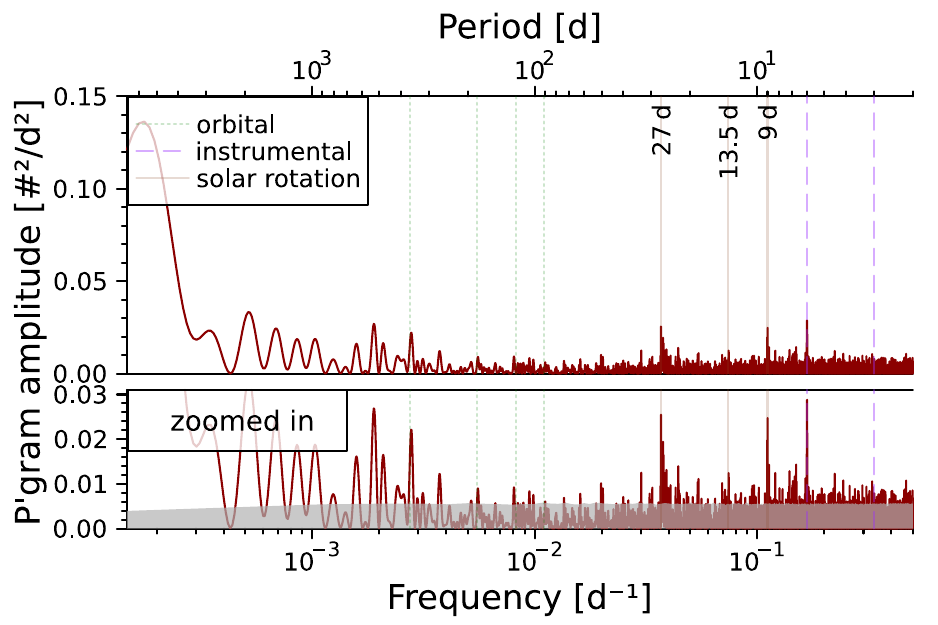}
    \caption{Periodogram of the daily number of dust impacts with morphological types C and D observed at $\mathrm{L}_1$, marked as in Fig.~\ref{fig:spectrum}. The bottom panel shows a zoomed-in view of the full spectrum with the estimated $95\%$ significance thresholds indicated by a grey shaded area.}
    \label{fig:spectrum-cd}
\end{figure}

Figure~\ref{fig:spectrum-cd} shows the periodogram of all dust impacts with types C and D observed at $\mathrm{L}_1$ (see Fig.~\ref{fig:spectrum} for the periodogram of dust impacts with types A and B). Compared to the spectrum of types A and B, the dust impacts of types C and D are less powerful because there are fewer dust impacts with types C and D than with types A and B. The spectrum of types C and D features a significantly smaller peak at $365\,\si{d}$; the seasonal variation is much less apparent, once again indicating the likely interplanetary nature of these dust particles. The solar rotation signatures, however, are clearly evident for impacts of types C and D. While the solar rotation harmonics occur at periods of $13.5\,\si{d}$ and $9\,\si{d}$ for types C and D, same as for types A and B, the primary occurs at $27.1\,\si{d}$ for types C and D, compared to $26.4\,\si{d}$ for types A and B. The physical origin of this frequency shift is unknown and requires further investigation.

One hypothesis for the origin of impact signals of types C and D is that they stem from extremely abrupt signals that are smoothed to the C and D morphologies by the Wind/WAVES instrument. It is possible that only the highest-amplitude impacts can be smoothed to C and D impact signals, which would explain the odd amplitude distribution (see Fig.~\ref{fig:distr-ch2-againstwith-cd}). This motivates excluding C and D impact signals for all analyses where the amplitude is of relevance. Another possibility is that C and D impact signals are not caused by dust signals at all; this motivates excluding C and D impacts outright. However, because the frequency analysis of C and D impacts (see Fig.~\ref{fig:spectrum-cd}) does show the solar rotation signatures, the veracity of this hypothesis is uncertain. Further investigation is required.

This analysis of dust impacts with morphological types C and D shows that both the time evolution, especially the seasonal variation, and the amplitude distribution of these dust impacts differs from those of types A and B. This motivates taking into account only dust impacts of types A and B for the analyses in Sects.~\ref{sec:theory} and~\ref{sec:fourier}. However, the solar rotation signatures are evident for types C and D as well as for types A and B.

\FloatBarrier
\section{Methodology of the scaled superposed epoch analysis}
\label{app:depletion}

The systematic reduction of dust impact observations during SIRs since 2005 has been investigated in Sect.~\ref{sec:depletion} by a superposed epoch analysis \citep{chree1913}. 
SIRs were taken from the datasets by \citet{jian+2006,jian2009} and \citet{hajra+2022}. The dataset by \citet{jian+2006,jian2009} differentiates between SIRs that are part of CIRs and SIRs that do not recur; only the SIRs that are associated with CIRs were taken into account. All events from the \citet{hajra+2022} dataset were included. The SIR datasets contain an initial and a final timestamp of each SIR, which determine the duration of the respective SIR.

By superposition, the rate of dust impacts of all morphological types during all SIRs was compared to the rate of dust impacts in the preceding and following time intervals. The superposed epoch analysis was performed as follows:
\begin{enumerate}[1.]
    \item The time intervals that precede (follow) the SIRs were each $2\,\si{d}$ long, ending (beginning) with the initial (final) timestamp of the respective SIR.
    \item The duration of each individual SIR was given by the initial and final timestamps from the datasets. Because these durations vary for the different SIRs, they were rescaled to the average duration, which is ${\sim}1.42\,\si{d}$ (see Fig.~\ref{fig:depletion_histoscaled}a).
    \item The timestamps of all dust impacts were reduced to the same time interval, i.e.\ all SIRs were superposed, beginning and ending at the same rescaled time.
    \item To calculate the number of dust impacts per day, first the number of dust impacts during a given time interval must be known. Therefore, a moving average, $N_{\mathrm{d}}(t)$, with a window width of either $\Delta t = 4\,\si{h}$ or $\Delta t = 30\,\si{min}$ was calculated over the superposed SIR, containing the total number of dust impacts during the sliding window, summed over all rescaled SIRs.
    \item The extrapolated number of dust impacts per day, $n(t)=N_{\mathrm{d}}(t)/\qty(N_{\mathrm{SIR}}\Delta t)$, is the total number of dust impacts, $N_{\mathrm{d}}(t)$, divided by the number of SIRs, $N_{\mathrm{SIR}}$, and the duration of the sliding window, $\Delta t$. 
    \item Measurement gaps were taken into account by rescaling: if a measurement gap of duration $t_{\mathrm{gap}}$ occurred during an event of total duration $t_{\mathrm{dur}}$, the duration of the sliding window was rescaled to $\qty(\Delta t)_{\mathrm{scal}}=\Delta t \cdot \qty(t_{\mathrm{dur}}-t_{\mathrm{gap}})/t_{\mathrm{dur}}$.
\end{enumerate}

Figure~\ref{fig:depletion_histoscaled}a shows the results of this scaled superposed epoch analysis of the dust impacts during SIRs for sliding windows of $30\,\si{min}$ and $4\,\si{h}$, including intervals of $95\%$ confidence. This confidence interval corresponds to the interval within $\pm 1.95\sigma$ around the distribution's mean, $n(t)$, where $\sigma(t)=\sqrt{n(t)\Delta t/N_{\mathrm{SIR}}}/(\Delta t)$ is the standard deviation of the average number of dust counts within the respective time window. 

The strength of the reduction or enhancement of dust impacts was quantified by dividing the average rate of dust impacts during the superposed SIR, $n_{\mathrm{dur}}$, by the average rate of dust impacts in the preceding and following time intervals, $n_{\mathrm{prc+flw}}$. Its uncertainty corresponds to $\pm 1.95\sigma$, where $\sigma=\sqrt{\sigma_{\mathrm{dur}}^2+\sigma_{\mathrm{prc+flw}}^2}$ with $\sigma_i=\sqrt{N_i/N_{\mathrm{SIR}}}/N_i$ for $i\in\{\mathrm{dur};\, \mathrm{prc{+}flw}\}$, and $N_i$ is the total amount of dust impacts measured during all SIRs or in all preceding and following time intervals. This method has been validated with synthetic data and applied in-depth to the dust impacts observed by Wind by Péronne et al.\ (in prep.).

\end{document}